\documentclass{article}
\usepackage{mathptmx}
\usepackage{amsmath}
\usepackage[scaled=.92]{helvet}
\usepackage{courier}
\usepackage[dvipsnames,svgnames,x11names]{xcolor}
\usepackage{underscore}

\usepackage{xspace}
\usepackage{enumitem}
\usepackage[normalem]{ulem}

\usepackage{listings}
\newcommand{\NaN}{\texttt{NaN}\xspace}
\newcommand{\NaNs}{\texttt{NaN}s\xspace}
\newcommand{\Inf}{\texttt{Inf}\xspace}
\newcommand{\Infs}{\texttt{Inf}s\xspace}
\newcommand{\ignore}[1]{}
\usepackage{url}
\usepackage[breaklinks,colorlinks,linkcolor=DarkGreen,citecolor=DarkRed,urlcolor=white!0!blue!99]{hyperref}

\usepackage{todonotes}
\makeatletter
\if@todonotes@disabled
    
\else
    
\fi
\makeatother

\begin{document}
\title{Proposed Consistent Exception Handling \\
for the BLAS and LAPACK}

\author{James Demmel, EECS and Math Depts., UC Berkeley \linebreak 
\and Jack Dongarra, ICL, U. of Tennessee, Knoxville \linebreak
\and \phantom{sometext}Mark Gates, ICL, U. of Tennessee, Knoxville\phantom{sometext} \linebreak 
\and \phantom{sometextsometext}Greg Henry, Intel Corp.\phantom{sometextsometext} \linebreak
\and Julien Langou, Dept. of Math. and Stat. Sci., U. Colorado Denver \linebreak
\and Xiaoye Li, AMCR Division, Lawrence Berkeley National Lab \linebreak 
\and Piotr Luszczek, ICL, U. of Tennessee, Knoxville \linebreak
\and Weslley Pereira, Dept. of Math. and Stat. Sci., U. Colorado Denver \linebreak
\and \phantom{sometextsometext}Jason Riedy, Lucata Corp.\phantom{sometextsometext}  \linebreak 
\and Cindy Rubio-Gonz\'alez, CS Dept., UC Davis
}

\date{\today}

\maketitle

\begin{abstract}
Numerical exceptions, which may be caused by overflow, operations like
division by 0 or sqrt($-1$), or convergence failures, are unavoidable in
many cases, in particular when software is used on unforeseen and
difficult inputs. As more aspects of society become automated, e.g.,
self-driving cars, health monitors, and cyber-physical systems more
generally, it is becoming increasingly important to design software that
is resilient to exceptions, and that responds to them in a consistent
way. Consistency is needed to allow users to build higher-level software
that is also resilient and consistent (and so on recursively). In this
paper we explore the design space of consistent exception handling for
the widely used BLAS and LAPACK linear algebra libraries, pointing out a
variety of instances of inconsistent exception handling in the current
versions, and propose a new design that balances consistency,
complexity, ease of use, and performance. Some compromises are needed,
because there are preexisting inconsistencies that are outside our
control, including in or between existing vendor BLAS implementations,
different programming languages, and even compilers for the same
programming language. And user requests from our surveys are quite diverse.
We also propose our design as a possible model for other
numerical software, and welcome comments on our design choices.
\end{abstract}

\tableofcontents

\section{Introduction}

Sometimes it takes an event like the crash of the Ariane 5
rocket~\cite{arnold2000ariane5}, a naval propulsion
failure~\cite{slabodkin1998navy}, or a crash in a robotic car
race~\cite{reddit2020driverless} to make people aware of the importance
of handling exceptions correctly in numerical
software~\cite{demmel2021ieee754,HuckleNeckel2019}! As applications like
self-driving cars, health monitors, and cyber-physical systems more
generally become widespread, society's dependence on the correctness of
these applications will only become more apparent. Since many of these
applications are built using lower-level building blocks, often
including linear algebra, it is clear that these building blocks need to
be resilient to exceptions (e.g., still terminate), and respond to
exceptions in a predictable, consistent way (e.g., report certain
exceptions on exit) to allow higher-level applications using them to be
resilient as well.

In this paper, we will explore the design space of ways to make
exception handling more resilient and consistent, in particular for the
widely used BLAS~\cite{blasnetlib} and LAPACK~\cite{lapacknetlib} linear
algebra libraries.
While these have been widely used for decades, it turns out they do not
handle exceptions consistently in a number of ways that we will describe
later. We explore this design space because there is no single best
solution for a number of reasons:

\begin{enumerate}


\item Based on our user surveys, in which 67\% of respondents said
exception handling was important or very important, there is no single
approach that meets all needs. These needs range from users who want the
behavior of interfaces to change as little as possible (because of large
amounts of existing code) to users who want more fine-grained control over the
way exceptions are handled or reported.
 And not all users may agree on the definition of
``consistency''. Consider an upper triangular matrix with an \Inf~or \NaN~
entry above the diagonal. Are its eigenvalues well-defined, being just
the diagonal entries, or not? What about a diagonal matrix with an \Inf~
or \NaN~on the diagonal? Matlab currently chooses to return with a
warning, and no eigenvalues are reported.
Roughly speaking, user wishes for consistency fall into 3 (related) categories:
correct mathematical behavior (eg, what an
eigenvalue means, as above), propagation (eg,
a \NaN that is input to or created during a
subroutine call should propagate to the output unless there is a mathematical reason
that it shouldn't), and reporting (eg, using LAPACK's INFO parameter,
or new mechanisms, to flag exceptions).
We present these user requests in more detail later.
Some of these needs can be met
by having different ``wrappers'' that offer different interfaces and
semantics to the users who want them, some by new subroutine arguments, and some by new routines
that let users choose options
for subsequent LAPACK calls.

\item The BLAS and LAPACK are in turn based on lower-level
building blocks, on whose (hopefully mostly) consistent behavior theirs
depend. We will describe these building blocks, and their potential
inconsistencies that we need to account for. The first building block is
the IEEE 754 floating point standard, which has evolved over time. The
standard recommends (but does not require) an extended precision format. For example,  an 80 bit format
is still in the x86 architecture.
Inconsistencies could arise if the compiler chooses to
keep some intermediate results in this longer format, so our solution
must accommodate this and other uncertainties. A correction was made
in the faulty definition of min and max in the most recent 2019 version
of the standard~\cite{ieee754standard2019,hough2019ieee754}, to assure
that min and max are associative
when some arguments are \NaNs. Higher-level language standards (Fortran
and C), on which BLAS and LAPACK depend, are currently modifying their
specifications to include these new definitions, but it will take a
while to propagate into compilers. More generally, the BLAS and LAPACK
are evolving to use multiple programming languages that may define
operations differently; besides min and max, this could include the
product or quotient of complex scalars. Indeed, different compilers for
the same language may also generate different code, as could the same compiler provided different optimization leeway.

\item There are vendor-tuned BLAS implementations that may have
different exceptional behavior. In some cases, this will mean that we
will need to update the reference BLAS (\textit{e.g.}, see IxAMAX below) to
assure consistency, and encourage vendors to adopt these new BLAS,
including providing updated test code. In other cases that may depend on
architecture-specific optimizations, it means that we will compromise on
what consistency means, \textit{e.g.}, accepting that an exception can propagate
either as an \Inf~or a \NaN, as long as it propagates somewhere in the
result.


\item There is a cost/consistency tradeoff, with the most rigorous
definitions of ``consistency'' potentially taking much more runtime.
For example, the most rigorous definition of consistency would insist on
bit-wise reproducibility, of numerical results and exceptions, from run
to run of the same code on the same platform. On modern parallel
architectures, where operations may be scheduled dynamically and so
their order, \textit{e.g.} summation order, may change from run to run, bit-wise
reproducibility is not guaranteed. A simple example is summing the 4
finite numbers {[}x,x,-x,-x{]}, where x+x overflows; depending on the
order of summation, the result could be +\Inf, -\Inf, \NaN,~or 0 (the
correct result). Solutions for reproducible summation have been
proposed, which work independently of the order of summation, but at a
cost of being several times slower~\cite{ahrens2020reproblas}, although
with opportunities for hardware
acceleration~\cite{ieee754standard2019,hough2019ieee754}.  Intel also
provides a version of
their Intel(R) Math Kernel Library (Intel(R) MKL) with CNR = Conditional
Numerical Reproducibility, which guarantees deterministic, and so
reproducible, execution order for their multicore platforms, also with a
potential performance slowdown. We leave bit-wise reproducibility to
future work building on \emph{consistent} handling of exceptional cases.

\end{enumerate}

Section~\ref{sec:explore} explores a variety of existing exception handling inconsistencies and possible solutions.
Section~\ref{sec:ieeearith} considers IEEE arithmetic and its differing implementations.
Section~\ref{sec:proglang} considers 
programming languages and compilers.
Both sections~\ref{sec:ieeearith} and
\ref{sec:proglang} explore inconsistencies
over which we have no control, and must
accommodate.
Section~\ref{sec:blas} considers the BLAS,
and discusses both situations in which \NaNs
and \Infs are not expected to propagate, and
where they propagate incorrectly in the
reference BLAS, with proposed corrections.

Sections~\ref{sec:LAPACKex} and \ref{sec:LAPACK} 
address LAPACK. Section~\ref{sec:LAPACKex}
includes existing propagation
failures (which are the result of 
propagation failures in the reference BLAS),
an example where exceptions are deliberately
and correctly not propagated, failures
caused by incorrectly checking for 
exceptional inputs by the LAPACKE C-interface
for LAPACK,
and failures caused by integer overflow. 
Next, section~\ref{sec:LAPACK} presents 
a detailed proposal for error 
reporting using INFO and other new proposed
mechanisms. This is by far the most
complicated part of the proposed design,
having gone through many design iterations,
so we also list the many ``inconsistent''
user requests for consistency that we
received, sketch the previous design iterations 
to see why we ended up with the latest version,
and point out some potential small changes to
the current design on which user feedback is
welcome.

Section~\ref{sec:testconsist} describes how to generate test code to verify that
our solutions work, including using the concept of ``fuzzing'', which in
our case means (perhaps randomly) inserting exceptional values in the middle of
execution, even if they would not appear during a regular execution.
``Fuzzing'' will help test whether exceptions are propagated and handled
correctly no matter where they appear during an execution. 
We also discuss the
challenges of using proof-based techniques to verify consistency.

Finally, in Section~\ref{sec:tasks} we lay out a sequence of proposed
improvements to make, sorted in order of priority. Comments are welcome.

We provide appendices with more details:
Appendix~A provides a corrected implementation of
I\{C,Z\}AMAX, Appendix~B gives test cases for the BLAS,
including examples where several current
BLAS implementations do not comply with
the proposed new standard,
and Appendix~C gives the source code for the
new version of SGESV with error checking (and all the
routines in its call tree).

Here is some common notation that we will use later. OV is the overflow
threshold in IEEE 754 arithmetic, \textit{i.e.}, the largest finite number (its
magnitude depends of course on whether we are using single or double
precision arithmetic, but the discussions below apply to either). UN is
the underflow threshold, i.e., the smallest positive normalized number.

\section{Exploring existing inconsistencies, obstacles, and possible solutions}
\label{sec:explore}

We present a variety of examples of inconsistencies, possible
solutions, and obstacles at various levels in the stack, from the
computer arithmetic to LAPACK. We mentioned a few of these above but go
into more detail here. This list is not exhaustive, and indeed rigorous
testing (or proofs) are required to have confidence that nearly all
possibilities have been found. We return to that topic in
Section~\ref{sec:testconsist}.

\subsection{IEEE Arithmetic}
\label{sec:ieeearith}

Nearly all numerical software depends on the semantics of IEEE 754
floating point arithmetic, including the BLAS and LAPACK. The growing
use of shorter formats motivated by machine learning, \textit{e.g.},
bfloat16~\cite{bfloat16wikipedia}~\cite{intel-bf16}, and their adoption for mixed
precision linear algebra algorithms~\cite{doi:10.1177/10943420211003313}
are an important exception, but we leave that to
future work since the BLAS and LAPACK do not yet use these formats. In
addition to changes affecting exception handling in the latest standard \cite{ieee754standard2019},
the standard allows some flexibility in the semantics of operations used
to evaluate arithmetic expressions, which can affect exception handling.

Section 3.7 of the 754 standard recommends, but does not require, that
extended precision formats, those wider than the usual single or double
formats, be available, and gives lower bounds on how many more exponents
and mantissa bits they should have, but does not specify exactly how
many. A common implementation of this was the 80-bit format, which is
still in the x86 architecture, but most compilers (not all) will
generate code using the faster SSE instructions using the standard
64-bit format. Obviously, if a compiler decides to execute some subset
of the operations in extended precision with different underflow and overflow thresholds, along with additional operations to round intermediate results to
and from the 64-bit format, there may be different exceptions generated.
This makes reproducibility of exceptions intractable across different
compilers, compiler flags, and/or architectures, but should not prevent
us from still reporting exceptions that do occur in a consistent
way.

The same comments apply to other semantic possibilities, including 
\begin{enumerate}
\item using fused-multiply-add $a\times b+c$, where an overflow exception depends
on the final value but not $a\times b$, 
\item whether underflow is
detected before or after rounding, 
\item the use of the default gradual
underflow vs flush-to-zero, which could change whether $c/(a-b)$ signals
divide-by-zero or not, and 
\item changing rounding modes, which may impact
whether a final result is rounded down to the overflow threshold OV or
up, causing an overflow.
\end{enumerate}

One important change (a bug fix) in the 2019 standard are the added
definitions of the operations min(x,y) and max(x,y), which are specified
to return \NaN~if either operand is a \NaN, so that they are associative
and propagate \NaNs. The 2008 standard did not define these, just the
related operations minNum, maxNum, minNumMag, and maxNumMag, but these
were defined in a way that was not associative, and might or might not
propagate \NaNs~\cite{ieee2019nanpropag}. The 2019 standard changed
the definitions of these related operations to make them associative
too. Eventually, these new definitions will find their way into language
standards (the C and Fortran standards committee are working on it) and
compilers, but this will take a while, so in the meantime, we need to
make sure we don't rely on them to propagate \NaNs.

Finally, we will {\em not} depend on the five IEEE754 exception flags, which
indicate whether an exception (invalid operation like sqrt(-1), division
by 0, overflow, underflow, inexact) has occurred since the corresponding
flag was the last reset. The 2008 and 2018 Fortran standards defined how
to access these flags, but said that whether they were provided was
implementation and hardware dependent. 
Similarly, the C99 standard\cite{ANSI:1999:AII} provides for a floating-point
environment within which exceptions can be examined, but again
its availability depends on the implementation.
And while they may be available on one processor,
if some routines (like the BLAS) are implemented on an accelerator (like
a GPU) without access to the flags, then we can't depend on them. If
available, these flags can provide useful but different information than
our proposed checks, and users are of course welcome to use them to see
if any exception has occurred during the execution of a routine of
interest. But they will not signal whether the user is passing an input
containing an \Inf~or (quiet) \NaN to a routine, or necessarily whether
such a value propagates to the output or not, since operations like
3+(quiet)\NaN~(or 3+\Inf) return \NaN~(or \Inf) without signaling an
exception. Of course, we rely on semantics like 3+\NaN=\NaN to guarantee
that exceptions propagate to the output, i.e., are not ``lost''.

For further discussion of IEEE 754 and exceptions,
see~\cite{ieee2019except,hauser1996numexcept}.

\subsection{Programming languages and compilers}
\label{sec:proglang}

In addition to different programming languages and compilers choosing
different ways to provide IEEE 754 features as described above,
different programming languages and compilers may also implement basic
mathematical operations in different ways, with different exception
behavior. Beyond min and max as mentioned above (which may or may not
currently propagate \NaNs), there are also the absolute value, product,
and quotient of complex numbers. When implemented using their textbook
definitions, they are susceptible to over/underflow even when the final
result is innocuous. The Fortran standard explicitly does not specify
how to implement intrinsic arithmetic operations. However, the 2008 and
2018 Fortran standards do say that the absolute value of a complex
number should be ``computed without undue overflow or
underflow''~\cite{fortran2008standard}.

A discussion generated by a recent bug report~\footnote{https://github.com/Reference-LAPACK/lapack/issues/575} proved that some of the LAPACK tests may fail depending on the intrinsic Fortran ``abs'' and ``/'' functions for complex numbers. For instance, a test for ZLAHQR fails due to underflow depending on the algorithm used for complex ``abs''.
%
Regarding complex division, there is already an LAPACK routine
\{C,Z\}LADIV to divide two complex numbers
carefully~\cite{priest2004scalecomplex,baudin2012complexdiv}.
{ZLADIV}~is used 44 times in 10 different LAPACK routines (in the SRC
directory of LAPACK version 3.10.0). But there are a number of complex divisions done in
routines using the built-in ``/'' operator. Two ways forward are (1)
adding test code to the LAPACK installation package to see whether ``/''
behaves correctly when dividing very large or very small complex
arguments to provide a warning if it does not, and (2) converting
all complex divisions to use \{C,Z\}LADIV, with a possible performance
penalty. One could also have two versions of these routines, one using
\{C,Z\}LADIV and one using ``/'', and decide which one to install based
on the results of the tests.
The analogous reasoning applies to the LAPACK \{C,Z\}LAPY2 that is at least as accurate as the intrinsic Fortran ``abs''.
LAPACK 3.10.1 applies the strategy (1).
Since version 3.10.1, LAPACK has a set of programs, executed at build time, that is used to verify the accuracy of the Fortran compiler when computing absolute value, product, quotient, min and max of complex numbers. Those programs also verify if Inf and NaN propagate to the output.

There is also disagreement on how to represent an ``infinite'' complex number.
The C99 and C11 standards~\cite[Annex G.3]{iso2011c} define a complex number to be
``infinite'' if either component is infinite, even if one component is a
\NaN, and define multiplication so that infinite*(finite-nonzero or
infinite) is infinite, even if the 754 rules would yield both components
of the product being \NaN. For example, straightforward evaluation of
complex multiplication yields (\Inf~+ 0*i)(\Inf~+ \Inf*i) = \NaN~+ \NaN*i,
but the C-standard-conforming compiler yields (\Inf~+ 0*i)(\Inf~+ \Inf*i) =
\Inf~+ \Inf*i. The C standard includes a 30+ line procedure for complex
multiplication~\cite[Annex G.5.1]{isoiec2011cstd}.  While
compilers will adopt this as a default, the compilers also
provide options for using other definitions (\textit{e.g.}
\texttt{gcc}'s \texttt{-fcx-limited-range}\footnote{\url{https://gcc.gnu.org/onlinedocs/gcc/Optimize-Options.html\#Optimize-Options}}).
There are similar rules for
complex division provided in the C standard working
draft~\cite[Annex G.5.1]{isoiec202xn2478} although no
procedure is provided in the C standard document. In contrast, the IEEE
754 standard and 2008 and 2018 Fortran standards say nothing about
handling exceptions in complex arithmetic, or how intrinsic operations
like complex multiplication and division should be implemented. This is
obviously a challenge for consistent exception handling as well. To
accommodate this, we propose to allow exceptions to propagate either as
\Infs~or \NaNs, since either one provides a warning to the user.

The C++ standard included complex numbers as a template class available for
three floating-point types float, double, and long double.  The behavior of
that class for other types is unspecified but most implementations permit
integral template arguments which results in compile-time errors when a
floating-point complex value is combined with an integral one.
{For instance, std::abs applied to \texttt{std::complex<T>}( \Inf~+ \NaN*i ) returns \Inf~if T is either float, double, or long double. However, the same function call returns a \NaN~if T is the multiprecision type mpfr::mpreal from \cite{mprealgithub}. We obtain the same pattern for (-\Inf~+ \NaN*i), (\NaN~+ \Inf*i) and (\NaN~- \Inf*i). Moreover, if T = mpfr::mpreal, std::abs returns a \NaN~for any of the inputs ($\pm$\Inf~+ \Inf*i), (\Inf~$\pm$\Inf*i), ($\pm$\Inf~+ 0*i), and (0 $\pm$\Inf*i), and it returns 0 for the input (0 + \NaN*i).
Another curious operation is complex division. For each of the standard types, float, double and long double, the divisions ( 0 + 0*i )/( $\pm$\Inf~+ \NaN*i ) and ( 0 + 0*i )/( \NaN $\pm$\Inf*i ) results in the complex ( 0 + 0*i ).\footnote{Behavior of complex abs and complex division. Tests on Ubuntu 18.04.5 LTS (5.4.0-70-generic) with GNU compilers, GCC version 7.5.0.}    }
Even the most
recent C++ standard draft specifies the complex number constructor
postcondition in terms of the equals operator which will never be satisfied for \NaN~
inputs~\cite[\S26.4.4]{isoiecn4842}. Neither of the standard C++ library
complex template classes nor its specializations include the details of
handling of exceptional floating-point values. One of the ways to address the
lack of compiler support for built-in complex data types is to provide a custom
implementation to be used instead of the standard
\textless{}complex\textgreater\ header.

And if one were to use Gauss's algorithm to multiply complex numbers
using 3 multiplies and 5 additions instead of 4 multiplies and 2
additions, exceptions could occur differently again. While Gauss's
algorithm is unlikely to provide a speedup when multiplying complex
scalars 
 it could when multiplying complex matrices, since the cost of matrix
multiplication is much larger than matrix
addition~\cite{scott1992oocintel}.
This leads us to consider the BLAS.

\subsection{BLAS}
\label{sec:blas}

We consider just the standard BLAS~\cite{blasnetlib}, which perform a single
operation, not the batched BLAS~\cite{BatchBLAS2021}, which may perform many operations with
a single call. We propose that analogous consistency requirements apply
to the batched BLAS as well. 
Our goals are to identify consistent rules for propagating (or not) \Infs and \NaNs
from the input or intermediate results to
the output, and to avoid creating those values by appropriate scaling when possible
\cite{anderson2017l1scaling,bindel2002givens}. The BLAS do
check for illegal integer and character
inputs (\textit{e.g.} negative dimensions) and report
these using XERBLA, which then terminates
execution. Termination is not the right
response to floating point exceptions,
so we will perform reporting at the LAPACK
level, as described in section~\ref{sec:LAPACK}.

\subsubsection{How to interpret alpha = 0 or beta = 0 in C = alpha*A*B + beta*C, and other BLAS}

GEMM in the reference BLAS, and presumably optimized versions,
interprets alpha=0 as meaning that the operation to be performed is C =
beta*C. In other words, A and B are not accessed, and no \Infs~or \NaNs
that could be present in A or B are propagated. This is the expected
semantics, and so it is ``consistent'' not to propagate \Infs~and \NaNs~in
this case. Similar comments apply to beta=0, where the intended
semantics are C = alpha*A*B, so that \Infs~and \NaNs~in C are not expected
to be propagated. Analogous comments apply to many other BLAS routines,
including various versions of matrix-matrix multiplication,
matrix-vector multiplication (not all), low-rank updates, triangular
solve (only TRSM, not TRSV), and AXPY. Interestingly, SCAL (x = alpha*x)
does not check for alpha = 0 or 1.

The GraphBLAS~\cite{buluc2019graphblasapi}, a specification supporting
graph algorithms using sparse
linear algebra (more or less), demonstrates one alternative design. The
GraphBLAS does not have alpha or beta scalar parameters or the transpose
parameters. Instead, the GraphBLAS includes masks and descriptors in
each operation, e.g.~GrB\_mxm (SpGEMM). The mask controls updates
on each matrix element / graph edge and does not fit into the dense BLAS
well. The descriptor is an opaque object that has fields controlling
how C is updated and whether A and B are transposed. C can be
completely cleared before the final result is written or updated ``in
place.'' In both cases, the input entries of C are accumulated
through a user-defined function. Setting that function to GrB\_NULL
ignores the entries. So setting the descriptor field GrB\_OUTP to
GrB\_REPLACE and passing GrB\_NULL for the accumulator is the
GraphBLAS equivalent to beta = 0 in the BLAS. There is no equivalent
for alpha = 0. The GraphBLAS specification spends much more text on
describing the options than the dense BLAS's documentation of alpha = 0
and beta = 0 above.

\subsubsection{I\{S,D,C,Z\}AMAX and \{S,D,C,Z\}NRM2}
\label{sec:amaxnrm2}

The AMAX routines take a vector as input and return the index of the
(first) entry of the largest absolute value. Instead of using abs(z),
the complex versions use abs(real(z)) + abs(imag(z)), because it is
cheaper to compute, less susceptible to over/underflow, and adequate for
many purposes, \textit{e.g.}, pivot selection in Gaussian elimination.

The straightforward reference implementation of the real (single
precision / binary32) version used in the reference BLAS

\begin{lstlisting}
isamax = 1
smax = abs(A(1))
do i = 2:n
   if (abs(A(i)) > smax) then
     isamax = i
     smax = abs(A(i))
   end if
end do
\end{lstlisting}

fails to behave consistently on the following permuted inputs:

\begin{lstlisting}
ISAMAX([0,NaN,2]) = 3
\end{lstlisting}
and
\begin{lstlisting}
ISAMAX([NaN,0,2]) = 1
\end{lstlisting}

This is because a comparison like x \textgreater{} y always returns
False if either argument is a \NaN. There are various ways to define the
semantics to behave in a consistent manner despite exceptional inputs.
Since we want to both propagate exceptions, and point to the ``same
value'' independent of the order of the inputs (we explain why ``same
value'' is in quotes below), we propose the following semantics:
I\{S,D\}AMAX should return

\begin{enumerate}
    \item  the index of the first \NaN, if the input contains a \NaN, else
    \item the index of the first \Inf~or -\Inf, if the input contains an \Inf~or
    -\Inf, else
    \item the index of the first finite value of the largest absolute value.
\end{enumerate}
``Same value'' is in quotes because all \NaNs~are treated as equal,
\textit{i.e.}, it ignores the value of their mantissa fields, which could in
principle contain information useful for debugging. But since this
feature is seldom used, and there is no way to prioritize one \NaN~over
another, we treat them all as equal. Here is a possible implementation
with the desired semantics:

\begin{lstlisting}
smax = abs(A(1))
isamax = 1
if (isnan(smax)), return ! return index of first NaN
do i = 2:n
   if (.not.(abs(A(i)) <= smax)) then
      ! either A(i) is a NaN or abs(A(i)) > smax
      smax = abs(A(i)), isamax = i
      if (isnan(smax)), return
      ! return index of first NaN
      end if
end do
\end{lstlisting}

The complex versions I\{C,Z\}AMAX are more problematic because even in
the absence of exceptional inputs, overflow can cause inconsistent
outputs. For example, if z1 and z2 both have the property that
abs(real(zi))+abs(imag(zi)) overflows, then they will be treated as
equal (to \Inf) even if they are not, with I\{C,Z\}AMAX({[}z1,z2{]}) and
I\{C,Z\}AMAX({[}z2,z1{]}) both returning 1. See Appendix~\ref{AppA} for a
``simple'' ($\approx$30 lines) correct implementation, and some
performance data comparisons with the existing ICAMAX.

We note that there are analogous LAPACK routines I\{C,Z\}MAX1 with
similar functionality (and inconsistencies) that use abs(z) instead of
abs(real(z))+abs(imag(z)).

See Section~\ref{sec:sgesv} for an example where ISAMAX helps cause a
\NaN~to fail to propagate in Gaussian elimination. See Appendix~\ref{AppB} for test results
showing how I\{S,D,C,Z\}AMAX 
does not comply with the proposed new standard for
several existing BLAS implementations.

The \{S,D,SC,DZ\}NRM2 routines, which compute the 2-norm of a vector,
have similar but less serious issues
\cite{Blue1978}.
If the input vector contains two
or more \Infs~but no \NaNs, the routines will divide \Inf/\Inf~and return a
\NaN. So an exceptional value does propagate, but not the expected
one. Version 3.9.1 of LAPACK routines \{S,D,C,Z\}LASSQ have the same
problem, but this is fixed in LAPACK version 3.10, based on safe scaling
method in Level~1 BLAS~\cite{anderson2017l1scaling}. 
Again, see Appendix~\ref{AppB} for test
results.

We note that \cite{anderson2017l1scaling} proposes a different way to handle \NaNs~``consistently'' in I\{S,D,C,Z\}AMAX, returning the index of the largest non-\NaN~entry, or 1 if all entries are \NaNs. This is proposed for consistency with Fortran's MAXLOC and MAXVAL, and with Matlab and R, or more generally when \NaN~is interpreted as ``missing data''. This differs from our approach, in which we want to make sure that exceptions propagate.

\subsubsection{TRSV and TRSM}

The TRSV routine in the reference BLAS solves a triangular system of
equations $Tx = b$ for $x$; $T$ may be upper or lower triangular, and unit
diagonal ($T(i,i)=1$) or not. One may also ask TRSV to solve the
transposed linear system $T^Tx=b$. The reference TRSV returns $x =
\begin{bmatrix}
    1 \\
    0
\end{bmatrix}
$ when asked to solve $U*x = b$ with $U=
\begin{bmatrix}
    1 & \mathrm{NaN} \\
    0 & \mathrm{NaN}
\end{bmatrix}
$ and $b=
\begin{bmatrix}
    1 \\
    0
\end{bmatrix}$, because it checks for trailing zeros in $b$ and does not
access the corresponding columns of $U$ (which it would multiply by zero).
More generally, TRSV overwrites $b$ with $x$ and checks for zeros appearing
anywhere in the updated $x$, to avoid multiplying the corresponding
columns of $U$ by zero. This means solving $Ux = b$ with $U =
\begin{bmatrix}
    1 & \mathrm{NaN} & 1 \\
    0 & 1   & 1 \\
    0 & 0   & 1
\end{bmatrix}
$
and $b = 
\begin{bmatrix}
    2 \\
    1 \\
    1
\end{bmatrix}$ yields $x =
\begin{bmatrix}
    1 \\
    0 \\
    1
\end{bmatrix}$. So in
both cases, the \NaNs~in U do not propagate to the result $x$ (neither
would an \Inf, which if multiplied by zero should also create a \NaN).
However, if $L$ is the 2-by-2 transpose of the first $U$ above, then calling
TRSV to solve $L^Tx=b$, with $b =
\begin{bmatrix}
    1 \\
    0
\end{bmatrix}$ returns $x=
\begin{bmatrix}
    \mathrm{NaN} \\
    \mathrm{NaN}
\end{bmatrix}$.
This is the same linear system as above, but the reference
implementation does not check for zeros in this case; this is
inconsistent. Again, one could imagine vendor TRSV behaving differently
(in Matlab, the \NaN~does propagate to the solution in these examples).
In these cases, if \NaN~were interpreted to mean ``some unknown but
finite number'', so that 0*\NaN~was always 0, then not propagating the
\NaN~would be reasonable. But if \NaN~meant ``some unknown, and possibly
infinite number'', then 0*\NaN~should be a \NaN~(the default IEEE 754
behavior, which we assume), and not propagating the \NaN~is incorrect.
Analogous comments apply to TRSM.

There are several ways to make exception handling consistent. The first
and simplest way is to disallow all checking for zeros. The second way
is to allow checking only for leading zeros in $b$ when solving $Lx=b$
(or $U^Tx=b$, or trailing zeros in $b$ when solving $Ux=b$, or
$L^Tx=b$). The reason for the second option is that some users may
expect the semantics of solving $Lx=b$ to mean solving a smaller (and
possibly much cheaper) linear system when b has trailing zeros, much as
they expect alpha = 0 or beta = 0 to affect the semantics of C =
alpha*A*B + beta*C. In contrast, zeros appearing after the first nonzero
entry in $x$ could be the result of cancellation, and so not something the
user can expect in general. But they could also be a result of the
sparsity pattern of $L$ and $b$, for example, if $L$ is block diagonal, and
$b(i)$ is nonzero only for indices $i$ corresponding to a subset of the
diagonal blocks. On the other hand, optimized versions of TRSM, that
apply operations like GEMM to subblocks of matrices, may check for zeros
in different (blocked) ways, or not at all (e.g., in MKL), for
performance reasons.

We propose to take the first approach, disallowing all zero checking,
because it is the simplest, and ensures consistency between TRSM and
TRSV. This leaves the user the option of checking for leading or
trailing zeros themselves and simply changing the size of the linear
system in the call to TRSV or TRSM. To support this we will provide
simple routines to return the index of the first or last nonzero entry
of a 1D-array, or the first or last nonzero row of a 2D-array; the
latter is most compatible with blocked optimizations of TRSM. Some
versions of these already exist, for the last nonzero row or column of a
2D-array, ILA\{S,D,C,Z\}L\{R,C\}.

Analogous comments apply to TPSV and TBSV.

See Section~\ref{sec:sgesv} for an example where TRSV helps cause a \NaN
to fail to propagate in Gaussian elimination.

\subsubsection{GER, SYR and related routines}

The GER routine computes $A = A + \mathrm{alpha}*x*y^T$, where $A$ is a matrix,
alpha is a scalar, and $x$ and $y$ are column vectors. In the reference
implementation of GER, if alpha = 0, the code returns immediately, so no
\Infs~or \NaNs~in $x$ or $y$ propagate to $A$. If alpha is nonzero, and if any
$y(i) = 0$, the code skips multiplying $y(i)$ by alpha and $x$. But it does
not check for zeros in $x$. So an \Inf~or \NaN~in $y(i)$ will propagate to all
entries in column $i$ of $A$, but an \Inf~or \NaN~in $x(j)$ may not propagate to
all entries in row $j$ of $A$. And if all $y(i)=0$, then no \Inf~or \NaN~in $x(j)$
will propagate; this is inconsistent. Vendor GER may again be different.
Our proposal for consistent exception handling would allow checking for
alpha=0, but not checking for zeros inside $x$ or $y$. A mathematically
equivalent but possibly faster implementation would scan $x$ for any \Infs
or \NaNs~when the first $y(i)=0$ is encountered, to decide whether it is ok
to skip multiplications by zero entries of y.

The SYR routine, for $A = A + \mathrm{alpha}*x*x^T$ for symmetric $A$, has a
worse inconsistency: The user needs to choose whether to update the
upper or lower half of A. Since the code only checks for zero entries of
$x^T$, this means a numerically different answer can be computed if $x$
contains 0s and \NaNs, depending on whether the upper or lower half is
updated.

Searching for IF statements that compare to zero indicate that the
above comments also apply to the following BLAS routines (and their
complex counterparts): \{S,D\}SPR\{,2\}, \{S,D\}SYR\{,2,K,2K\},
\{S,D\}\{TB,TP\}\{M,S\}V, and \{S,D\}TR\{M,S\}\{M,V\}.

There are various other BLAS1 (DOT), BLAS2 (GEMV, GBMV, S\{Y,B,P\}MV)
and BLAS3 (GEMM,SYMM) routines that could check for zeros, but currently
do not, and should continue not to do so.

See Section~\ref{sec:sgesv} for an example where GER helps cause a \NaN
to fail to propagate in Gaussian elimination.

\subsubsection{Givens rotations}

Both the BLAS (xROTG, xROTGM) and LAPACK (xLARTG, \{S,D\}LARTGP) have
routines to compute Givens rotations. LAPACK introduced its own for two
reasons. First, the semantics are slightly different, \textit{e.g.} how signs of
the output are chosen, as needed by the routines that call them. Second,
they are designed more carefully to avoid
over/underflow~\cite{bindel2002givens}. The
BLAS versions should be changed in analogous ways to avoid
over/underflow when possible, while assuring that \NaNs~also propagate.
The paper~\cite{anderson2017l1scaling} also provides more reliable
versions of Givens rotations, and claims these are better than the
versions in~\cite{bindel2002givens}, which do not pass all the tests
in~\cite{anderson2017l1scaling}, but the version of CLARTG
currently in LAPACK differs from the one tested
in~\cite{anderson2017l1scaling}.

Here are two possible specifications for consistency for generating a
Givens rotation. Let x and y be the two inputs, and c, s, r and z be the
four outputs (cosine, sine, length of the vector, and a single scalar
from which it is possible to reconstruct both c and s). The simplest
requirement would be that if either x or y is an \Inf~or \NaN, then at
least one output must be an \Inf~or \NaN. A more rigorous requirement for
\{S,D\}ROTG could be the following:

\begin{enumerate}
    \item  x = +-\Inf~and y = finite =\textgreater{} c = 1, ~s = 0, r = x, and
    z = 0 ~
    \item x = finite and y = +-\Inf~=\textgreater{} c = 0, ~s = 1, r = y, and
    z = 1 ~
    \item x = +-\Inf~and y = +-\Inf~=\textgreater{} c = \NaN, s = \NaN, r =
    +-\Inf, and z = \NaN~(signs may not match)
    \item either x = \NaN~or y = \NaN~=\textgreater{} c = \NaN, s =\NaN, r =
    \NaN~and z = \NaN
\end{enumerate}

For \{C,Z\}ROTG, which does not compute z, we would need to distinguish
the cases where the real and/or imaginary parts of x and y are finite:

\begin{enumerate}
    \item  x contains an \Inf~and y = finite =\textgreater{} c = 1, s = 0, and
    r = x
    \item x = finite and y = +-\Inf~+ i*finite =\textgreater{} c = 0, s = +-1
    + i*0, and r = \Inf
    \item x = finite and y = finite +-i*\Inf ~ ~=\textgreater{} c = 0, s = 0
    -+i, and r = \Inf
    \item x = finite and y = +-\Inf~+-i*\Inf ~ ~=\textgreater{} c = 0, s =
    \NaN, and r = \Inf
    \item Both x and y contain an \Inf~but no \NaN~=\textgreater{} c = \NaN, s
    = \NaN, r = \Inf
    \item Either x or y contains a \NaN~=\textgreater{} c ~= \NaN, s = \NaN, r
    = \NaN
\end{enumerate}

\subsubsection{Level 3 BLAS routines}

Level 3 BLAS routines have classical implementations that perform
$O(n^3)$ flops when all input dimensions are n. This means both that
reducing the arithmetic cost by up to 25\% is possible for complex
inputs using Gauss's algorithm
as discussed at the end of Section~\ref{sec:proglang}, and that
methods like
Strassen's algorithm \cite{Strassen69} exist that can reduce the arithmetic cost to
$O \left(n^{log_2 7} \right)$ or less. Both methods (or their combinations) can cause
different exceptions to occur, but as long as they propagate to the
output as expected, they should be reported consistently by the routines
that call them.

\subsection{LAPACK - Examples of Inconsistent Exception Handling}\label{sec:LAPACKex}

The consistency issues for LAPACK are more complicated than
those in the BLAS for several reasons. First, in addition to much more,
and more complicated, code, LAPACK has many layers of subroutine
calls, so that consistency must be preserved through these layers.
Second, in addition to the issues of correct mathematical behavior
and propagation of \Infs and \NaNs, there is reporting of exceptions.
The existing INFO parameter can be used for this, but some users
have expressed a desire for more detailed reporting than can
be provided by a single scalar. Third, user requests vary significantly
regarding desired functionality and interfaces, including removing
all checks to speedup execution on small matrices.

We begin in section~\ref{sec:sgesv} with an example, SGESV for solving
a linear system, that fails to propagate exceptions correctly because
of the BLAS routines discussed in section~\ref{sec:blas}; this would
be corrected by our proposed changes in the BLAS.
Section~\ref{sec:sstemr} discusses a tridiagonal eigensolver, SSTEMR,
that deliberately does not propagate some exceptions, in order to compute
the correct answer more quickly.
Section~\ref{sec:LAPACKE} discusses the C interface to LAPACK provided by
LAPACKE. LAPACKE does provide optional checking for \NaN inputs, but not
\Infs; we show that this can unintentionally create \NaNs.
Section~\ref{sec:intoverflow} discusses an example of integer overflow
in workspace size calculation, and a proposal to report it using INFO.


\subsubsection{SGESV -- Incorrectly not propagating exceptions}
\label{sec:sgesv}

The purpose of this example is to show how inconsistent exception
handling in the BLAS and lower level LAPACK routines leads to
inconsistencies in higher-level drivers. The challenge of finding all
such examples motivates our proposal to report exceptional inputs and
outputs.

SGESV is the LAPACK driver for solving A*x=b using Gaussian
Elimination. For this example, we assume we use the original
non-recursive version which calls SGETF2 internally. We give a 2x2
example that shows how the inconsistencies described before in ISAMAX,
GER and TRSV interact to cause a \NaN~in input A not to propagate to
output x. Let $A = 
\begin{bmatrix}
    1 & 0 \\
    \mathrm{NaN} & 2
\end{bmatrix}$ and $b = 
\begin{bmatrix}
    0 \\
    1
\end{bmatrix}$. First, in LU
factorization ISAMAX is called on 
$\begin{bmatrix}
    1 & \mathrm{NaN}
\end{bmatrix}$ to identify the pivot,
and returns 1. Next GER is called to update the Schur complement, i.e.
replace 2 by 2 -- \NaN*0. But GER notes the 0 factor, does not multiply
it by the \NaN, and leaves 2 unchanged, instead of replacing it by \NaN.
This yields the LU factorization $A =
\begin{bmatrix}
    1 & 0 \\
    \mathrm{NaN} & 1
\end{bmatrix} \times
\begin{bmatrix}
    1 & 0 \\
    0 & 2
\end{bmatrix}$.
The first call to TRSV solves L*y=b, correctly setting y(1) = 0,
and then chooses not to multiply 0*\NaN~when updating y(2) = 1 --
0*\NaN, leaving y(2)=1. Finally, TRSV is called again to solve U*x=y,
yielding $x=
\begin{bmatrix}
    0 \\
    0.5
\end{bmatrix}$, with no \NaN~appearing in the final output.

If one calls the
recursive
of SGESV introduced in release 3.6.0, then 2 -- 0*\NaN will be computed
by a call to SGEMM, which may be more likely to compute a \NaN.

The many other one-sided factorization routines should be examined to
see if they are susceptible to similar problems and amenable to similar fixes.

\subsubsection{SSTEMR -- correctly not propagating exceptions}
\label{sec:sstemr}

The purpose of this example is to show that some algorithms are
designed with the expectation that there will be exceptions, and handle
them internally. In such cases, there is no reason to report them. Such
examples are uncommon and need to be carefully documented.

SSTEMR computes selected eigenvalues, and optionally eigenvectors, of a
symmetric tridiagonal matrix T. One internal operation is counting the
number of eigenvalues of T that are \textless{} s, for various values of
s. Letting D be the array of diagonal entries of T, and E be the array
of offdiagonal entries, the inner loop (which appears in SLANEG) that
does the counting looks roughly like this (an
analysis is provided elsewhere~\cite{lawn172}):

\begin{lstlisting}
DPLUS = D(i) + T
IF (DPLUS .LT. 0) COUNT = COUNT + 1
T = (T/DPLUS) * LLD(i) -- s
\end{lstlisting}

As described in the symmetric tridiagonal eigensolvers report~\cite{lawn172},
it is possible for a tiny DPLUS to cause T to
overflow to \Inf, which makes the next DPLUS equal to \Inf, which makes
the next T = \Inf/\Inf~= \NaN, which then continues to propagate. Checking
for this rare event in the inner loop would be expensive, so SLANEG only
checks for T being a \NaN every 128 iterations, yielding significant
speedups in the most common cases. The value 128 is a tuning parameter.

Similar examples are also discussed
elsewhere~\cite{demmel1994fasterexcept,demmel2007gpusymtrieig},
including other significant speedups.

\subsubsection{LAPACKE and matrix norms}
\label{sec:LAPACKE}

LAPACKE provides a C language interface to LAPACK. It currently has an
option for the LAPACK driver routines to check input matrices for \NaNs,
and return with an error flag. This makes sense for the drivers, such as
eigensolvers, for which the mathematical problem may be considered
ill-posed (but see the earlier discussion that not everyone may agree on
what ``ill-posed'' means, e.g., computing the eigenvalues of a diagonal
matrix with an \Inf~or \NaN~on the diagonal). However, \Infs~are also
problems for these routines, for the following reason: All these drivers
begin by computing the norm of the matrix, checking to see whether it is
so large or small that squaring numbers of that size would
over/underflow, and if needed scaling the matrix to be in a ``safe''
range to avoid this problem. But if the matrix contains an \Inf, scaling
would multiply each entry by x/\Inf~= 0, where x is a finite number (the
target norm), resulting in a matrix containing only zeros and \NaNs, on
which computation would continue, defeating the purpose of avoiding
\NaNs. Since these drivers already compute the matrix norm, it is easy to
detect whether the input contains an \Inf~or \NaN~and return with INFO
indicating this. If more than one input contains \Infs~or \NaNs, only the
first one would be reported, analogous to the previous usage of INFO.

We note that the norm routines for real matrices compute max\_ij
abs(A(i,j)) in a way that is guaranteed to propagate \NaNs, by not
depending on the built-in max() function, which may not propagate \NaNs.
However, in the routine for complex matrices abs(A(i,j)) can overflow
even if A(i,j) is finite (same issue as I\{C,Z\}MAX1), so these routines
need to be modified to add the option of computing max\_ij
max(abs(real(A(i,j))),abs(imag(A(i,j)))), which is finite if and only if
no \Infs~or \NaNs~appear.

For these driver routines that already compute norms, checking inputs
for \Infs~or \NaNs~adds a trivial O(1) additional cost. For other routines
that may do only O(1) arithmetic operations per input word, adding
input argument checking could add a constant factor additional cost. 
Our design in section~\ref{sec:LAPACK} discusses our proposal for
optional input argument checking.

\subsubsection{Integer Overflow}\label{sec:intoverflow}

Several LAPACK routines require workspace supplied
by the user. The user has to provide a workspace (the WORK array) and the length of this work array
(the integer LWORK). Because the integer LWORK is the dimension of an array, it is INPUT only in the 
interface (similarly to integers such as M, N, or LDA).  The minimum amount of workspace
needed is given by a formula in the leading comments of the LAPACK routine, but a larger workspace often
enables LAPACK to perform better, leading to the concept of ``optimal'' workspace,
meaning ``optimal for better performance``. (We acknowledge that ``optimality'' is
determined by a high-level performance model, not detailed benchmarking.) 
In LAPACKv1 (1992) and LAPACKv2 (1994), the source code read ``For optimum performance \verb+LWORK+ \verb+>=+\verb+N*NB+, where \verb+NB+ is the optimal blocksize.'' To know the optimal blocksize, \verb+NB+, the user would call ILAENV. Starting with LAPACKv3 (1999), LAPACK introduced the concept of workspace query by allowing 
the user to query the optimal amount of workspace needed, by passing in 
the integer LWORK with the value -1. 
When LWORK=-1, this is called a ``workspace query'', and the LAPACK subroutine does not perform any numerical computation
and only returns the 
optimal workspace size in WORK(1), as an integer stored as a floating point number (rounded up slightly if necessary).
Once the user knows the necessary size,
the user can then pass on the necessary workspace to LAPACK by calling again LAPACK with LWORK 
set to the size of the provided workspace.
The optimal workspace can 
depend on the LAPACK implementations and versions used, 
which is why the user must query it. Also, computing the optimal workspace can be quite long and complicated, 
for example for the subroutine DGESDD, it takes 273 lines of code, including 20 function calls 
to, in turn, query their own optimal workspace sizes. 

Now comes the issue of integer overflow. 
LWORK is an integer and so can overflow. LWORK is often the largest integer in the LAPACK interface.
It can be much larger than N for example. And so LWORK is often the most integer overflow prone quantity in the interface.
The integer overflow threshold depends on whether LAPACK is compiled using 32-bit or 64-bit integers.
(For example, the threshold is $2^{31}-1$ for signed 32-bit integers.)
As an example, the largest optimal workspace for all LAPACK subroutines is for the routines xGESDD
and can be at least $4*N^2$, where $N$ is the input matrix dimension (assuming square for simplicity). 
If LWORK is a signed 32-bit integer, then $4*N^2$ will overflow when $4*N^2 \geq 2^{31}$, or $N \geq 23,171$. 
It is therefore not feasible to correctly call SGESDD with 32-bit integers, $N \geq 23,171$, and use the optimal workspace.
To inform the user of this infeasibility,
during a workspace query, when WORK(1) is such that LWORK would overflow, 
we propose using INFO to let the user know.
Even if we can return a correct WORK(1), and
even if the user can successfully allocate this much memory,
they will not be able to call LAPACK again with LWORK equal to the true length of WORK. 
To address this potential exception, we propose to test if LWORK can be set 
correctly, by setting INTTMP = WORK(1), testing if INTTMP .EQ. WORK(1), and if not, setting INFO to point to
LWORK on exit, to indicate the problem. The test INTTMP .EQ. WORK(1) will depend (as it should) on whether INTEGER is 32-bit or 64-bit.
Note for developers: When computing the optimal workspace size during a workspace query, 
we will also need (1) to compute the optimal workspace returned
in WORK(1) carefully, to avoid integer overflow, and (2) to round it up slightly, if needed, if we compute it using floating point arithmetic. 

\subsection{LAPACK - Proposed Exception Handling Interface}\label{sec:LAPACK}

This section contains the most complicated part of this
document, our proposal for a new LAPACK interface for (optionally)
reporting exceptions. Section~\ref{sec:argcheckuser} begins with
a summary of all the user requests we received over the course of
the design process. Section~\ref{sec:argcheckD9} presents our current design. 
Section~\ref{sec:exampleSGESVEC} illustrates our design as
applied to SGESV (a model implementation
of the new SGESV, and all the routines in its
call tree, appears in Appendix~\ref{AppC}).
Section~\ref{sec:argcheckDold} summarizes the
previous designs, and why we evolved them, finally arriving at
our current design. 

\subsubsection{User requests}\label{sec:argcheckuser}

Our latest design (ninth in a sequence) tries to balance
user requests ranging from not wanting to change any legacy code, to adding
significant new exception handling capabilities in multithreaded environments,
all with allowing the LAPACK developers to continue maintaining one core
implementation. Specifically, we want to maintain just
\begin{enumerate}
\item one core that can be called from multiple languages, including C, C++ and Python, 
\item one ``wrapper'' providing the legacy interface, and 
\item the existing LAPACKE wrapper for C and C++ programmers. 
\end{enumerate}
Item (1) means that we decided not to use some features of modern Fortran, 
like optional arguments, because Fortran's optional arguments don't have 
counterparts in other languages and compilers don't implement them 
in a uniform way. Similarly, we chose to pass routine names (for error
reporting) as arrays of characters with an additional length argument,
since different languages may represent character strings differently.

We start with a list of all the user requests we have received, and more thoughts
about who may want control over exception handling and what they might want.
These requests and related thoughts occurred at different points in our design
process, which is why our design has gone through multiple versions. 

Here is a list of all the user requests so far:
\begin{description}
\item [(R1)]  I’m happy with my legacy code, which calls LAPACK from Fortran/C/C++/NumPy…,
         don’t make me change it!
\item [(R2)]  Ok if you want to help other folks with debugging, but don’t slow down.
\item [(R3)]  I’d like help debugging, when I find it necessary; I might need different kinds
         of information for this, depending on the situation, and just in the parts of
         the code where I suspect the problem to be. I’m willing to modify my code to do this,
         i.e. set the kind of exception handling I want, and get a report back, for each call
         that I make to an LAPACK routine.
\item [(R4)]  I want to be able to set a flag at the beginning of execution that selects the kind of 
         exception-handling I need for every LAPACK call and how to report them. I might 
         want reporting done by returning information in a subroutine argument (analogous 
         to LAPACK’s INFO), or by collecting reports for multiple subroutine calls in some 
         common data structure that I can inspect later.  I don’t want to modify my legacy code 
         beyond setting this flag, and possibly inspecting the common data structure.
\item [(R5)]  Same as (R4), except the flag should be settable (and changeable) at run-time.
\item [(R6)]  Same as (R5), except I program in a multi-threaded/multi-task environment, so different 
         threads/tasks may need to independently control how they handle and report exceptions,   
         i.e. depending on ``context''.
\item [(R7)]  I want to write bullet proof code, i.e. that won’t crash or give surprising wrong answers.
         I’m willing to slow down a little for this, but hopefully not much. All the approaches
         above from (R3) to (R6) are relevant.
\item [(R8)]  I want to be able to turn off all error checking (eg $N < 0$) and exception handling, to run faster.
\end{description}

In addition to these requests, we considered which stakeholders might want to “control”
the ways exceptions can be handled, including
\begin{description}
\item [(C1)] The user  
\item [(C2)] A library calling LAPACK internally 
\item [(C3)] Vendors of LAPACK equivalents  
\item [(C4)] Core LAPACK team, perhaps just prescribing what happens in “model” error handlers, 
          with changes allowed in downstream customizations. 
          \end{description}
We want our interface design to be flexible enough to support all of these. Of course, if 
a user links versions of different routines that have been built with different assumptions 
about who is ``in control,'' this could cause bugs which are beyond our control.

Regarding item C2 above, we considered the DOE xSDK (Extreme-scale Scientific 
Software Development Kit) project\footnote{\url{https://github.com/xsdk-project/}}
which is a large LAPACK user, and maintains a set of “community policies” for library
development\footnote{\url{https://github.com/xsdk-project/xsdk-community-policies/tree/master/package_policies}}.
Recommendation R3.md and policies M11.md, M12.md and M16.md are particularly relevant.
We briefly summarize these: 
\begin{description}
\item[R3]    Adopt and document a consistent system for propagating/returning 
             error conditions/exceptions and provide an API for changing this behavior.
             (This is clearly consistent with our goals.)
\item[M11] No hardwired print or I/O statements that cannot be turned off via an API.
             (This also impacts LAPACK’s default use of XERBLA, which prints an error message
              and stops. Our design is independent of how XERBLA is implemented.)
\item[M12] If a package imports software that is externally developed and maintained, then
             it must allow installing, building, and linking with an outside copy of that software.
             (This refers specifically to the BLAS and LAPACK.) 
\item[M16] Any xSDK-compatible package that compiles code should have a configuration  
             option to build in Debug mode.
\end{description}
    
We believe that our design is consistent with this (long) list of requests and recommendations,
and solicit comments.

As we considered the programming effort required to satisfy all these requests, we decided 
that we wanted there to be one core version of the LAPACK code to maintain that offers all
these new features, and that can be called from all the different languages from which
LAPACK is called, including C, C++, Python and perhaps others. We also decided that the
interface should be simple enough to allow significant code reuse across different
LAPACK routines.

\subsubsection{LAPACK Interface Proposal}\label{sec:argcheckD9}

Each LAPACK routine that already has an INFO argument will be modified as follows.
The subroutine name will be changed to add \_EC (for ``error checking'') to the end,
allowing the original name to be retained for a wrapper providing the ``legacy'' interface
and functionality.

In the new \_EC version, following INFO (currently the last argument), 3 more arguments will 
be added: 
\begin{enumerate}
  \item FLAG\_REPORT.
For terseness and clarity, in the descriptions below we will use the abbreviations
FLAG\_REPORT(1) = WHAT (since it specifies what errors and exceptions to report), and
FLAG\_REPORT(2) = HOW (since it specifies how to report them). 
  \item INFO\_ARRAY is an array used
for more detailed reporting than possible 
using a single scalar INFO.
  \item CONTEXT is an ``opaque'' argument
  that can be used to identify errors and exceptions associated with different threads or tasks.
\end{enumerate}

Before giving the detailed semantics of these new arguments, we give a high level summary
of the choices WHAT and HOW offer the user.
The possible choices of WHAT errors or exceptions to report are as follows:
\begin{itemize}
\item WHAT $\leq -1$: turn off all error checking
\item WHAT = 0: ``legacy’’ error checking only
\item WHAT = 1: also check input and output arguments for Infs and NaNs
\item WHAT $\geq 2$: also check input and output arguments throughout the call tree of the subroutine being called
\end{itemize}
The user must independently choose HOW to report this information:
\begin{itemize}
\item HOW $\leq 0$: only report using the scalar INFO
\item HOW = 1: also report more details using the array INFO\_ARRAY
\item HOW = 2:  also, if INFO $\neq 0$, call the routine REPORT\_EXCEPTIONS, which
   can provide a customized way of reporting this information, eg a print statement, or
   recording this information in a data structure for later inspection. We will provide some
   simple model implementations of REPORT\_EXCEPTIONS, but leave further
   customization to other software providers.
\item HOW = 3:  do all the above reporting throughout the call tree of the subroutine being called
\item HOW $\geq 4$:  call GET\_FLAGS\_TO\_REPORT, to get values of WHAT and HOW,
   thus allowing users to choose WHAT and HOW with less modification of source code
\end{itemize}
We note that the choice WHAT = HOW = 0 corresponds to the legacy LAPACK interface.
We leave details of how other choices of WHAT and HOW interact with one another to
later in this section.

Now we provide a more detailed description of
these new arguments.
\newline

\noindent {\bf (1) FLAG\_REPORT(1:2), integer array of length 2, input only.} 

\noindent {\bf WHAT}  = FLAG\_REPORT(1) will be used to select among the following options
for what possible errors to check in an LAPACK routine:
\begin{description}
 \item[WHAT $\leq -1$]: Check nothing. INFO = 0 is always returned.
            If WHAT $< -1$, we replace it by -1 in the subsequent text.
            When WHAT = -1 we pass WHAT = -1 in internal LAPACK calls.
 \item[WHAT $= 0$:] Legacy INFO checks only (eg test for $N < 0$, zero pivots, etc.).
            If any of these checks results in a nonzero INFO, this has priority to report
            over any other errors detected when WHAT $> 0$.
            More generally, an error that would be found with a lower value of WHAT has
            priority to report using INFO than an error that would only be found with
            a higher value of WHAT.
            When WHAT = 0 we pass WHAT = 0 in internal LAPACK calls.
 \item[WHAT $= 1$]: Also check input and output arguments for \Infs and \NaNs. If HOW = 0,
            we might stop checking early after discovering the first \Inf or \NaN, or 
            continue checking all inputs and outputs if HOW$>0$ (to report all problematic 
            inputs and outputs using INFO\_ARRAY, as described below); 
            this comment applies to  higher values of WHAT too.
            If the checks performed for WHAT = 0 detect no errors, INFO will point
            to the first input argument containing an \Inf or \NaN, i.e. INFO = -k points
            to argument k. In general computation will continue if an input contains
            an \Inf or \NaN, unless the problem is mathematically ill-defined (eg computing
            eigenvalues and/or eigenvectors). If no input contains an \Inf or \NaN but an output
            does, INFO will return a unique positive value (details later) to identify the first
            such output.  (Input-only arguments are only checked on input, and output-only 
            arguments are only checked on output.)
            When WHAT = 1 we pass WHAT = 0 in internal LAPACK calls.
 \item[WHAT $\geq 2$]: Also check input and output arguments of LAPACK routines called
            internally (i.e. throughout the call tree). If WHAT $> 2$ we replace it by 2
            in the subsequent text. Only LAPACK routines that
            have INFO parameters themselves will be checked.
            If the checks performed for WHAT = 0 or 1 detect no errors, and
            HOW $\geq 1$, INFO will point to the first internal LAPACK call with an input
            or output argument containing an \Inf or \NaN, or that itself calls an LAPACK
            routine with an input or output containing an \Inf or \NaN, if one exists.
            INFO will not be used to report such events if HOW = 0,  instead INFO will 
            be set the same as with WHAT = 1.   
            We describe how this works in more detail below.
            When WHAT = 2 we pass WHAT = 2 in internal LAPACK calls, to let the
            internal LAPACK routines do their argument checking.
\end{description}

\noindent {\bf HOW} = FLAG\_REPORT(2)  will be used to select among the following options for how to report
the errors described above. If WHAT = -1, nothing is reported, and HOW
is ignored. Otherwise:
\begin{description}
 \item[HOW $\leq 0$]: Report only using INFO, returning the legacy value if nonzero, otherwise
        the first error found among the checks determined above by WHAT.
        If HOW $< 0$, we replace it by 0 in the subsequent text.
        This is attained by passing HOW = 0 in internal LAPACK calls.
        Note that to return the legacy INFO only, use WHAT = HOW = 0.
 \item[HOW $=1$]: In addition to reporting using INFO, return array INFO\_ARRAY,
        with a more complete description of all the errors found by the checks 
        determined above by WHAT (details below). In particular, all floating point arguments 
        will be checked when WHAT = 1 or 2, rather than just reporting the first exception 
        found. This is attained by passing HOW = 1 in internal LAPACK calls.
 \item[HOW $=2$]: In addition to the reporting using INFO and INFO\_ARRAY, the routine 
        called by the user will call REPORT\_EXCEPTIONS on exit, with INFO\_ARRAY 
        as an argument, to report this information in an implementation dependent way
        (details below).
        REPORT\_EXCEPTIONS will only be called if there are any errors to report,
        i.e. INFO is nonzero. Again, all arguments selected by WHAT will be checked, 
        rather than just reporting the first exception found.
        This is attained by passing HOW = 1 in internal LAPACK calls.
 \item[HOW $=3$]: In addition to the reporting when HOW = 2, all routines in the call tree will
        be treated the same way, in particular REPORT\_EXCEPTIONS will be called 
        by all LAPACK routines in the call tree that have an INFO parameter.
        As above, REPORT\_EXCEPTIONS will only be called if there are any errors to 
        report, i.e. INFO is nonzero.
        This is attained by passing HOW = 3 in internal LAPACK calls.
 \item[HOW $\geq 4$]: The LAPACK routine called directly by the user will call GET\_FLAGS\_TO\_REPORT
        to get values of WHAT and HOW, which the user must have set by calling
        SET\_FLAGS\_TO\_REPORT before calling the LAPACK routine. If HOW $> 4$, we replace
        it by 4 in the subsequent text.  Setting HOW = 4 gives users 
        more flexibility to change WHAT and HOW at run-time. The returned value of 
        HOW will be replaced by max(0, min(HOW, 3)) to ensure that the value of HOW
        is in \{0,1,2,3\}. Other LAPACK routines in the call tree will not call
        GET\_FLAGS\_TO\_REPORT, but just use the values of WHAT and HOW passed to
        them by the calling LAPACK routine.
        This is attained by using previous rules to choose the values of WHAT and HOW
        to use in internal LAPACK calls.
        Details of the routines GET\_FLAGS\_TO\_REPORT, SET\_FLAGS\_TO\_REPORT and
        REPORT\_EXCEPTIONS are described later.
\end{description}

Here is how we use INFO to report \Infs and \NaNs in inputs and outputs.
We note that argument INFO\_ARRAY can be used to report errors
in more detail as described later, since using one scalar argument INFO to
report errors limits the amount of information that can be returned. The values
of INFO we use for reporting will necessarily be unique for each LAPACK 
routine, for the following reasons. If the k-th argument is the first one to 
contain an \Inf or \NaN on input, and INFO would otherwise be 0 based on 
standard error checking, then INFO is set to -k.  For output, each LAPACK 
routine with an INFO argument currently defines what a positive return value
of INFO means, e.g. INFO = k means pivot k is zero in SGESV. 
Therefore we will use the first positive unused values of INFO to point to the 
first output argument containing an \Inf or \NaN. For example, for SGESV, 
INFO = N+1 will mean argument A contains an \Inf or \NaN on output (and no 
input arguments contain an \Inf or \NaN, which would have been reported using 
INFO = -3 for A and INFO = -6 for B instead), and INFO = N+2 will mean 
argument B contains an \Inf or \NaN on output (and none of A on input, 
B on input or A on output contains an \Inf or \NaN). 

We choose not to distinguish \Infs~from \NaNs~in this reporting
for 3 reasons. First, \Infs~and \NaNs~can be equally problematic, and we
leave it to the user to decide how to react. Second, previous examples
of inconsistencies in underlying operations (like complex division) mean
that some operations could generate \Infs, \NaNs, or both, so we again
leave it to the user to decide how to react. Third, this lets us
continue to use INFO as before, with INFO=-k indicating that the k-th
input argument is problematic, i.e., contains either an \Inf~or \NaN.

After the values of INFO used to report \Infs and \NaNs in inputs and outputs,
the subsequent unused values of INFO will be used to point to internal LAPACK 
subroutine calls, when HOW $\geq 1$. Each consecutive appearance of an LAPACK call 
(with an INFO parameter) will be assigned a unique positive value of INFO to use for 
reporting. Note that each such appearance may be called multiple times, eg if it is a 
loop, so INFO will report if any call to that routine reported an error. For example, for 
SGESV, INFO = N+3 will mean the call to SGETRF reported an error, and INFO = N+4 
will mean the call to SGETRS reported an error. Here, ``reported an error'' refers to 
the case WHAT = 2, so WHAT = 2 and INFO = N+3 could mean 
either SGETRF reported an \Inf or \NaN as an input or output, or SGETRF2, which 
SGETRF calls internally, reported an \Inf or \NaN as an input or output.
If separate reporting for each call is desired, including when there are 
multiple calls (eg in a loop), more detailed reporting can be provided using
HOW = 2 as described below. 

We summarize the pairs of possible values of WHAT 
and HOW in the table below, and their interpretations:

\begin{center}
    \begin{tabular}{|c|c||c|c|c|c|}
    \hline
    \multicolumn{2}{|c||}{} & \multicolumn{4}{c|}{WHAT to report} \\ \hline
   \multicolumn{2}{|c||}{}   & WHAT $\leq -1$ & WHAT $=0$ & WHAT $=1$ & WHAT $\geq 2$ \\
   \multicolumn{2}{|c||}{HOW to report} & report        & legacy    & + input/output  & + throughout  \\
   \multicolumn{2}{|c||}{}  & nothing       & INFO      & variables & call tree     \\
    \hline
HOW $\leq 0$ & use INFO                    & ignore & Yes & Yes & same as \\
             &                             &  HOW   & (legacy    &     & WHAT $=1$ \\ 
             &                             &        & interface) &     &           \\ \hline
HOW $=1$     & + use INFO\_ARRAY           & ignore & Yes & Yes & Yes \\
             &                             &  HOW   &     &     &     \\ \hline
HOW $=2$     & + call                      & ignore & Yes & Yes & Yes \\
             & REPORT\_EXCEPTIONS          &  HOW   &     &     &     \\ \hline
HOW $=3$     & + call throughout call tree & ignore & Yes & Yes & Yes \\ 
             &                             &  HOW   &     &     &     \\ \hline  
HOW $\geq 4$ & call                        & ignore & ignore & ignore & ignore \\
             & GET\_FLAGS\_TO\_REPORT      &  HOW   &  WHAT  &  WHAT  &  WHAT \\  \hline
    \end{tabular}
\end{center}

We point out a variation on the proposed use of INFO above,
and its pros and cons. Instead of using previously unused
positive values of INFO to report Infs and NaNs in
inputs, outputs, and in internal subroutine calls, we
could use negative values. While the positive values may
need to depend on the problem size (eg N+1 through N+4 for 
the SGESV example above), the negative values could always
be assigned fixed values (say -100, -101 etc). 
The advantage of this
is that INFO is easier to interpret: one does not need
N in addition to INFO to interpret INFO. 
A disadvantage
is that it violates the LAPACK convention of INFO $<0$
meaning that an input is ``incorrect,'' and INFO $>0$
meaning that an error occurred during execution.
It is likely that many users may simply test for INFO $<0$
in their code (if they test at all) to detect input errors,
and so this variation would invalidate that assumption.
Also, to completely interpret INFO requires knowing
WHAT and HOW, eg to know whether inputs and outputs were
checked for Infs and NaNs. This is why we include WHAT and HOW in
INFO\_ARRAY below, which is designed to provide more
complete and interpretable information in one data structure.

Finally, we note that interpreting INFO currently requires
knowing another argument, like the dimension N, for a number
of subroutines, eg SGEES and SLAEBZ. So we could consider adding an
entry to INFO\_ARRAY below to return this additional parameter, if
it is needed; comments welcome. \newline

\noindent {\bf (2)  INFO\_ARRAY: integer array, input/output}
      This is accessed only if WHAT $\geq 0$ and HOW = 1, 2 or 3.
      The length of INFO\_ARRAY is customized for each routinename,
      with detailed reporting as requested above, as follows:
      \begin{itemize}
      \item INFO\_ARRAY(1) = legacy INFO
      \item INFO\_ARRAY(2) = value of FLAG\_report(1) = WHAT
               that was used to determine the other entries of INFO\_ARRAY.
      \item INFO\_ARRAY(3) = value of FLAG\_report(2) = HOW
               that was used to determine the other entries of INFO\_ARRAY.
      \item INFO\_ARRAY(4) = value of INFO depending on WHAT as described above
      \item INFO\_ARRAY(5) = number of routine arguments reported on
      \item INFO\_ARRAY(6) = number of internal LAPACK calls reported on
      \item INFO\_ARRAY(7:) contains a fixed number of entries, depending on the
              LAPACK routine, and on FLAG\_report
      \end{itemize}
      Here are more details on the values reported in INFO\_ARRAY(7:).
                
      Locations INFO\_ARRAY(7:6+INFO\_ARRAY(5)) contain one entry per floating point argument
      of the routine, with values
      \begin{itemize}
      \item -1 if not checked (default)
      \item 0 if checked and ok (no \Inf or \NaN in input or output)
      \item 1 if it contains an input \Inf or \NaN, but not output
      \item 2 if it contains an output \Inf or \NaN, but not input
      \item 3 if it contains both an input and output \Inf or \NaN
      \end{itemize}
                Input-only arguments are only checked on input, with possible return values
                in \{0,1\}, and output-only arguments are only checked on output, with possible
                return values in \{0,2\}. If an input argument has already been checked
                before calling the routine, this is indicated by setting INFO\_ARRAY(*) = 0 
                on input (if checked and ok) or 1 (if it contains an \Inf or \NaN), otherwise 
                INFO\_ARRAY(*) should be set to -1 on input. For example, when calling 
                SGESV\_EC, the matrix A may have been checked by SGETRF\_EC on output, so it does not 
                need to be checked again by SGETRS\_EC on input, saving work. Similarly, B may 
                have been checked by SGESV\_EC on input, so it does not need to be checked 
                again by SGETRS\_EC on input. Input values of INFO\_ARRAY(*) less than -1 
                or greater than 1 will be treated the same as -1, i.e. not checked.

        Locations INFO\_ARRAY(7+INFO\_ARRAY(5) : 6+INFO\_ARRAY(5)+INFO\_ARRAY(6)) contain
        one entry per LAPACK call (with an INFO parameter) appearing in the source code, with values
        \begin{itemize}
        \item -1 if not checked (default)
        \item 0 if checked and ok
        \item 1 if no input or output contains an \Inf or \NaN, but some LAPACK call
                        deeper in the call chain signaled.
        \item 2 if an argument contains an input \Inf or \NaN, but not an output
        \item 3 if an argument contains an output \Inf or \NaN, but not an input
        \item 4 if an argument contains both an input and output \Inf or \NaN
        \end{itemize}

           As before, we do not distinguish multiple calls to the same LAPACK routine at the
           same location (say inside a loop) in the source code, instead all their reports are 
           combined into one, by taking the maximum of all the reporting values described above.
           Note that we do not attempt to report details about exceptions throughout the
           call chain of LAPACK routines. This is done by setting HOW = 3 so that the routine
           REPORT\_EXCEPTIONS is called, as described below.

           We note that INFO\_ARRAY has a length customized for each routine.
           To make programming easier, we will document the maximum length of all these
           arrays, so that users can simply declare all INFO\_ARRAY arrays to have 
           this maximum length.

\noindent {\bf (3)     CONTEXT: input only}
          This opaque argument may be accessed only if WHAT $\geq 0$ and HOW = 2, 3 or 4. 

          When HOW = 2 or 3, it is used as an argument when LAPACK calls the routine
               REPORT\_EXCEPTIONS (CONTEXT, SIZE\_ROUTINENAME, ROUTINENAME, INFO\_ARRAY)
          to report exceptional information in INFO\_ARRAY immediately before 
          returning from an LAPACK routine. REPORT\_EXCEPTIONS will only be called if there
          is an exception to report, i.e. INFO is nonzero. CONTEXT is meant to accommodate application 
          or architecture specific reporting methods, for example dealing with multithreaded 
          programming environments, as in user request R6 (this argument can be ignored if
          it is not relevant).
          Here ROUTINENAME is a character array of length 
          SIZE\_ROUTINENAME, to indicate which routine is being reported on.  
          All arguments are input only.

          When HOW = 4, the LAPACK routine calls
              GET\_FLAGS\_TO\_REPORT(CONTEXT, FLAG\_REPORT)
          to get the (output-only) integer array FLAG\_REPORT(1:2), which the user should have
          set by calling
              SET\_FLAGS\_TO\_REPORT(CONTEXT, FLAG\_REPORT)
          before calling the LAPACK routine with HOW = 4. This
          allows the user to report differently in different CONTEXTs. The default value of FLAG\_REPORT
          should be [0, 0], i.e. legacy INFO reporting, in case the user has not called 
          SET\_FLAGS\_TO\_REPORT. 

          Since the semantics of CONTEXT are system dependent, we will only supply
          2 placeholder versions of the 3 routines above:

\begin{itemize}
             \item Placeholder 1 (“verbose”):
             \begin{itemize}
             \item GET\_FLAGS\_TO\_REPORT(CONTEXT,FLAG\_REPORT) will always return
                   WHAT = 4 and HOW = 3, meaning that throughout the call chain, the most
                   complete checking for exceptions will be done. CONTEXT will not be accessed.
             \item SET\_FLAGS\_TO\_REPORT(CONTEXT,FLAG\_REPORT) will simply return, not accessing
                   its arguments. In other words, ``verbose reporting'' will always be turned on,
                   as long as the user calls LAPACK routines with WHAT $\geq 0$ and HOW = 3.
             \item REPORT\_EXCEPTIONS(CONTEXT, SIZE\_ROUTINENAME, ROUTINENAME, INFO\_ARRAY)
                   will print a report containing the information ROUTINENAME and
                   INFO\_ARRAY (only if there are any exceptions or errors to report),
                   and always continue execution. CONTEXT will not be accessed. The question
                   of whether to continue execution given an error like N $<0$ is left to XERBLA
                   as it is now. REPORT\_EXCEPTIONS will simply print the routine name, and 
                   the first 6+INFO\_ARRAY(5)+INFO\_ARRAY(6) entries of INFO\_ARRAY; 
                   since their interpretation is routine specific, we leave it to the user to interpret 
                   them. We note that the longest allowed name in Fortran or C is 63 characters,
                   so that we can safely declare all ROUTINENAME arrays to be of length 63.
                   We also note that the use of character array ROUTINENAME and its length
                   SIZE\_ROUTINENAME avoids problems caused by different languages possibly
                   representing string arguments differently.
             \end{itemize}

             \item Placeholder 2 (“terse”):
                These placeholders will simply return, not accessing their arguments.
\end{itemize}

          We leave it to other developers (eg vendors, software library builders, etc)
          to provide custom versions of the above 3 routines for their programming 
          environments as they see fit. For example, REPORT\_EXCEPTIONS could collect 
          information in a common data structure that a user could later inspect, or invoke 
          a debugger, or print a report and then halt if an exception occurred (a la XERBLA), 
          or other possibilities.
          
          Finally, the wrapper providing the legacy interface will call the \_EC version
          with WHAT = HOW = 0, INFO\_ARRAY an array of length 1 (it will not be
          accessed), and CONTEXT a null pointer (it will also not be accessed).
          
\subsubsection{Example: SGESV\_EC}\label{sec:exampleSGESVEC}

We illustrate the proposed interface from the last
section by summarizing how it applies to SGESV,
yielding SGESV\_EC.
A complete implementation (not yet tested) of SGESV\_EC (and all the routines in its call tree)
appears in Appendix~\ref{AppC}. The calling sequence of SGESV\_EC is as follows, where the 3 
new arguments appear at the end: \newline

\noindent{SGESV_EC( N, NRHS, A, LDA, IPIV, B, LDB, INFO, FLAG_REPORT, INFO_ARRAY, CONTEXT )
}\newline

First, we explain how to interpret INFO;
the possible values are listed in decreasing
priority order (only the first error found
is reported):

\begin{enumerate}
\item ``Legacy'' values of INFO:
\begin{description}
    \item[= 0:] if successful execution, else
    \item[= -1:] if N $<0$, else
    \item[= -2:] if NRHS $<0$, else
    \item[= -4:] if LDA $<$ min(1,N), else
    \item[= -7:] if LDB $<$ min(1,N), else 
    \item[1 $\leq$ INFO $\leq$ N:] if U(INFO,INFO)=0, else ...
\end{description}
\item Possible values of INFO if checking input/output arguments for Infs/NaNs is 
requested (WHAT $\geq 1$):
\begin{description}
    \item[= -3:]  if A contains an Inf/NaN on input, else
    \item[= -6:] if B contains an Inf/NaN on input, else
    \item[= N+1:] if A contains an Inf/NaN on output, else
    \item[= N+2:] if B contains an Inf/NaN on output, else ...
\end{description}
\item Possible values of INFO if checking internal subroutine calls for Infs/NaNs is 
requested (WHAT $\geq 2$):
\begin{description}
    \item[= N+3:] if SGETRF\_EC had an Inf/NaN
    in an input/output, or a subroutine in its call tree did, else
    \item[= N+4:] ditto for SGETRS\_EC
\end{description}
\end{enumerate}

Next, we explain how to interpret the entries of
INFO\_ARRAY, an array of length 10:

\begin{description}
    \item[(1):] INFO from ``legacy'' argument checking only
    \item[(2):] FLAG\_REPORT(1) = WHAT to report
    \item[(3):] FLAG\_REPORT(2) = HOW to report
    \item[(4):] INFO as determined by WHAT, as 
    explained above
    \item[(5):] $\leq 2$; the number of arguments reported on (0 or 2, i.e. A and B)
    \item[(6):] $\leq 2$; the number of internal calls reported on (0 or 2, i.e. SGETRF\_EC and
    SGETRS\_EC)
    \item[(7):] Reports on Infs/NaNs in A (if WHAT $\geq$ 1)
    \begin{description}
        \item[= -1:] if not checked (default)
        \item[=  0:] if checked and no Infs/NaNs on input/output
        \item[=  1:] if checked and contains Inf/NaN on input
                 but not output
        \item[=  2:] if checked and contains Inf/NaN on output 
                 but not input
        \item[=  3:] if checked and contains Inf/NaN on input and output
    \end{description}
    \item[(8):] Ditto for B
    \item[(9):] Reports on exceptions in call to SGETRF\_EC 
    (if WHAT $\geq$ 2)
    \begin{description}
        \item[= -1:] if not checked (default)
        \item[=  0:] if checked and no Infs/NaNs
        \item[=  1:] if checked and no input/output of SGETRF\_EC contains an Inf/NaN, but some LAPACK call
        deeper in the call chain signalled an Inf/NaN
        \item[=  2:] if checked and an input contains an 
        Inf/NaN, but not an output
        \item[=  3:] if checked and an output contains an
        Inf/NaN, but not an input
        \item[=  4:] if checked and both an input and output
        contain an Inf/NaN
    \end{description}
    \item[(10):] Ditto for call to SGETRS\_EC
\end{description}

\subsubsection{Evolution of the Current Proposal}\label{sec:argcheckDold}

As stated above, the design in the last section is the 9th in a sequence.
It kept evolving as new user requests were received, and the complexity of
implementation was evaluated, motivating us to simplify the design.
Here we sketch the changes, and why we made them, starting with the first
design (D1), which also appeared in the first version of this 
document\footnote{\url{https://people.eecs.berkeley.edu/~demmel/Exception\_Handling\_for\_the\_BLAS\_and\_LAPACK\_12Aug2021.pdf}}.

\paragraph{Design 1 (D1)}

Our initial design was to provide 3 wrappers to LAPACK routines, to
provide 3 kinds of exception handling, all just using INFO:
(1) ``legacy'' behavior of INFO, plus fixes discussed in section~\ref{sec:LAPACKE},
(2) inputs and outputs checked for \Infs and \NaNs, reporting the first one found, and
(3) in addition to input and outputs, reports of \Infs and \NaNs from internal
subroutine calls would be reported. This approach was conceived before requests
(R4) through (R7) in section~\ref{sec:argcheckuser} were received, and did not 
satisfy these requests for not modifying code to decide what to do at
run-time, or for more detailed reporting than possible just with INFO.

\paragraph{Design 2 (D2)}
This design proposed adding 4 optional arguments to LAPACK routines already
using INFO: WHAT, HOW, INFO\_ARRAY and CONTEXT, to accommodate requests (R4)
through (R7). It turns out that the use of Fortran optional arguments is
not necessarily consistent with all ``legacy'' uses of LAPACK when it is
called from other programming languages, such as C, C++ and Python.

\paragraph{Design 3 (D3)}
This design proposed having LAPACK routines with \_EC (for ``error checking'')
appended to their names, with the 4 optional arguments of D2. The original
LAPACK names would be maintained as wrappers around the \_EC versions, 
to accommodate users who do not want to change anything. Since requests
(R4) through (R7) require at least modifying the legacy interface to add
a CONTEXT argument, also adding \_EC to each call seems like a small
additional burden.

\paragraph{Design 4 (D4)}
This is very similar to D3, except all new optional arguments of \_EC
routines became conventional (required) arguments, to make sure the
to make sure the error checking options are available to callers from
other programming languages that do not accommodate Fortran optional
arguments. D4 proposed passing a NULL argument to get the same
intended semantics as a missing argument.

\paragraph{Design 5 (D5)}
First, it again turned out that different programming languages do not all
treat Fortran NULL arguments in the same way, leading us to 
(1) merge the WHAT and HOW arguments into a single array argument
FLAG\_REPORT(1:2), and (2) use values of HOW to decide whether
to access the INFO\_ARRAY and CONTEXT arguments.
Second, based on one more user request (R8), we also introduced the
option of turning off all error checking, including traditional ones
like $N<0$. This allows higher efficiency on small matrices,
and has been independently included in \cite{BatchBLAS2021}.

\paragraph{Design 6 (D6)}
We introduced the two sample implementations of routine REPORT\_EXCEPTIONS
described in section~\ref{sec:argcheckD9}, ``verbose'' and ``terse''.
We also chose to pass the routine name as a character array, plus
a length argument, to REPORT\_EXCEPTIONS, to accommodate different
programming languages that might represent character string
arguments differently.

\paragraph{Design 7 (D7)}
After trying to implement D6 in SGESV\_EC, we decided to simplify
the definition of the array argument FLAG\_REPORT, to make the options
it provided easier to understand, and to better separate the roles
of its two components, WHAT and HOW. 

\paragraph{Design 8 (D8)}
After implementing D7 in SGESV\_EC and the other routines it calls,
we decided to simplify the set of choices offered to users, from those originally in D6:
\begin{itemize}
     \item WHAT = -1:  no error checking
     \item WHAT = 0:  “legacy” checking only
     \item WHAT = 1:  also check inputs for Infs and NaNs
     \item WHAT = 2:  also check outputs for Infs and NaNs
     \item WHAT = 3:  also check inputs and outputs of LAPACK routines directly called by the routine
     \item WHAT = 4:  also check inputs and outputs of all LAPACK routines throughout the call tree
\end{itemize}
by merging 1 with 2 and 3 with 4:
\begin{itemize}
    \item WHAT = -1 (or smaller): no error checking
    \item WHAT = 0:  “legacy” checking only
    \item WHAT = 1:  also check inputs and outputs for Infs and NaNs
    \item WHAT = 2 (or larger):  also check inputs and output of all LAPACK routines throughout the call tree
\end{itemize}
The choices defined by HOW in D7 remain unchanged, except for treating all values
less than 0 the same as 0, and all values  greater than 4 the same as 4, to simplify error handling.

\paragraph{Design 9 (D9)}

We make more simplifications for the following reason. Our initial implementation of 
SGESV\_EC involved embedding all the necessary logic into the source code. Then, to make
the software engineering more manageable when extending to the rest of LAPACK, we
tried to embed as much of the complicated logic as possible into a few routines that could 
be reused by all LAPACK routines. We ran into an obstacle for the case of WHAT $\geq 2$
(checking input and output routines of all internally called LAPACK routines) and
HOW = 0 (using only INFO to report, not INFO\_ARRAY); this required too much
logic unique to each call to easily embed it into reusable routines. As a consequence,
in the case of WHAT $\geq 2$ and HOW = 0, we decided to not use INFO to report that an 
internal subroutine call generated an exception (that did not propagate to an output 
variable, which would be reported anyway). To get more details, the user will need to set 
HOW $\geq 1$ to report using INFO\_ARRAY, in which case INFO will report as before.
In other words, WHAT $\geq 2$ and HOW=0 behaves the same as  WHAT = 1 (check input/output
arguments) and HOW = 0.

Our second simplification is to treat all values of WHAT $\leq 1$ the same as WHAT = -1,
rather than reporting an error when WHAT $< -1$. We make analogous simplifications
for the cases WHAT $\geq 2$, HOW $\leq 0$ and HOW $\geq 4$.

Also, INFO\_ARRAY was modified so that instead
of INFO\_ARRAY(5) indicating the total length of the array,  INFO\_ARRAY(5) indicates 
the number of entries of the array reporting on input and output arguments, and 
INFO\_ARRAY(6) indicates the number of entries of the array reporting on
internal calls. Thus input and output argument reports are always stored at locations
INFO\_ARRAY(7 : 6+INFO\_ARRAY(5)), and internal call reports are always stored at
locations INFO\_ARRAY(7+INFO\_ARRAY(5) : 6+INFO\_ARRAY(5)+INFO\_ARRAY(6)).
The total length of INFO\_ARRAY is 6+INFO\_ARRAY(5)+INFO\_ARRAY(6).
This makes interpreting INFO\_ARRAY more self-contained. Another cosmetic change
is to have all the new arguments appear at the end of the current calling sequence,
so after INFO instead of before.  It would also be possible to merge these last 3
arguments into a single struct containing them all, but we have not done this yet.

Finally, we tried to optimize the code to eliminate redundant error checking, eg
if SGESV\_EC checks A for \Infs and \NaNs on input, it is not necessary for 
SGETRF\_EC to do so as well. And there are similar possible optimizations for
checking outputs. But doing this to eliminate as many redundancies as
possible made the code more complicated, and made it harder to reuse code,
so we only did some simple cases.

To summarize our new error-checking routines that we can reuse in all new 
LAPACK routines to hide the complicated but shared logic (hopefully):
\begin{itemize}
   \item CHECKINIT1: called at the beginning to initialize error checking flags
   \item CHECKINIT2: called after computing the ``legacy'' INFO on inputs,
       to initialize error flags
   \item xyyCHECKARG: called on an input or output array of type xyy (just SGE so far)
       to test for \Infs and \NaNs
   \item CHECKCALL: called after each internal LAPACK call to check for exceptions
   \item UPDATE\_INFO: called before returning to compute the final value of INFO
\end{itemize}

Source code for all the above routines, as well as SGESV\_EC and all the 
routines that it calls, are in Appendix~\ref{AppC}.


\section{How to test consistency}
\label{sec:testconsist}

Ideally, an implementation of an algorithm would come with a proof that
it handles exceptions consistently, \textit{e.g.}, terminates, given various
assumptions about underlying building blocks. This sounds like solving
the halting problem, but it should be much easier in practice for the
algorithms we are talking about, because of their structure: loops with
finite loop bounds, or iterations that can be confirmed to include a
condition that the number of iterations cannot exceed a certain finite
limit (\textit{i.e.} do not simply iterate until a condition like
Error\_Estimate \textless{} Threshold holds, which will be false if
Error\_Estimate is a \NaN). The structure of (most) LAPACK algorithms
(and certainly the BLAS) is simple enough to either do this by hand or
perhaps automate it, at least enough to identify those routines that
require a human expert to analyze them. We leave this for future work.

In the meantime, we propose expanding our LAPACK test code to test both
termination and error reporting. It is simplest to insert \Infs
and \NaNs~into random locations in the inputs of a routine and see what
happens, and this should be our first step. But this will not test what
happens if an \Inf~or \NaN is generated in an unpredictable location in
the middle of an execution, which is certainly needed to test all the
ways exceptions could be reported as discussed in
Section~\ref{sec:LAPACK}. 
One approach is to use ``fuzzing''~\cite{wikipedia2021fuzzing}, which in our
context means introducing an \Inf~or a \NaN into one or more
(randomly) chosen variables during the course of an execution. We
propose fuzzing because it is very difficult to devise an input without
\Infs~or \NaNs~that will, or should, generate an \Inf~or \NaN~during some
intermediate calculation. Since our exception handling should be
impervious to when and where exceptions are generated, fuzzing is a
suitable approach. In addition to choosing (random) locations to introduce
\Infs~and \NaNs, we may want to introduce them into subroutines
called by the routine being tested (and so on recursively) to make sure
that our reporting mechanisms work (or identify what needs to be fixed).


There are (at least) two challenges to implementing this testing approach:
(1) how to insert \Infs and \NaNs into selected variables, and
(2) the combinatorially large number of combinations of locations where exceptions 
might occur in different phases of an execution, in particular when many subroutines are called.

Here is a possible solution for (1): At the end of Section~\ref{sec:argcheckDold}
we mentioned that we had created the routine xyyCHECKARG, which can be reused
by any LAPACK routine to check whether an input or output array A contains an
\Inf or \NaN.
We can create a ``malicious'' version of xyyCHECKARG that, on request, deliberately inserts an \Inf or \NaN into A.
The simplest way to do this is to add a subroutine call at the beginning of xyyCHECKARG to
a new routine (call it xyyMALARG for now), which takes A as an argument, perhaps
along with other xyyCHECKARG arguments saying whether A is input or output, etc.,
and decides whether to insert an \Inf or \NaN in some location of A. This could 
depend, for example, on counting how many times xyyCHECKARG is called, so we could choose to insert an \Inf or \NaN on the first call, second call, etc. 
(the counter would live in some global memory, like a common block, so the test harness can reset it between test calls).
This malicious version of xyyCHECKARG would only be used for testing.

This might solve problem (1), but not (2). As with our other LAPACK testing, a layered
approach seems feasible, testing routines at each level of the call tree from bottom to
top. But to make sure that exception reports propagate upwards in the call tree as
desired (via routine CHECKCALL), we would (at least) need to use xyyMALARG to insert
\Infs and \NaN in the routines called by the routine at the root of the call tree, 
i.e. the children of the root. Just doing this for the children seems to avoid the 
potentially combinatorially large number of possible places exceptions could occur, 
at the cost of some smaller test coverage (which is dealt with by testing all possible roots = subroutines). This begs the question of how we can design xyyMALARG
to know when it should insert an \Inf or \NaN. This seems to require knowing how to
label the call tree so that we know whether the j-th call to xyyCHECKARG on an input
(or an output) corresponds to a desired child of the root. We could also potentially 
pass the name of the routine (always available in array ROUTINE\_NAME) to xyyCHECKARG
(which doesn’t otherwise need it, so just perhaps for the malicious version),
which could then pass it to xyyMALARG, which could then perhaps more
straightforwardly ask whether this is the child of the root. Of course if the same
routine is called at multiple locations in the call tree, we would still need to
use the counter to distinguish these.

Or we could accept less rigorous testing and just insert \Infs and \NaNs at the root
of the call tree. Comments welcome. 

\section{Proposed tasks to improve consistency, in priority order}
\label{sec:tasks}

We propose a list of tasks to improve the exception handling
consistency of the BLAS and LAPACK. They are sorted in priority order,
with ``easy'' tasks with potentially larger impact first. Comments are
welcome.

\begin{enumerate}

\item Modify the LAPACK test code to see whether complex division
(including real/complex) and complex absolute value are implemented in a
way that avoids over/underflow, and issue a warning if they are not, as
described in Section~\ref{sec:proglang}. Even though we are trying to avoid
dependence on min/max, we should test these as well, as described in
Section~\ref{sec:ieeearith}.  Finally, we could test complex multiplication
as described in
Section~\ref{sec:proglang}, and simply report whether the semantics is
based on the textbook definition, or the C standard (this would just be
informative, not an error check).
This task was recently completed and the new tests appear in LAPACK 3.10.1. Those tests run during the build process of LAPACK, and display useful information when unexpected results are identified.~\footnote{\url{http://www.netlib.org/lapack/lapack-3.10.1.html\#_2_6_notes_about_compiler_dependency}}   

\item Modify LAPACKE to provide the option to return with an error flag
if inputs of selected driver routines contain either \NaNs~or \Infs. The
current version only checks for \NaNs, but as described in
Section~\ref{sec:LAPACKE} this is inadequate.

\item Modify the LAPACK routines that compute complex matrix norms
(e.g., \{C,Z\}LANGE) to provide a norm that correctly signals whether or
not the input matrix contains an \Inf~or \NaN, as in
Section~\ref{sec:LAPACKE}. For
the driver routines that already start by computing norms of the input
matrix or matrices, modify them to use INFO to report whether an input
contains an \Inf~or \NaN, and return immediately if the mathematical
problem is not ``well-posed,'' consistently with LAPACKE.
Correspondingly change the LAPACK test code to confirm they behave as
desired.

\item Fix the reference BLAS routines as described in Section~\ref{sec:blas}.
Update the BLAS test code
accordingly
to make sure \Infs~and \NaNs~are propagated as expected. This includes
devising finite inputs that create \Infs~and \NaNs~internally, which  should be possible given the relative simplicity of the
operations performed. Encourage vendors to adopt the new BLAS. Provide
a simple
routine
to return the index of the first or last nonzero entry of a 1D-array, or
the first or
last nonzero row of a 2D-array, to help users accelerate xTRS\{V,M\} also as
described in Section~\ref{sec:blas}.

\item Confirm that the draft implementation
in Appendix~\ref{AppC}
of the proposed LAPACK interface for
SGESV\_EC is correct, making any necessary
corrections. Extend the implementation to
a few other drivers, eg SGEEV\_EC to make
sure that the 6 routines intended to be
reusable across LAPACK are indeed reusable.

\item Build testcode, following the outline
in Section~\ref{sec:testconsist}, and confirm
that this approach works for the draft
implementations in Task~5.

\item Extend Tasks~5 and 6 to the rest of
LAPACK.


\end{enumerate}

\section*{Acknowledgements}
This work was supported in part by the
National Science Foundation under the project
Basic ALgebra LIbraries for Sustainable
Technology with Interdisciplinary Collaboration
(BALLISTIC), 
Grant Nos. 2004763 (UC Berkeley), 
2004541 (U. Tennessee) and
2004850 (U Colorado Denver). MathWorks also provided support.

\appendix\newpage

\section{More details on I\{C,Z\}AMAX}\label{AppA}

The following is a consistent implementation of I\{C,Z\}AMAX, according
to Section \ref{sec:amaxnrm2}. It is written so that its correctness should be
apparent, but one can imagine various ways to optimize its performance,
depending on the architecture and length n of the input. For example,
the algorithm below does just one pass over the data. Also, checking for
exceptions only periodically, as in Section~\ref{sec:sstemr}, would
likely be faster.

\begin{lstlisting}
noinfyet = 1 ! no Inf has been found yet
scaledsmax = 0
! indicates whether A(i) finite but
! abs(real(A(i))) + abs(imag(A(i))) = Inf
smax = -1
do i = 1:n
   if (isnan(real(A(i))) .or. isnan(imag(A(i)))) then
      ! return when first NaN found
      icamax = i, return
   elseif (noinfyet == 1) ! no Inf found yet
      if (isinf(real(A(i))) .or. isinf(imag(A(i)))) then
         ! record location of first Inf,
         ! keep looking for first NaN
         icamax = i, noinfyet = 0
      else ! still no Inf found yet
         if (scaledsmax == 0)
            ! no abs(real(A(i))) + abs(imag(A(i))) = Inf yet
            x = abs(real(A(i))) + abs(imag(A(i)))
            if (isinf(x))
               scaledsmax = 1
               smax = (.25*abs(real(A(i)))) + (.25*abs(imag(A(i))))
               icamax = i
            elseif (x > smax) ! everything finite so far
               smax = x
               icamax = i
            endif
         else ! scaledsmax = 1
            x = (.25*abs(real(A(i)))) + (.25*abs(imag(A(i))))
            if (x > smax)
               smax = x
               icamax = i
            endif
         endif
      endif
   endif
enddo
\end{lstlisting}

We also did timings for ICAMAX on an Intel(R) Xeon(R) Platinum 8180 CPU
@ 2.50GHz ("2S Skylake"), 1 core/thread out of 28/56, with 38.5MB L2 and
12x16GB DDR4-2666 memory, comparing the old reference BLAS
code~\cite{blasnetlib} vs. the new code above vs. the latest Intel(R)
MKL. Not surprisingly, Intel(R) MKL's performance in the presence of no
NaNs was fastest. The reference code vs. the above new code was compiled
with the Intel(R) Fortran Intel(R) 64 Compiler for applications running
on Intel(R) 64, Version 19.0.8, compiled with "-O3 -fprotect-parens -mp1
-mieee-fp". Performance in the presence of a \NaN~being inserted depended
on where the \NaN~was inserted. It appears that Intel(R) MKL does not do
a \NaN~check first or while scanning the data. Instead, the performance
suggests it does the operation and then checks if a \NaN~was present, and
then calculates the position of the first \NaN~(that is, Intel(R) MKL's
ICAMAX is already conformant to this new standard).
So this new code could be made to look arbitrarily faster than Intel(R)
MKL just by inserting a \NaN~into the beginning. The new code would exit
immediately regardless of how large the problem size N is, whereas
Intel(R) MKL would look at the entire array. But if the \NaN~were
inserted at the end, then the extra performance of the vendor library
would dominate.

\newpage
\section{BLAS Test Cases}\label{AppB}

This is a draft description of how to generate BLAS test cases of
length n (possible values for n: 1, 2, 3, 10, 128, ...).


\paragraph{ISAMAX test cases:} Default entries: A(k) = (-1)\^{}k*k

\begin{enumerate}
    \item  At least 1 \NaN, no \Inf~s (\textless=15 cases)
    \begin{enumerate}
        \item 1 \NaN, at location:
        \begin{enumerate}
            \item  1;~
            2;
            ~n/2; ~n
        \end{enumerate}
        \item 2 \NaNs~(if possible, i.e. n\textgreater1, ditto later)
        \begin{enumerate}
            \item 1,2; ~ 1,n/2; ~1,n; ~ 2,n/2; ~2,n; ~n/2,n
        \end{enumerate}
        \item 3 \NaNs
        \begin{enumerate}
            \item 1,2,n/2; ~1,2,n; ~1,n/2,n; ~2,n/2,n
        \end{enumerate}
        \item All \NaNs
    \end{enumerate}
    \item At least 1 \NaN~and at least 1 \Inf.
    For each example above (\textless=7*15 cases):
    \begin{enumerate}[label={\roman*.}]
        \item  Insert \Inf~in first non-\NaN
        ~location
        \item Insert -\Inf~in first non-\NaN~location
        \item Ditto for last non-\NaN~location
        \item Ditto for first and last non-\NaN~locations
        \item Insert (-1)\^{}k*\Inf~in all non-\NaN~locations
    \end{enumerate}
    \item No \NaNs, at least 1 \Inf~(15 cases).
    Same pattern as (1) above, inserting (-1)\^{}k*\Inf~into A(k).
\end{enumerate}

Total \#cases \textless{} 5*(15 + 7*15 + 15) = 615

\paragraph{ICAMAX test cases:}
In addition to the above, we need additional cases
because A(k) can be finite but 

abs(real(A(k)))+abs(imag(A(k))) can overflow.

\noindent \textbf{New cases}:

\begin{enumerate}
\setcounter{enumi}{3}
\item A(k) is finite but abs(real(A(k)))+abs(imag(A(k))) can overflow.
\begin{enumerate}[label={\Alph*.}]
    \item  A(k) = -k + i*k for k even, A(k) = OV*((k+2)/(k+3)) +
    i*OV*((k+2)/(k+3)) for k odd.\\ Correct answer = last odd k.
    \item Swap odd and even.\\ Correct answer = last even k.
    \item A(k) = -k + i*k for k even, A(k) = OV*((n-k+2)/(n-k+3)) +
    i*OV*((n-k+2)/(n-k+3)) for k odd.\\ Correct answer = 1.
    \item Swap odd and even.\\ Correct answer = 2.
\end{enumerate}
\item
For each of the above 4 cases, insert \NaNs~and/or \Infs~in same way as for ISAMAX.
\end{enumerate}

\paragraph{xNRM2 test cases:} ~\\
%

Some of the following tests use the Blue's constants b and B for the sum of squares from \cite{Blue1978}. Any floating-point number $x$ satisfying $b \leq x \leq B$ has its square ($x^2$) guaranteed to not over- nor underflow.

\begin{enumerate}
\item Finite input which expects a correct output.
\begin{enumerate}
\item A(k) = (-1)\^{}k*b/2, where b is the Blue's min constant. A(k)\^{}2 underflows but the norm is (b/2)*sqrt(n).

\item A(k) = (-1)\^{}k*x, where x is the underflow threshold. A(k)\^{}2 underflows but the norm is x*sqrt(n).

\item A(k) = (-1)\^{}k*x, where x is the smallest subnormal number. A(k)\^{}2 underflows but the norm is x*sqrt(n).

* Mind that not all platforms might implement subnormal numbers.

\item A(k) = (-1)\^{}k*2*B/n, where B is the Blue's max constant, n greater than 1. A(k)\^{}2 and the norm are finite but sum\_k A(k)\^{}2 underflows.

\item A(k) = (-1)\^{}k*2*B, where B is the Blue's max constant. A(k)\^{}2 overflows but the norm is (2*B)*sqrt(n).

\item A(k) = b for k even, and A(k) = -7*b for k odd, where b is the Blue's min constant. The norm is 5*b*sqrt(n).

\item A(k) = B for k even, and A(k) = -7*B for k odd, where B is the Blue's max constant. The norm is 5*B*sqrt(n).

\item A(k) = (-1)\^{}k*2*OV/sqrt(n), n greater than 1. A(k) is finite but the norm overflows.
\end{enumerate}
\item Input contains 1 or more \Infs, but no \NaNs, so correct output is \Inf.
	Test cases add 1 or more \Infs~to all the cases under (1), in addition to adding \Infs~to the ``harmless" example:
    \begin{enumerate}
        \item[(i)] A(k) = (-1)\^{}k*k.
    \end{enumerate}
    \Inf~locations:
\begin{enumerate}[label={\roman*.}]
\item 1 \Inf~at: 1;  2; n/16; n/2;  n.
\item 2 \Infs~at: 1,2; 1,n/16;    1,n/2;    1,n;   
                      2,n/16;    2,n/2;    2,n;
                              n/16,n/2; n/16,n;
                                         n/2,n.
\item 3 \Infs~at: 1,2,n/16; 1,2,n/2;    1,2,n;
                        1,n/16,n/2; 1,n/16,n;
                                     1,n/2,n;
                        2,n/16,n/2; 2,n/16,n;
                                    2,n/16,n;
                                  n/16,n/2,n.
\item All \Infs.
\end{enumerate}
    \item Input contains 1 or more \NaNs, so correct output is a \NaN.
	Test cases add 1 or more \NaNs~to all the cases under (1) and (2). \NaN~locations: Same as the \Inf~locations in (2). For each \NaN~location scenario:
\begin{enumerate}
\item Input contains no \Infs.
\item Input contains \Infs:
\begin{enumerate}
\item
Insert \Inf~in first non-\NaN~location;
\item
Insert -Inf in first non-\NaN~location;
\item
Ditto for last non-\NaN~location;
\item
Ditto for first and last non-\NaN~locations.
\item
A(k) = (-1)\^{}k*\Inf.
\end{enumerate}
\item All \NaNs.
\end{enumerate}
\end{enumerate}

\subsection{Proof of Concept}

The tests described above, for both IxAMAX and xNRM2, were implemented in the testBLAS test suite~\footnote{\url{https://github.com/tlapack/testblas}}. This is a C++ library that contains tests for corner cases and Inf and NaN propagation for BLAS routines. It currently uses the $\langle$T$\rangle$LAPACK~\footnote{\url{https://github.com/tlapack/tlapack}} legacy interface to access BLAS implementations such as Reference BLAS, MKL, OpenBLAS, Apple BLAS and Arm Performance Libraries. Briefly speaking, each test in testBLAS calls routines in $\langle$T$\rangle$LAPACK, using a C++ legacy interface, which translates the Fortran BLAS interface. The $\langle$T$\rangle$LAPACK routine works as either a built-in C++ implementation or, if available, a wrapper to optimized BLAS code. The link between C++ wrappers and optimized BLAS is done by the library LAPACK++~\footnote{\url{https://bitbucket.org/icl/lapackpp}}.

We ran those two sets of tests (for IxAMAX and xNRM2) on seven BLAS libraries: Apple Accelerate 12.2.1, IBM ESSL 6.3.0, Intel MKL 2022.1.0, LAPACK 3.9.1, LAPACK 3.10.1, LAPACK 3.11-beta and OpenBLAS 0.3.8. For that, we use the machines:
\begin{itemize}
\item darwin-clang, macOS Monterey 12.2.1 using Apple clang version 13.1.6.
\item summit-gnu, Summit node using GNU compilers v9.1.0.
\item ubuntu-gnu, Ubuntu 20.04.4 LTS using GNU compilers v9.4.0.
\end{itemize}
~
LAPACK 3.11-beta is a copy of LAPACK 3.10.1 with I\{C,Z\}AMAX as proposed in Appendix~\ref{AppA}. We show a summary of the results in Tables \ref{tab:failingtestsISAMAX}, \ref{tab:failingtestsICAMAX} and \ref{tab:failingtestsSNRM2}.
~
The proposed test cases were able to identify several inconsistencies in the BLAS libraries we analyze; see the following paragraphs.

\begin{table}[!htb]
    \centering
    \begin{tabular}{|c|c|c|c|}
        \hline
        Test case: & 1 & 2 & 3 \\\hline
        Apple Accelerate 12.2.1 (darwin-clang) & (a)-(c) & (a)-(c), i-v & - \\\hline
        IBM ESSL 6.3.0 (summit-gnu) & (a)-(c) & (a)-(c), i-v & - \\\hline
        Intel MKL 2022.1.0 (ubuntu-gnu) & - & - & - \\\hline
        LAPACK 3.9.1 (ubuntu-gnu) & (a)-(c) & (a)-(c), i-v & - \\\hline
        LAPACK 3.10.1 (ubuntu-gnu) & (a)-(c) & (a)-(c), i-v & - \\\hline
        LAPACK 3.11-beta (ubuntu-gnu) & - & - & - \\\hline
        OpenBLAS 0.3.8 (ubuntu-gnu) & (a)-(d) & (a)-(c), i-v & - \\\hline
    \end{tabular}
    \caption{Unsatisfied tests for I\{S,D\}AMAX.}
    \label{tab:failingtestsISAMAX}
\end{table} 

\begin{table}[!htb]
    \centering
    \begin{tabular}{|c|c|c|c|c|c|}
        \hline
        Test case: & 1 & 2 & 3 & 4 & 5 \\\hline
        Apple Accelerate 12.2.1 (darwin-clang) & (a)-(d) & (a)-(c), i-v & - & A, B & A-D \\\hline
        IBM ESSL 6.3.0 (summit-gnu) & (a)-(c) & (a)-(c), i-v & - & A, B & A-D \\\hline
        Intel MKL 2022.1.0 (ubuntu-gnu) & - & - & - & A, B & A-D \\\hline
        LAPACK 3.9.1 (ubuntu-gnu) & (a)-(c) & (a)-(c), i-v & - & A, B & A-D \\\hline
        LAPACK 3.10.1 (ubuntu-gnu) & (a)-(c) & (a)-(c), i-v & - & A, B & A-D  \\\hline
        LAPACK 3.11-beta (ubuntu-gnu) & - & - & - & - & - \\\hline
        OpenBLAS 0.3.8 (ubuntu-gnu) & (a)-(d) & (a)-(c), i-v & - & A, B & A-D  \\\hline
    \end{tabular}
    \caption{Unsatisfied tests for I\{C,Z\}AMAX.}
    \label{tab:failingtestsICAMAX}
\end{table}

\begin{table}[!htb]
    \centering
    \begin{tabular}{|c|c|c|c|}
        \hline
        Test case: & 1 & 2 & 3\\\hline
        Apple Accelerate 12.2.1 (darwin-clang) & (a),(d)-(h) ** & (a)-(i) ** & - \\\hline
        IBM ESSL 6.3.0 (summit-gnu) & (c) * & - & - \\\hline
        Intel MKL 2022.1.0 (ubuntu-gnu) & - & - & - \\\hline
        LAPACK 3.9.1 (ubuntu-gnu) & (f)-(h) & (a)-(i) & - \\\hline
        LAPACK 3.10.1 (ubuntu-gnu) & - & - & - \\\hline
        LAPACK 3.11-beta (ubuntu-gnu) & - & - & - \\\hline
        OpenBLAS 0.3.8 (ubuntu-gnu) & - & - & - \\\hline
    \end{tabular}
    \caption{Unsatisfied tests for xNRM2. (*) Fails for double and complex double only. (**) Fails for float and complex float only.}
    \label{tab:failingtestsSNRM2}
\end{table} 

First, only I\{S,D\}AMAX from Intel MKL 2022.1.0 and LAPACK 3.11-beta conform to the proposed exception handling semantics. The other BLAS packages do not return the first \NaN as the output of I\{S,D\}AMAX. Notice that this also happens for vectors with only a single \NaN. In case 1(d), ``All \NaNs", I\{S,D\}AMAX from OpenBLAS 0.3.8 returns the first \NaN for the sizes $n = 1, 2, 3$, and returns the last \NaN for $n = 10, 128$. All implementations are able to identify the first \Infs when no \NaN is present.
The implementations of I\{C,Z\}AMAX in all tested packages, excluding LAPACK 3.11-beta, do not satisfy several tests. If the input is real, I\{C,Z\}AMAX returns the same value as I\{S,D\}AMAX as expected. All these implementations of I\{C,Z\}AMAX do not pass tests using the new input cases A, B, and also do not pass for the inputs A-D when the input contains \Infs. This happens because those routines are not able to compare the numbers x, y and \Inf, when both $|Re(x)| + |Im(x)|$ and $|Re(y)| + |Im(y)|$ overflow. I\{C,Z\}AMAX proposed in Appendix~\ref{AppA} pass all the proposed tests.

The implementation of xNRM2 in Intel MKL 2022.1.0, LAPACK 3.10.1 and OpenBLAS 0.3.8 pass all the tests.
Apple Accelerate 12.2.1 fails in tests (a) and (d)-(h) when no \Inf is present, and (a)-(i) when there is at least one \Inf in the input. This only happens on tests using single precision.
IBM ESSL 6.3.0 fails in the case with subnormal numbers in double precision and no \Infs. 
LAPACK 3.9.1 fails in tests (f)-(h) when no \Inf is present, and (a)-(i) when there is at least one \Inf input.
Note that the algorithms for xNRM2 from LAPACK 3.9.1 differ from the one in LAPACK 3.10.0; the latter comes from \cite{anderson2017l1scaling}. The new version, introduced as a safe-scaling NRM2, addresses all the proposed new
exception handling semantics.


\newpage
\section{Sample implementations of SGESV and routines in its call tree}\label{AppC}

This appendix include model implementations of our proposed LAPACK interface~\footnote{\url{https://github.com/BallisticLA/exception-handling/releases/tag/D9}}.
The 6 routines 
CHECKINIT1, CHECKINIT2, 
SGECHECKARG, SGEINFNAN, 
CHECKCALL, 
and 
UPDATE\_INFO are meant to be reusable across most LAPACK routines, encapsulating
most of the logic described in section~\ref{sec:argcheckD9}. The 4 routines
SGESV\_EC, SGETRF\_EC, SGETRF2\_EC and SGETRS\_EC are all the routines in the
call tree of SGESV\_EC, modified according to section~\ref{sec:argcheckD9}. 
(Note: These compile correctly, but have not been run to verify correctness.) 

\begin{itemize}
    \item CHECKINIT1 - initializes variables used to
    determine how to check and report errors.
    \item CHECKINIT2 - initializes variables used to report errors.
    \item SGECHECKARG - checks whether argument A contains \Infs or \NaNs, depending on the input variable FLAG\_REPORT\_INTERNAL, and accordingly updates reporting variables INFO\_INTERNAL and INFO\_ARRAY.
    \item SGEINFNAN - returns INFNAN = 1 if the matrix A contains an \Inf or \NaN, and 0 otherwise.          The location of the \Inf or \NaN is at A(I,J).
    \item CHECKCALL - checks whether an internal LAPACK call signaled an exception and updates reporting variables INFO\_INTERNAL and INFO\_ARRAY accordingly.
    \item UPDATE\_INFO - updates INFO and INFO\_ARRAY before an LAPACK routine returns.
    \item SGESV\_EC - SGESV with error checking
    \item SGETRF\_EC - SGETRF with error checking
    \item SGETRF2\_EC - SGETRF2 with error checking
    \item SGETRS\_EC - SGETRS with error checking
\end{itemize}

As a simple metric of how much more
complicated the code becomes to
incorporate error checking, the table
below has line counts for the original
LAPACK routines and their \_EC versions,
broken down into lines of comments,
and lines of code (declarations and executable).
The ``\%growth'' is the ratio of
the number of new lines (total\_ec) to the original number of lines (total). The 6 reusable routines total 1059 lines.

\begin{center}
    \begin{tabular}{|l|r|r|r|}
    \hline
         Routine    & Total Lines & Comments & Code \\ \hline
         sgesv      & 176         & 149      & 27  \\
         sgesv\_ec  & 396         & 322      & 74  \\
         difference & 220         & 173      & 47  \\
    \hline
        sgetrf      & 222         & 167      & 55  \\
        sgetrf\_ec  & 433         & 331      & 102 \\
        difference  & 211         & 164      & 47  \\
    \hline
        sgetrf2     & 269         & 194      & 75  \\
        sgetrf2\_ec & 477         & 358      & 119 \\
        difference  & 208         & 164      & 44  \\
    \hline
        sgetrs      & 222         & 175      & 47  \\
        sgetrs\_ec  & 394         & 313      & 81  \\
        difference  & 172         & 138      & 34  \\
    \hline \hline
        total       & 889         & 685      & 204 \\
        total\_ec   & 1700        & 1324     & 376 \\
        difference  & 811         & 639      & 172 \\
        \%growth& 191\%        & 193\%     & 184\%\\
    \hline
    \end{tabular}
\end{center}
We append below the call graph of all these routines.

\begin{center}
    \includegraphics[scale=.7]{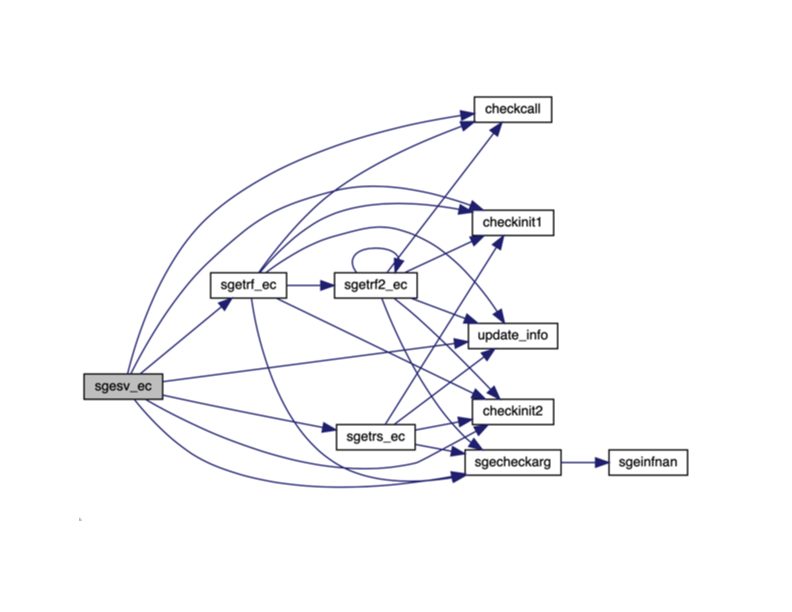}
\end{center}
 
\newpage
\begin{verbatim}
*> \brief \b CHECKINIT1
*
*  =========== DOCUMENTATION ===========
*
* Online html documentation available at
*            http://www.netlib.org/lapack/explore-html/
*
*  Definition:
*  ===========
*
*       SUBROUTINE CHECKINIT1( FLAG_REPORT, FLAG_REPORT_INTERNAL, 
*                              FLAG_REPORT_CALL, 
*                              CALL_REPORT_EXCEPTIONS, CONTEXT )
*
*       .. Scalar Arguments ..
*       LOGICAL            CALL_REPORT_EXCEPTIONS 
*       ..
*       .. Array Arguments ..
*       INTEGER            FLAG_REPORT( 2 ), FLAG_REPORT_INTERNAL( 2 )
*       INTEGER            FLAG_REPORT_CALL( 2 )
*       ..
*       .. Pointer Arguments ..
*       POINTER            CONTEXT ... advice requested
*
*> \par Purpose:
*  =============
*>
*> \verbatim
*>
*> CHECKINIT1 initializes variables used to determine how
*> to check and report errors.
*>
*> \endverbatim
*
*  Arguments:
*  ==========
*
*> \param[in] FLAG_REPORT
*> \verbatim
*>          FLAG_REPORT is INTEGER array, dimension(2)
*>          FLAG_REPORT(1) defines what kinds of exceptions to report,
*>          using INFO and possibly also INFO_ARRAY for more details.
*>          FLAG_REPORT(1)
*>           <= -1 turns off all error checking
*>            =  0 standard error checks only (eg LDA < 0)
*>            =  1 also check inputs and outputs for Infs and NaNs
*>           >=  2 also check input and output arguments of internal 
*>               LAPACK routines (not performed unless 
*>               FLAG_REPORT(2) >= 1)
*>           FLAG_REPORT(2) determines how errors are reported:
*>           FLAG_REPORT(2)
*>           <= 0 only returns INFO
*>            = 1 also returns INFO_ARRAY with more details
*>            = 2 also calls REPORT_EXCEPTIONS if there are errors
*>            = 3 also calls REPORT_EXCEPTIONS in internal LAPACK
*>                routines
*>           >= 4 will call GET_FLAGS_TO_REPORT(CONTEXT,FLAG_REPORT)
*>                to get values of FLAG_REPORT to use, overriding
*>                input values. The user needs to have called 
*>                SET_FLAGS_TO_REPORT(CONTEXT,FLAG_REPORT) 
*>                before calling SCHECKINIT in order to set 
*>                FLAG_REPORT, otherwise the default is 
*>                FLAG_REPORT = [0, 0]. The input array FLAG_REPORT 
*>                will not be overwritten.
*> \endverbatim
*>
*> \param[out] FLAG_REPORT_INTERNAL
*> \verbatim
*>          FLAG_REPORT_INTERNAL is INTEGER array, dimension(2)
*>          FLAG_REPORT_INTERNAL contains updated values of
*>          FLAG_REPORT on return.
*> \endverbatim
*>
*> \param[in] FLAG_REPORT_CALL
*> \verbatim
*>          FLAG_REPORT_CALL is INTEGER array, dimension(2)
*>          FLAG_REPORT_CALL contains the values of FLAG_REPORT
*>          to be used in internal LAPACK calls.
*>
*> \param[out] CALL_REPORT_EXCEPTIONS
*> \verbatim
*>          CALL_REPORT_EXCEPTIONS is LOGICAL
*>          CALL_REPORT_EXCEPTIONS is true if REPORT_EXCEPTIONS should 
*>          be called, and false if not.
*> \endverbatim
*>
*> \param[in] CONTEXT
*> \verbatim
*>          CONTEXT is POINTER to an "opaque object"
*> \endverbatim
*>
*  Authors:
*  ========
*
*> \author Univ. of Tennessee
*> \author Univ. of California Berkeley
*> \author Univ. of Colorado Denver
*> \author NAG Ltd.
*
*> \ingroup realGEcomputational
*
*  =====================================================================
      SUBROUTINE CHECKINIT1( FLAG_REPORT, FLAG_REPORT_INTERNAL, 
     $                       FLAG_REPORT_CALL,
     $                       CALL_REPORT_EXCEPTIONS, CONTEXT )
*
*  -- LAPACK computational routine --
*  -- LAPACK is a software package provided by Univ. of Tennessee,    --
*  -- Univ. of California Berkeley, Univ. of Colorado Denver and NAG Ltd..--
*
*     .. Scalar Arguments ..
      LOGICAL            CALL_REPORT_EXCEPTIONS 
*     ..
*     .. Array Arguments ..
      INTEGER            FLAG_REPORT( 2 ), FLAG_REPORT_INTERNAL( 2 )
      INTEGER            FLAG_REPORT_CALL( 2 )
*     ..
*     .. Pointer Arguments ..
      POINTER            CONTEXT 
*
*  =====================================================================
*
*     .. Parameters ..
      INTEGER, DIMENSION(4), PARAMETER ::
     $   WHAT_NEXT = (/ -1, 0, 0, 2 /)
      INTEGER, DIMENSION(4), PARAMETER ::
     $   HOW_NEXT = (/ 0, 1, 1, 3 /)
*
*     .. Local Scalars ..
      INTEGER            WHAT, HOW
*
*     .. Local Arrays ..
      INTEGER            FLAG_REPORT_TMP( 2 )       
*     ..
*     .. External Subroutines ..
      EXTERNAL           GET_FLAGS_TO_REPORT
*     ..
*     .. Intrinsic Functions ..
      INTRINSIC          MAX, MIN
*     ..
*     .. Executable Statements ..
*
*     WHAT records what errors are to be reported
      WHAT = MAX( -1, MIN( FLAG_REPORT(1), 2 ) )
*     HOW records how those errors are to be reported
      HOW = 0
      CALL_REPORT_EXCEPTIONS = .FALSE.
      FLAG_REPORT_INTERNAL( 1 ) = WHAT
      FLAG_REPORT_INTERNAL( 2 ) = HOW
      FLAG_REPORT_CALL( 1 ) = WHAT_NEXT(WHAT+2)
      FLAG_REPORT_CALL( 2 ) = HOW_NEXT(HOW+1)
*     Check if error reporting turned off
      IF (WHAT .EQ. -1) RETURN
      HOW = MAX( 0, MIN(FLAG_REPORT(2), 4 ) )
      FLAG_REPORT_INTERNAL( 2 ) = HOW
      IF (HOW .EQ. 4) THEN
*        Get updated FLAG_REPORT
         CALL GET_FLAGS_TO_REPORT(CONTEXT, FLAG_REPORT_TMP)
         WHAT = MAX( -1, MIN( FLAG_REPORT_TMP(1), 2 ) )
         FLAG_REPORT_INTERNAL( 1 ) = WHAT
         FLAG_REPORT_CALL( 1 ) = WHAT_NEXT(WHAT+2)
         IF (WHAT .EQ. -1) THEN
*          Error reporting turned off
           HOW = 0
           FLAG_REPORT_INTERNAL( 2 ) = HOW
           FLAG_REPORT_CALL( 2 ) = HOW_NEXT(HOW+1)
           RETURN
         END IF
         HOW = MAX( 0, MIN(FLAG_REPORT_TMP(2), 3 ) )
         FLAG_REPORT_INTERNAL( 2 ) = HOW
         FLAG_REPORT_CALL( 2 ) = HOW_NEXT(HOW+1)
      END IF
*     Error reporting not turned off
*     Decide whether to call REPORT_EXCEPTIONS
      IF (HOW .GE. 2) CALL_REPORT_EXCEPTIONS = .TRUE.
*     
      RETURN
*
*     End of CHECKINIT1
*
      END
\end{verbatim}

\newpage
\begin{verbatim}
*> \brief \b CHECKINIT2
*
*  =========== DOCUMENTATION ===========
*
* Online html documentation available at
*            http://www.netlib.org/lapack/explore-html/
*
*  Definition:
*  ===========
*
*       SUBROUTINE CHECKINIT2( FLAG_REPORT_INTERNAL, INFO, 
*      $                       INFO_INTERNAL, INFO_ARRAY, NUMARGS, 
*      $                       NUMCALLS )
*
*       .. Scalar Arguments ..
*       INTEGER            INFO, NUMARGS, NUMCALLS
*       ..
*       .. Array Arguments ..
*       INTEGER            FLAG_REPORT_INTERNAL( 2 ) 
*       INTEGER            INFO_INTERNAL( 2 )
*       INTEGER            INFO_ARRAY( * )
*
*> \par Purpose:
*  =============
*>
*> \verbatim
*>
*> CHECKINIT2 initializes variables used to report errors.
*>
*> \endverbatim
*
*  Arguments:
*  ==========
*
*> \param[in] FLAG_REPORT_INTERNAL
*> \verbatim
*>          FLAG_REPORT_INTERNAL is INTEGER array, dimension(2)
*>          FLAG_REPORT_INTERNAL(1) defines what kinds of exceptions 
*>          to report, using INFO and possibly also INFO_ARRAY for 
*>          more details.
*>          FLAG_REPORT_INTERNAL(1)
*>           <= -1 turns off all error checking
*>            =  0 standard error checks only (eg LDA < 0)
*>            =  1 also check inputs and outputs for Infs and NaN
*>           >=  2 also check input and output arguments of internal 
*>               LAPACK routines
*>           FLAG_REPORT_INTERNAL(2) determines how errors are 
*>           reported:
*>           FLAG_REPORT_INTERNAL(2)
*>           <= 0 only returns INFO
*>            = 1 also return INFO_ARRAY with more details
*>            = 2 also calls REPORT_EXCEPTIONS if there are errors
*>           >= 3 also calls REPORT_EXCEPTIONS in internal LAPACK
*>                routines
*> \endverbatim
*>
*> \param[in] INFO
*> \verbatim
*>          INFO is INTEGER
*>          INFO is used to report errors, and contains the value
*>          of INFO from standard input argument checking.
*> \endverbatim
*>
*> \param[out] INFO_INTERNAL
*> \verbatim
*>          INFO_INTERNAL is INTEGER array, dimension(2)
*>          INFO_INTERNAL( 1 ) is used to track potential changes
*>          to INFO from errors detected by checking input and output
*>          arguments.
*>          INFO_INTERNAL( 2 ) is used to track potential changes
*>          to INFO from errors detected by internal subroutine calls.
*>          Both entries are initialized to 0.
*> \endverbatim
*>
*> \param[in,out] INFO_ARRAY
*> \verbatim
*>          INFO_ARRAY is INTEGER array, dimension( * )
*>          INFO_ARRAY is used to report errors, and is initialized.
*> \endverbatim
*>
*> \param[in] NUMARGS
*> \verbatim
*>          NUMARGS is INTEGER
*>          NUMARGS is the number of input and output arguments
*>          reported in INFO_ARRAY, in locations 
*>          INFO_ARRAY(7:6+NUMARGS). 
*>          NUMARGS is used to initialize INFO_ARRAY.
*> \endverbatim
*>
*> \param[in] NUMCALLS
*> \verbatim
*>          NUMCALLS is INTEGER
*>          NUMCALLS is the number of internal LAPACK calls
*>          reported in INFO_ARRAY, in locations
*>          INFO_ARRAY(7+NUMARGS:6+NUMARGS+NUMCALLS).
*>          NUMCALLS is used to initialize INFO_ARRAY.
*> \endverbatim
*>
*  Authors:
*  ========
*
*> \author Univ. of Tennessee
*> \author Univ. of California Berkeley
*> \author Univ. of Colorado Denver
*> \author NAG Ltd.
*
*> \ingroup realGEcomputational
*
*  =====================================================================
      SUBROUTINE CHECKINIT2( FLAG_REPORT_INTERNAL, INFO, INFO_INTERNAL, 
     $                       INFO_ARRAY, NUMARGS, NUMCALLS )
*
*  -- LAPACK computational routine --
*  -- LAPACK is a software package provided by Univ. of Tennessee,    --
*  -- Univ. of California Berkeley, Univ. of Colorado Denver and NAG Ltd..--
*
*     .. Scalar Arguments ..
      INTEGER            INFO, NUMARGS, NUMCALLS
*     ..
*     .. Array Arguments ..
      INTEGER            FLAG_REPORT_INTERNAL( 2 ), INFO_INTERNAL( 2 )
      INTEGER            INFO_ARRAY( * )
*  =====================================================================
*
*     .. Local Scalars ..
      INTEGER            WHAT, HOW
*
*     ..
*     .. Executable Statements ..
*
      INFO_INTERNAL( 1 ) = 0
      INFO_INTERNAL( 2 ) = 0
      WHAT = FLAG_REPORT_INTERNAL( 1 )
      HOW = FLAG_REPORT_INTERNAL( 2 )
      IF (HOW .GE. 1) THEN
*        Initialize INFO_ARRAY
         INFO_ARRAY(1) = INFO
         INFO_ARRAY(2) = WHAT
         INFO_ARRAY(3) = HOW
         INFO_ARRAY(4) = INFO
         INFO_ARRAY(5) = NUMARGS
         INFO_ARRAY(6) = NUMCALLS
         IF (NUMARGS .GT. 0) THEN
            DO 10 I = 7, 6+NUMARGS
*              If argument already marked as checked, do not
*              reinitialize.
               IF (INFO_ARRAY(I).NE.0 .AND. INFO_ARRAY(I).NE.1)
     $            INFO_ARRAY(I) = -1    
10          CONTINUE
         ENDIF
         IF (NUMCALLS .GT. 0) THEN
            DO 20 I = 7+NUMARGS, 6+NUMARGS+NUMCALLS
               INFO_ARRAY(I) = -1
20          CONTINUE
         ENDIF
      END IF
*     
      RETURN
*
*     End of CHECKINIT2
*
      END    
\end{verbatim}

\newpage
\begin{verbatim}
*> \brief \b SGECHECKARG
*
*  =========== DOCUMENTATION ===========
*
* Online html documentation available at
*            http://www.netlib.org/lapack/explore-html/
*
*  Definition:
*  ===========
*
*       SUBROUTINE SGECHECKARG( FLAG_REPORT_INTERNAL, M, N, A, LDA, 
*      $                        INFO, INFO_INTERNAL, INFO_ARRAY,
*      $                        ARGNUM, INOUT, ERRFLAG, LOC )
*
*       .. Scalar Arguments ..
*       INTEGER            M, N, LDA, ARGNUM, INOUT, INFO, ERRFLAG 
*       INTEGER            LOC
*       ..
*       .. Array Arguments ..
*       REAL               A( LDA, * )
*       INTEGER            FLAG_REPORT_INTERNAL( 2 ) 
*       INTEGER            INFO_INTERNAL( 2 ), INFO_ARRAY( * )
*
*> \par Purpose:
*  =============
*>
*> \verbatim
*>
*> SGECHECKARG checks whether argument A contains Infs or NaNs,
*> depending on the input variable FLAG_REPORT_INTERNAL, and 
*> accordingly updates reporting variables INFO_INTERNAL and 
*> INFO_ARRAY.
*>
*> \endverbatim
*
*  Arguments:
*  ==========
*
*> \param[in] FLAG_REPORT_INTERNAL
*> \verbatim
*>          FLAG_REPORT_INTERNAL is INTEGER array, dimension(2)
*>          FLAG_REPORT_INTERNAL(1) defines what kinds of exceptions 
*>          to report, using INFO and possibly also INFO_ARRAY for 
*>          more details.
*>          FLAG_REPORT_INTERNAL(1)
*>           <= -1 turns off all error checking
*>            =  0 standard error checks only (eg LDA < 0)
*>            =  1 also check inputs and outputs for Infs and NaN
*>           >=  2 also check input and output arguments of internal 
*>               LAPACK routines (not performed unless 
*>               FLAG_REPORT_INTERNAL(2) >= 1)
*>           FLAG_REPORT_INTERNAL(2) determines how errors are 
*>           reported:
*>           FLAG_REPORT_INTERNAL(2)
*>           <= 0 only returns INFO
*>            = 1 also return INFO_ARRAY with more details
*>            = 2 also calls REPORT_EXCEPTIONS if there are errors
*>           >= 3 also calls REPORT_EXCEPTIONS in internal LAPACK
*>                routines
*> \endverbatim
*>
*> \param[in] M
*> \verbatim
*>          M is INTEGER
*>          The number of rows in A. M >= 0
*> \endverbatim
*>
*> \param[in] N
*> \verbatim
*>          N is INTEGER
*>          The number of columns in A. M >= 0
*> \endverbatim
*>
*> \param[in] A
*> \verbatim
*>          A is REAL array, dimension (LDA,N)
*>          Matrix to be checked for containing Infs and NaNs
*> \endverbatim
*>
*> \param[in] LDA
*> \verbatim
*>          LDA is INTEGER
*>          The leading dimension of the array A.  LDA >= max(1,M).
*> \endverbatim
*>
*> \param[in] INFO
*> \verbatim
*>          INFO is INTEGER
*>          INFO contains the standard value that LAPACK would return
*> \endverbatim
*>
*> \param[in,out] INFO_INTERNAL
*> \verbatim
*>          INFO_INTERNAL is INTEGER array, dimension(2)
*>          INFO_INTERNAL( 1 ) is used to track potential changes
*>          to INFO from errors detected by checking input and output
*>          arguments.
*>          INFO_INTERNAL( 2 ) is not accessed.
*> \endverbatim
*>
*> \param[in,out] INFO_ARRAY
*> \verbatim
*>          INFO_ARRAY is INTEGER ARRAY
*>          INFO_ARRAY can report error checks for each
*>          floating point input and output, and each 
*>          internal subroutine call.
*> \endverbatim
*>
*> \param[in] ARGNUM
*> \verbatim
*>          ARGNUM is INTEGER
*>          A is ARGNUM-th argument of routine calling SGECHECKARG
*> \endverbatim
*>
*> \param[in] INOUT
*> \verbatim
*>          INOUT is INTEGER
*>          INOUT = 0 if A is an input-only argument
*>                    in the routine calling SGECHECKARG.
*>          INOUT = 1 if A is an output-only argument.
*>          INOUT = 2 if A is both an input and output argument,
*>                    and is being checked on input.
*>          INOUT = 3 if A is both an input and output argument,
*>                    and is being checked on output.
*> \endverbatim
*>
*> \param[in] ERRFLAG
*> \verbatim
*>          ERRFLAG is INTEGER
*>          INFO_INTERNAL(1) = ERRFLAG is used to indicate that
*>          A contains an Inf or NaN on output, but not input.
*>          Not accessed if A is checked on input (INOUT = 0 or 2).
*> \endverbatim
*>
*> \param[in] LOC
*> \verbatim
*>          LOC is INTEGER
*>          LOC points to the entry of INFO_ARRAY used to report on A
*> \endverbatim
*>
*  Authors:
*  ========
*
*> \author Univ. of Tennessee
*> \author Univ. of California Berkeley
*> \author Univ. of Colorado Denver
*> \author NAG Ltd.
*
*> \ingroup realGEcomputational
*
*  =====================================================================
      SUBROUTINE SGECHECKARG( FLAG_REPORT_INTERNAL, M, N, A, LDA, 
     $                        INFO, INFO_INTERNAL, INFO_ARRAY,
     $                        ARGNUM, INOUT, ERRFLAG, LOC )
*
*  -- LAPACK computational routine --
*  -- LAPACK is a software package provided by Univ. of Tennessee,    --
*  -- Univ. of California Berkeley, Univ. of Colorado Denver and NAG Ltd..--
*
*     .. Scalar Arguments ..
      INTEGER            M, N, LDA, ARGNUM, INOUT, INFO, ERRFLAG 
      INTEGER            LOC
*     ..
*     .. Array Arguments ..
      REAL               A( LDA, * )
      INTEGER            FLAG_REPORT_INTERNAL( 2 ) 
      INTEGER            INFO_INTERNAL( 2 ), INFO_ARRAY( * )
*
*  =====================================================================
*
*     .. Local Scalars ..
      INTEGER            WHAT, HOW, INFO_A, INFNAN, II, JJ      
*     ..
*     .. External Subroutines ..
      EXTERNAL           SGEINFNAN
*     ..
*     .. Intrinsic Functions ..
      INTRINSIC          MAX, MIN
*     ..
*     .. Executable Statements ..
*
*     Check for exceptional entries in A 
      WHAT = FLAG_REPORT_INTERNAL( 1 )
      IF (WHAT .GE. 1) THEN
*        Check input or output for Infs and NaNs
         HOW = FLAG_REPORT_INTERNAL( 2 )
         IF (HOW .GE. 1) THEN
*           INFO_ARRAY used for reporting, as well as INFO
            INFO_A = INFO_ARRAY(LOC)
            IF (INOUT .EQ. 0 .OR. INOUT .EQ. 2) THEN
*              Checking A on input
               IF (INFO_A .NE. 0 .AND. INFO_A .NE. 1) THEN
*                 A not checked yet, so check
                  CALL SGEINFNAN( M, N, A, LDA, INFNAN, II, JJ )
                  INFO_ARRAY(LOC) = INFNAN
               ENDIF
*              Update INFO_INTERNAL to point to A if not already set
               IF (INFO_INTERNAL(1).EQ.0 .AND. INFO_ARRAY(LOC).EQ.1)
     $            INFO_INTERNAL(1) = -ARGNUM
            ELSEIF (INOUT .EQ. 1) THEN
*              Checking output-only variable A on output
               CALL SGEINFNAN( M, N, A, LDA, INFNAN, II, JJ )
               INFO_ARRAY(LOC) = 2*INFNAN
*              Update INFO_INTERNAL to point to A if not already set
               IF (INFO_INTERNAL(1).EQ.0 .AND. INFO_ARRAY(LOC).EQ.2)
     $            INFO_INTERNAL(1) = ERRFLAG
            ELSE
*              Checking input-output variable A on output
               CALL SGEINFNAN( M, N, A, LDA, INFNAN, II, JJ )
               IF (INFNAN .EQ. 1) INFO_ARRAY(LOC) = INFO_ARRAY(LOC)+2
*              Update INFO_INTERNAL to point to A if not already set
               IF (INFO_INTERNAL(1).EQ.0 .AND. INFO_ARRAY(LOC).EQ.2)
     $            INFO_INTERNAL(1) = ERRFLAG
            ENDIF
         ELSEIF (INFO_INTERNAL(1) .EQ. 0 .AND. INFO .EQ. 0) THEN
*           Only INFO used for reporting, not set yet
            CALL SGEINFNAN( M, N, A, LDA, INFNAN, II, JJ )
            IF (INOUT .EQ. 0 .OR. INOUT .EQ. 2) THEN
*              Checking A on input
               IF (INFNAN .EQ. 1) INFO_INTERNAL(1) = -ARGNUM
            ELSE
*              Checking A on output
               IF (INFNAN .EQ. 1) INFO_INTERNAL(1) = ERRFLAG
            ENDIF
         ENDIF
      ENDIF
*     
      RETURN
*
*     End of SGECHECKARG
*
      END   
\end{verbatim}

\newpage
\begin{verbatim}
*> \brief \b SGEINFNAN
*
*  =========== DOCUMENTATION ===========
*
* Online html documentation available at
*            http://www.netlib.org/lapack/explore-html/
*
*  Definition:
*  ===========
*
*       SUBROUTINE SGEINFNAN( M, N, A, LDA, INFNAN, I, J )
*
*       .. Scalar Arguments ..
*       INTEGER            LDA, M, N, INFNAN, I, J
*       ..
*       .. Array Arguments ..
*       REAL               A( LDA, * )
*       ..
*
*
*> \par Purpose:
*  =============
*>
*> \verbatim
*>
*> SGEINFNAN  returns INFNAN = 1 if the matrix A contains an Inf or NaN, and 0 otherwise,
*>            The location of the Inf or NaN is at A(I,J)
*> \endverbatim
*>
*> \return SGEINFNAN
*
*  Arguments:
*  ==========
*
*>
*> \param[in] M
*> \verbatim
*>          M is INTEGER
*>          The number of rows of the matrix A.  M >= 0.  When M = 0,
*>          INFNAN is set to 0.
*> \endverbatim
*>
*> \param[in] N
*> \verbatim
*>          N is INTEGER
*>          The number of columns of the matrix A.  N >= 0.  When N = 0,
*>          INFNAN is set to 0.
*> \endverbatim
*>
*> \param[in] A
*> \verbatim
*>          A is REAL array, dimension (LDA,N)
*>          The M by N matrix A.
*> \endverbatim
*>
*> \param[in] LDA
*> \verbatim
*>          LDA is INTEGER
*>          The leading dimension of the array A.  LDA >= max(M,1).
*> \endverbatim
*>
*> \param[out] INFNAN
*> \verbatim
*>          INFNAN is INTEGER
*>          INFNAN = 1 if A contains an Inf or NaN, and 0 otherwise.
*> \endverbatim
*>
*> \param[out] I
*> \verbatim
*>          I is INTEGER
*>          If A contains an Inf or NaN, one is located at A(I,J),
*>          otherwise I = 0.
*> \endverbatim
*>
*> \param[out] J
*> \verbatim
*>          J is INTEGER
*>          If A contains an Inf or NaN, one is located at A(I,J),
*>          otherwise J = 0. 
*
*  Authors:
*  ========
*
*> \author Univ. of Tennessee
*> \author Univ. of California Berkeley
*> \author Univ. of Colorado Denver
*> \author NAG Ltd.
*
*> \ingroup realGEauxiliary
*
*  =====================================================================
      SUBROUTINE SGEINFNAN( M, N, A, LDA, INFNAN, I, J )
*
*  -- LAPACK auxiliary routine --
*  -- LAPACK is a software package provided by Univ. of Tennessee,    --
*  -- Univ. of California Berkeley, Univ. of Colorado Denver and NAG Ltd..--
*
      IMPLICIT NONE
*     .. Scalar Arguments ..
      INTEGER            LDA, M, N, INFNAN, I, J
*     ..
*     .. Array Arguments ..
      REAL               A( LDA, * )
*     ..
*
* =====================================================================
*
*     .. Parameters ..
      REAL               ONE
      PARAMETER          ( ONE = 1.0E+0 )
*     ..
*     .. Local Scalars ..
      INTEGER            MAXEXP
*     ..
*     .. Intrinsic Functions ..
      INTRINSIC          EXPONENT, HUGE
*     ..
*     .. Executable Statements ..
*
      INFNAN = 0
      I = 0
      J = 0
      IF( MIN( M, N ).EQ.0 ) RETURN
*     MAXEXP is the exponent of an Inf or NaN
      MAXEXP = EXPONENT(HUGE(ONE)) + 1
      DO 20 J = 1, N
         DO 10 I = 1, M
*           It could be faster to use the function IEEE_IS_FINITE 
*           provided by the intrinsic module IEEE_ARITHMETIC, but 
*           whether this module is provided is processor dependent, 
*           according to the Fortran 2008 standard.
            IF( EXPONENT(A(I,J)) .EQ. MAXEXP ) THEN
*              A(I,J) is an Inf or NaN
               INFNAN = 1
               RETURN
            ENDIF
   10    CONTINUE
   20 CONTINUE
      I = 0
      J = 0
      RETURN
*
*     End of SGEINFNAN
*
      END
\end{verbatim}

\newpage
\begin{verbatim}
*> \brief \b CHECKCALL
*
*  =========== DOCUMENTATION ===========
*
* Online html documentation available at
*            http://www.netlib.org/lapack/explore-html/
*
*  Definition:
*  ===========    
*
*       SUBROUTINE CHECKCALL( FLAG_REPORT_INTERNAL, INFO_INTERNAL, 
*      $                      INFO_CALLARRAY, INFO_ARRAY,
*      $                      CALL_ID, LOC )
*
*       .. Scalar Arguments ..
*       INTEGER            CALL_ID, LOC
*       ..
*       .. Array Arguments ..
*       INTEGER            FLAG_REPORT_INTERNAL(2), INFO_INTERNAL(2)
*       INTEGER            INFO_CALLARRAY( * ), INFO_ARRAY( * )
*
*> \par Purpose:
*  =============
*>
*> \verbatim
*>
*> CHECKCALL checks whether an internal LAPACK call signaled an
*> exception and updates reporting variables INFO_INTERNAL and
*> INFO_ARRAY accordingly.
*>
*> \endverbatim
*
*  Arguments:
*  ==========
*
*> \param[in] FLAG_REPORT_INTERNAL
*> \verbatim
*>          FLAG_REPORT_INTERNAL is INTEGER array, dimension(2)
*>          FLAG_REPORT_INTERNAL(1) defines what kinds of exceptions 
*>          to report, using INFO and possibly also INFO_ARRAY for 
*>          more details.
*>          FLAG_REPORT_INTERNAL(1)
*>           <= -1 turns off all error checking
*>            =  0 standard error checks only (eg LDA < 0)
*>            =  1 also check inputs and outputs for Infs and NaN
*>           >=  2 also check input and output arguments of internal 
*>               LAPACK routines (not performed unless 
*>               FLAG_REPORT_INTERNAL(2) >= 1)
*>           FLAG_REPORT_INTERNAL(2) determines how errors are 
*>           reported:
*>           FLAG_REPORT_INTERNAL(2)
*>           <= 0 only returns INFO
*>            = 1 also return INFO_ARRAY with more details
*>            = 2 also calls REPORT_EXCEPTIONS if there are errors
*>           >= 3 also calls REPORT_EXCEPTIONS in internal LAPACK
*>                routines
*> \endverbatim
*>
*> \param[out] INFO_INTERNAL
*> \verbatim
*>          INFO_INTERNAL is INTEGER array, dimension(2)
*>          INFO_INTERNAL( 1 ) is not accessed.
*>          INFO_INTERNAL( 2 ) is used to track potential changes
*>          to INFO from errors detected by internal subroutine calls.
*> \endverbatim
*>
*> \param[in] INFO_CALLARRAY
*> \verbatim
*>          INFO_CALLARRAY is INTEGER array, dimension (*)
*>          When HOW >= 1, INFO_CALLARRAY is the INFO_ARRAY returned
*>          by the routine being checked.
*> \endverbatim
*>
*> \param[in,out] INFO_ARRAY
*> \verbatim
*>          INFO_ARRAY is INTEGER ARRAY
*>          INFO_ARRAY can report error checks for each
*>          floating point input and output, and each 
*>          internal subroutine call.
*> \endverbatim
*>
*> \param[in] CALL_ID 
*> \verbatim
*>          CALL_ID is INTEGER
*>          CALL_ID is a unique identifier for each internal
*>          LAPACK call that may be checked for exceptions.
*> \endverbatim
*>
*> \param[in] LOC
*> \verbatim
*>          LOC is INTEGER
*>          LOC points to the entry of INFO_ARRAY used to report
*>          on the internal call identified by CALL_ID
*> \endverbatim
*>
*>
*  Authors:
*  ========
*
*> \author Univ. of Tennessee
*> \author Univ. of California Berkeley
*> \author Univ. of Colorado Denver
*> \author NAG Ltd.
*
*> \ingroup realGEcomputational
*
*  =====================================================================
      SUBROUTINE CHECKCALL( FLAG_REPORT_INTERNAL, INFO_INTERNAL,
     $                      INFO_CALLARRAY, INFO_ARRAY,
     $                      CALL_ID, LOC )
*
*  -- LAPACK computational routine --
*  -- LAPACK is a software package provided by Univ. of Tennessee,    --
*  -- Univ. of California Berkeley, Univ. of Colorado Denver and NAG Ltd..--
*
*     .. Scalar Arguments ..
      INTEGER            CALL_ID, LOC
*     ..
*     .. Array Arguments ..
      INTEGER            FLAG_REPORT_INTERNAL(2), INFO_INTERNAL(2)
      INTEGER            INFO_CALLARRAY( * ), INFO_ARRAY( * )
*  =====================================================================
*
*     .. Local Scalars ..
      INTEGER            WHAT, HOW
      INTEGER            TMP, TMPINOUT, TMPCALLS, I, NUMARGS, NUMCALLS                
*     ..
*     .. Intrinsic Functions ..
      INTRINSIC          MAX, MIN
*     ..
*     .. Executable Statements ..
*
      WHAT = FLAG_REPORT_INTERNAL( 1 )
      HOW = FLAG_REPORT_INTERNAL( 2 )
      IF (WHAT .GE. 2 .AND. HOW .GE. 1) THEN
*     Check inputs, outputs and internal calls of LAPACK call
*        Update INFO_ARRAY( LOC )
*        Determine whether called routine had any Inf or NaN
*        inputs or outputs
         TMP = INFO_ARRAY( LOC )
*        Determine whether called routine had any Infs or NaNs
*        reported by its own internal calls
         TMPCALLS = 0
*        NUMARGS = number of arguments reported in INFO_CALLARRAY
         NUMARGS = INFO_CALLARRAY(5)
*        NUMCALLS = number of internal subroutine calls reported in
*        INFO_ALLARRAY
         NUMCALLS = INFO_CALLARRAY(6)
         IF (NUMCALLS .GT. 0) THEN
            DO 10 I = 7+NUMARGS, 6+NUMARGS+NUMCALLS
               TMPCALLS = MAX( TMPCALLS, INFO_CALLARRAY( I ) )
10          CONTINUE
         ENDIF
         IF (TMPCALLS .GE. 1) TMP = MAX( TMP, 1 )
*        Determine whether called routine had any inputs or outputs
*        containing Infs or NaNs
         TMPINOUT = 0
         IF (NUMARGS .GT. 0) THEN
            DO 20 I = 7, 6+NUMARGS
               TMPINOUT = MAX( TMPINOUT, INFO_CALLARRAY( I ) )
20          CONTINUE
         ENDIF
         IF (TMPINOUT .GT. 0) TMP = MAX( TMP, TMPINOUT+1 )
         INFO_ARRAY( LOC ) = TMP
*        Update INFO_INTERNAL
         IF (INFO_INTERNAL( 2 ) .EQ. 0 .AND. TMP .GT. 0) 
     $      INFO_INTERNAL( 2 ) =  CALL_ID 
      ENDIF
*     
      RETURN
*
*     End of CHECKCALL
*
      END
\end{verbatim}

\newpage
\begin{verbatim}
*> \brief \b UPDATE_INFO
*
*  =========== DOCUMENTATION ===========
*
* Online html documentation available at
*            http://www.netlib.org/lapack/explore-html/
*
*  Definition:
*  ===========
*
*       SUBROUTINE UPDATE_INFO( INFO, INFO_ARRAY, INFO_INTERNAL )
*
*       .. Scalar Arguments ..
*       INTEGER            INFO 
*       ..
*       .. Array Arguments ..
*       INTEGER            INFO_ARRAY( * ), INFO_INTERNAL( 2 )
*
*> \par Purpose:
*  =============
*>
*> \verbatim
*>
*> UPDATE_INFO updates INFO and INFO_ARRAY before an LAPACK routine 
*> returns.
*>
*> \endverbatim
*
*  Arguments:
*  ==========
*
*> \param[in,out] INFO
*> \verbatim
*>          INFO is INTEGER
*>          On input, INFO contains the standard value that LAPACK 
*>          would return.
*>          On output, INFO also reports any exceptional inputs or 
*>          outputs, or exceptions in internal subroutine calls, 
*>          as reported in INFO_INTERNAL
*> \endverbatim
*>
*> \param[in,out] INFO_ARRAY
*> \verbatim
*>          INFO_ARRAY(1) is set to the input value of INFO.
*>          INFO_ARRAY(4) is set to the updated value of INFO.
*>          
*> \endverbatim
*>
*> \param[in] INFO_INTERNAL
*> \verbatim
*>          INFO_INTERNAL is INTEGER array, dimension(2)
*>          INFO_INTERNAL( 1 ) is used to track potential changes
*>          to INFO from errors detected by checking input and output
*>          arguments.
*>          INFO_INTERNAL( 2 ) is used to track potential changes
*>          to INFO from errors detected by internal subroutine calls.
*> \endverbatim
*>
*>
*  Authors:
*  ========
*
*> \author Univ. of Tennessee
*> \author Univ. of California Berkeley
*> \author Univ. of Colorado Denver
*> \author NAG Ltd.
*
*> \ingroup realGEcomputational
*
*  =====================================================================
      SUBROUTINE UPDATE_INFO( INFO, INFO_ARRAY, INFO_INTERNAL )
*
*  -- LAPACK computational routine --
*  -- LAPACK is a software package provided by Univ. of Tennessee,    --
*  -- Univ. of California Berkeley, Univ. of Colorado Denver and NAG Ltd..--
*
*     .. Scalar Arguments ..
      INTEGER            INFO 
*     ..
*     .. Array Arguments ..
      INTEGER            INFO_ARRAY( * ), INFO_INTERNAL( 2 )
*
*  =====================================================================
*
*     ..
*     .. Executable Statements ..
*
      INFO_ARRAY(1) = INFO
      INFO_ARRAY(4) = INFO
      IF (INFO_INTERNAL(1) .GT. 0 .AND. INFO_ARRAY(4) .EQ. 0) THEN
*         Update INFO and INFO_ARRAY(4) to indicate exceptional
*         input or output
          INFO_ARRAY(4) = INFO_INTERNAL(1)
          INFO = INFO_INTERNAL(1)
      ENDIF
      IF (INFO_INTERNAL(2) .GT. 0 .AND. INFO_ARRAY(3) .GE. 1 .AND.
     $    INFO_ARRAY(4) .EQ. 0) THEN
*         Update INFO and INFO_ARRAY(4) to indicate exceptions
*         in internal LAPACK call
          INFO_ARRAY(4) = INFO_INTERNAL(2)
          INFO = INFO_INTERNAL(2)
      ENDIF
*     
      RETURN
*
*     End of UPDATE_INFO
*
      END
\end{verbatim}

\newpage
\begin{verbatim}
*> \brief <b> SGESV_EC computes the solution to system of linear equations 
*> A * X = B for GE matrices</b> (simple driver)
*
*  =========== DOCUMENTATION ===========
*
* Online html documentation available at
*            http://www.netlib.org/lapack/explore-html/
*
*  Definition:
*  ===========
*
*       SUBROUTINE SGESV_EC( N, NRHS, A, LDA, IPIV, B, LDB, 
*      $                     INFO, FLAG_REPORT, INFO_ARRAY, CONTEXT )
*
*       .. Scalar Arguments ..
*       INTEGER            INFO, LDA, LDB, N, NRHS
*       ..
*       .. Array Arguments ..
*       INTEGER            IPIV( * )
*       INTEGER            FLAG_REPORT( 2 ), INFO_ARRAY( * )
*       REAL               A( LDA, * ), B( LDB, * )
*       ..
*       .. Pointer Arguments ..
*       POINTER            CONTEXT ... advice requested
*
*
*
*> \par Purpose:
*  =============
*>
*> \verbatim
*>
*> SGESV_EC computes the solution to a real system of linear equations
*>    A * X = B,
*> where A is an N-by-N matrix and X and B are N-by-NRHS matrices.
*>
*> The LU decomposition with partial pivoting and row interchanges is
*> used to factor A as
*>    A = P * L * U,
*> where P is a permutation matrix, L is unit lower triangular, and U 
*> is upper triangular.  The factored form of A is then used to solve
*> the system of equations A * X = B.
*>
*> SGESV_EC also provides new exception handling and 
*> reporting capabilities.
*>
*> \endverbatim
*
*  Arguments:
*  ==========
*
*> \param[in] N
*> \verbatim
*>          N is INTEGER
*>          The number of linear equations, i.e., the order of the
*>          matrix A.  N >= 0.
*> \endverbatim
*>
*> \param[in] NRHS
*> \verbatim
*>          NRHS is INTEGER
*>          The number of right hand sides, i.e., the number of columns
*>          of the matrix B.  NRHS >= 0.
*> \endverbatim
*>
*> \param[in,out] A
*> \verbatim
*>          A is REAL array, dimension (LDA,N)
*>          On entry, the N-by-N coefficient matrix A.
*>          On exit, the factors L and U from the factorization
*>          A = P*L*U; the unit diagonal elements of L are not stored.
*> \endverbatim
*>
*> \param[in] LDA
*> \verbatim
*>          LDA is INTEGER
*>          The leading dimension of the array A.  LDA >= max(1,N).
*> \endverbatim
*>
*> \param[out] IPIV
*> \verbatim
*>          IPIV is INTEGER array, dimension (N)
*>          The pivot indices that define the permutation matrix P;
*>          row i of the matrix was interchanged with row IPIV(i).
*> \endverbatim
*>
*> \param[in,out] B
*> \verbatim
*>          B is REAL array, dimension (LDB,NRHS)
*>          On entry, the N-by-NRHS matrix of right hand side matrix B.
*>          On exit, if INFO = 0, the N-by-NRHS solution matrix X.
*> \endverbatim
*>
*> \param[in] LDB
*> \verbatim
*>          LDB is INTEGER
*>          The leading dimension of the array B.  LDB >= max(1,N).
*> \endverbatim
*>
*> \param[out] INFO
*> \verbatim
*>          INFO is INTEGER
*>          INFO is defined below depending on FLAG_REPORT         
*> \endverbatim
*>
*> \param[in] FLAG_REPORT
*> \verbatim
*>          FLAG_REPORT is INTEGER array, dimension(2)
*>          FLAG_REPORT(1) defines what kinds of exceptions to report,
*>          using INFO and possibly also INFO_ARRAY for more details.
*>          FLAG_REPORT(1) 
*>           <= -1 turns off all error checking, so INFO=0 is 
*>                 returned.
*>            =  0 does standard argument checking:
*>                 INFO = 0  means successful exit
*>                 INFO = -i means the i-th (non-floating point) 
*>                           argument had an illegal value 
*>                           (first error found is reported)
*>                 INFO = i means U(i,i) = 0. The factorization
*>                          has been completed, but the factor U is 
*>                          exactly singular, and division by zero 
*>                          will occur if it is used to solve a system 
*>                          of equations.
*>                 Using INFO to report the above errors has priority 
*>                 over reporting any of the errors described below. 
*>                 More generally, an error that would be found with a 
*>                 lower value of FLAG_REPORT(1) has priority to 
*>                 report using INFO than an error that would only
*>                 be found with a higher value of FLAG_REPORT(1).  
*>            =  1 also checks for Infs and NaNs in inputs and  
*>                 outputs, if INFO not already nonzero:
*>                 INFO = -3 means A contained an Inf or NaN on 
*>                           input; execution continues.
*>                 INFO = -6 means B contained an Inf or NaN on 
*>                           input, but not A; execution continues.
*>                 INFO = N+1 means that A contained an Inf or NaN on 
*>                        output but neither A nor B did on input.
*>                 INFO = N+2 means that B contained an Inf or NaN on 
*>                        output but neither A nor B did on input nor
*>                        A on output.
*>                 A and B will also be checked on input and output if 
*>                 FLAG_REPORT(2) = 1, 2 or 3 and reported in 
*>                 INFO_ARRAY as described below.
*>            >= 2 also checks for Infs and NaNs as inputs or outputs 
*>                 in all internal LAPACK calls in the call chain, if 
*>                 INFO is not already nonzero, and 
*>                 FLAG_REPORT(2) = 1, 2 or 3. In this case: 
*>                 INFO = N+3 means that either the call to
*>                        SGETRF_EC had an Inf or NaN as an 
*>                        input or output as above, or a subroutine
*>                        in its call tree did.
*>                 INFO = N+4 means that the call to SGETRS_EC
*>                        had an Inf or NaN as an 
*>                        input or output as above.
*>                 Each input, output and internal LAPACK call will
*>                 also be checked if FLAG_REPORT(2) = 1, 2 or 3 and 
*>                 reported in INFO_ARRAY as described below.
*>
*>          FLAG_REPORT(2) defines how to report the exceptions 
*>          requested by FLAG_REPORT(1).
*>            If FLAG_REPORT(1) <= -1, FLAG_REPORT(2) is ignored and 
*>            INFO=0 is returned. 
*>            Otherwise, FLAG_REPORT(2)
*>             <=  0 only returns the value of INFO described above.
*>              =  1 also returns INFO_ARRAY, as described below.
*>              =  2 means that SGESV_EC will also call 
*>                   REPORT_EXCEPTIONS to report INFO_ARRAY, if INFO 
*>                   is nonzero.
*>              =  3 means that all calls in the call tree will also
*>                   call REPORT_EXCEPTIONS, if the value of INFO they
*>                   return is nonzero.
*>             >=  4 means that SGESV_EC will call 
*>                   GET_FLAGS_TO_REPORT(CONTEXT,FLAG_REPORT) 
*>                   to get values of FLAG_REPORT to use, overriding 
*>                   input values. The user needs to have called 
*>                   SET_FLAGS_TO_REPORT(CONTEXT,FLAG_REPORT) 
*>                   before calling SGESV_EC in order to set 
*>                   FLAG_REPORT, otherwise the default is 
*>                   FLAG_REPORT = [0, 0]. The input array FLAG_REPORT 
*>                   will not be overwritten.
*> \endverbatim
*>
*> \param[out] INFO_ARRAY
*> \verbatim
*>          INFO_ARRAY is INTEGER array, dimension( 10 )
*>          If FLAG_REPORT(1) <= -1 or FLAG_REPORT(2) <= 0, 
*>          INFO_ARRAY is not accessed. Otherwise:
*>          INFO_ARRAY(1)
*>              = value of INFO from standard argument checking 
*>                (as defined by FLAG_REPORT(1) = 0)
*>          INFO_ARRAY(2)
*>              = value of FLAG_REPORT(1) used to determine the rest 
*>                of INFO_ARRAY
*>          INFO_ARRAY(3)
*>              = value of FLAG_REPORT(2) used to determine the rest 
*>                of INFO_ARRAY
*>          INFO_ARRAY(4)
*>              = value of INFO as specified by FLAG_REPORT(1) above
*>          INFO_ARRAY(5)
*>              = 2 = number of input/output arguments reported on
*>          INFO_ARRAY(6)
*>              = 2 = number of internal LAPACK calls reported on
*>          INFO_ARRAY(7), reports exceptions in A, as specified by 
*>            FLAG_REPORT
*>              = -1 if not checked (default)
*>              =  0 if checked and contains no Infs or NaNs
*>              =  1 if checked and contains an Inf or NaN on input
*>                   but not output
*>              =  2 if checked and contains an Inf or NaN on output 
*>                   but not input
*>              =  3 if checked and contains an Inf or NaN on input 
*>                   and output
*>          INFO_ARRAY(8), reports exceptions in B, analogously to 
*>            INFO_ARRAY(7)
*>          INFO_ARRAY(9), reports exceptions in SGETRF_EC, as 
*>            specified by FLAG_REPORT
*>              = -1 if not checked (default)
*>              =  0 if checked and no Infs or NaNs reported
*>              =  1 if checked and no input or output contains an Inf 
*>                   or NaN, but some LAPACK call deeper in the call 
*>                   chain signaled an Inf or NaN
*>              =  2 if checked and an input argument contains an Inf 
*>                   or NaN, but not an output
*>              =  3 if checked and an output argument contains an Inf 
*>                   or NaN, but not an input
*>              =  4 if checked and an argument contains an Inf or NaN 
*>                   on input and output
*>          INFO_ARRAY(10), reports exceptions in SGETRS_EC, 
*>            analogously to INFO_ARRAY(9)
*> \endverbatim
*>
*> \param[in] CONTEXT
*> \verbatim
*>           CONTEXT is POINTER to an "opaque object"
*> \endverbatim
*>
*
*  Authors:
*  ========
*
*> \author Univ. of Tennessee
*> \author Univ. of California Berkeley
*> \author Univ. of Colorado Denver
*> \author NAG Ltd.
*
*> \ingroup realGEsolve
*
*  =====================================================================
      SUBROUTINE SGESV_EC( N, NRHS, A, LDA, IPIV, B, LDB, 
     $                     INFO, FLAG_REPORT, INFO_ARRAY, CONTEXT ) 
*
*  -- LAPACK driver routine --
*  -- LAPACK is a software package provided by Univ. of Tennessee,    --
*  -- Univ. of California Berkeley, Univ. of Colorado Denver and NAG Ltd..--
*
*     .. Scalar Arguments ..
      INTEGER            INFO, LDA, LDB, N, NRHS
*     ..
*     .. Array Arguments ..
      INTEGER            IPIV( * )
      REAL               A( LDA, * ), B( LDB, * )
      INTEGER            FLAG_REPORT( 2 ), INFO_ARRAY( * )
*     ..
*     .. Pointer Arguments
      POINTER            CONTEXT
*
*  =====================================================================
*
*     .. Parameters ..
      CHARACTER, DIMENSION(5), PARAMETER :: 
     $   ROUTINE_NAME = (/ 'S','G','E','S','V' /)
*     ..
*     .. Local Scalars ..
      LOGICAL            CALL_REPORT_EXCEPTIONS
      INTEGER            WHAT, HOW
*     ..
*     .. Local Arrays ..
      INTEGER            FLAG_REPORT_INTERNAL(2), FLAG_REPORT_CALL(2)
      INTEGER            INFO_SGETRF_TMP(9), INFO_SGETRS_TMP(9)
      INTEGER            INFO_INTERNAL(2)
*     ..
*     .. External Subroutines ..
      EXTERNAL           SGETRF_EC, SGETRS_EC, XERBLA
      EXTERNAL           CHECKINIT1, CHECKINIT2
      EXTERNAL           SGECHECKARG, CHECKCALL
      EXTERNAL           UPDATE_INFO
*     ..
*     .. Intrinsic Functions ..
      INTRINSIC          MAX
*     ..
*     .. Executable Statements ..
*
*     Test the input parameters.
*
      INFO = 0
*     Initialize error checking flags
      CALL CHECKINIT1(FLAG_REPORT, FLAG_REPORT_INTERNAL, 
     $                FLAG_REPORT_NEXT, CALL_REPORT_EXCEPTIONS, 
     $                CONTEXT ) 
      WHAT = FLAG_REPORT_INTERNAL( 1 )
      HOW = FLAG_REPORT_INTERNAL( 2 )
      IF (WHAT .EQ. -1 ) GOTO 100
*
*     Check for standard input errors
      IF( N.LT.0 ) THEN
         INFO = -1
      ELSE IF( NRHS.LT.0 ) THEN
         INFO = -2
      ELSE IF( LDA.LT.MAX( 1, N ) ) THEN
         INFO = -4
      ELSE IF( LDB.LT.MAX( 1, N ) ) THEN
         INFO = -7
      END IF
*
*     Initialize error flags
      CALL CHECKINIT2( FLAG_REPORT_INTERNAL, INFO, INFO_INTERNAL, 
     $                 INFO_ARRAY, 2, 2) 
*
      IF( INFO.NE.0 ) THEN
         IF (CALL_REPORT_EXCEPTIONS) 
     $       CALL REPORT_EXCEPTIONS(CONTEXT,5,ROUTINE_NAME,INFO_ARRAY)
         CALL XERBLA( 'SGESV ', -INFO )
         RETURN
      END IF
*
*     Quick return if possible
*
      IF( N.EQ.0 ) RETURN
*
*     Check for exceptional inputs in A 
      CALL SGECHECKARG( FLAG_REPORT_INTERNAL, N, N, A, LDA, 
     $                  INFO, INFO_INTERNAL, INFO_ARRAY, 3, 2, 0, 7 )
*     Check for exceptional inputs in B 
      CALL SGECHECKARG( FLAG_REPORT_INTERNAL, N, NRHS, B, LDB, 
     $                  INFO, INFO_INTERNAL, INFO_ARRAY, 6, 2, 0, 8 )
*
100   CONTINUE
*
*     Quick return if possible
*
      IF( N.EQ.0 ) RETURN
*
*     Compute the LU factorization of A.
*
*     Indicate if input A already checked for Infs and NaNs
      IF (WHAT .GE. 1 .AND. HOW .GE. 1) 
     $    INFO_SGETRF_TMP(7) = INFO_ARRAY(7)
      CALL SGETRF_EC( N, N, A, LDA, IPIV, 
     $               INFO, FLAG_REPORT_CALL, INFO_SGETRF_TMP, CONTEXT)
*
*     Check inputs, outputs and internal calls of SGETRF_EC
      CALL CHECKCALL( FLAG_REPORT_INTERNAL, INFO_INTERNAL,
     $                INFO_SGETRF_TMP, INFO_ARRAY, N+3, 9 )
      IF( INFO.EQ.0 ) THEN
*
*        Solve the system A*X = B, overwriting B with X.
*
*        Indicate if input B already checked for Infs and NaNs
         IF (WHAT .GE. 1 .AND. HOW .GE. 1) 
     $      INFO_SGETRS_TMP(8) = INFO_ARRAY(8)
         CALL SGETRS_EC( 'No transpose', N, NRHS, A, LDA, IPIV, B,LDB,
     $                  INFO,FLAG_REPORT_CALL,INFO_SGETRS_TMP,CONTEXT)
*
*        Check inputs, outputs and internal calls of call to SGETRS_EC
         CALL CHECKCALL( FLAG_REPORT_INTERNAL, INFO_INTERNAL,
     $                   INFO_SGETRS_TMP, INFO_ARRAY, N+4, 9 )
      ENDIF
*
*     Check for errors before returning
*
      IF (WHAT.EQ.-1) RETURN
*     Check for exceptional outputs in A
      CALL SGECHECKARG(FLAG_REPORT_INTERNAL, N, N, A, LDA, 
     $                 INFO, INFO_INTERNAL, INFO_ARRAY, 3, 3, N+1, 7)
*     Check for exceptional outputs in B 
      CALL SGECHECKARG(FLAG_REPORT_INTERNAL, N, NRHS, B, LDB, 
     $                 INFO, INFO_INTERNAL, INFO_ARRAY, 7, 3, N+2, 8)
*
*     Update INFO and INFO_ARRAY
      CALL UPDATE_INFO( INFO, INFO_ARRAY, INFO_INTERNAL )
      IF (CALL_REPORT_EXCEPTIONS .AND. INFO .NE. 0) 
     $   CALL REPORT_EXCEPTIONS(CONTEXT, 5, ROUTINE_NAME, INFO_ARRAY)
      RETURN
*
*     End of SGESV_EC
*
      END
\end{verbatim}

\newpage
\begin{verbatim}
*> \brief \b SGETRF_EC
*
*  =========== DOCUMENTATION ===========
*
* Online html documentation available at
*            http://www.netlib.org/lapack/explore-html/
*
*  Definition:
*  ===========
*
*       SUBROUTINE SGETRF_EC( M, N, A, LDA, IPIV, 
*      $                   INFO, FLAG_REPORT, INFO_ARRAY, CONTEXT )
*
*       .. Scalar Arguments ..
*       INTEGER            INFO, LDA, M, N
*       ..
*       .. Array Arguments ..
*       INTEGER            IPIV( * )
*       INTEGER            FLAG_REPORT( 2 ), INFO_ARRAY( * )
*       REAL               A( LDA, * )
*       ..
*       .. Pointer Arguments ..
*       POINTER            CONTEXT ... advice requested
*
*
*
*> \par Purpose:
*  =============
*>
*> \verbatim
*>
*> SGETRF_EC computes an LU factorization of a general M-by-N matrix A
*> using partial pivoting with row interchanges.
*>
*> The factorization has the form
*>    A = P * L * U
*> where P is a permutation matrix, L is lower triangular with unit
*> diagonal elements (lower trapezoidal if m > n), and U is upper
*> triangular (upper trapezoidal if m < n).
*>
*> This is the right-looking Level 3 BLAS version of the algorithm.
*>
*> SGETRF_EC also provides new exception handling and 
*> reporting capabilities.
*>
*> \endverbatim
*
*  Arguments:
*  ==========
*
*> \param[in] M
*> \verbatim
*>          M is INTEGER
*>          The number of rows of the matrix A.  M >= 0.
*> \endverbatim
*>
*> \param[in] N
*> \verbatim
*>          N is INTEGER
*>          The number of columns of the matrix A.  N >= 0.
*> \endverbatim
*>
*> \param[in,out] A
*> \verbatim
*>          A is REAL array, dimension (LDA,N)
*>          On entry, the M-by-N matrix to be factored.
*>          On exit, the factors L and U from the factorization
*>          A = P*L*U; the unit diagonal elements of L are not stored.
*> \endverbatim
*>
*> \param[in] LDA
*> \verbatim
*>          LDA is INTEGER
*>          The leading dimension of the array A.  LDA >= max(1,M).
*> \endverbatim
*>
*> \param[out] IPIV
*> \verbatim
*>          IPIV is INTEGER array, dimension (min(M,N))
*>          The pivot indices; for 1 <= i <= min(M,N), row i of the
*>          matrix was interchanged with row IPIV(i).
*> \endverbatim
*>
*> \param[out] INFO
*> \verbatim
*>          INFO is INTEGER
*>          INFO is defined below depending on FLAG_REPORT         
*> \endverbatim
*>
*> \param[in] FLAG_REPORT
*> \verbatim
*>          FLAG_REPORT is INTEGER array, dimension(2)
*>          FLAG_REPORT(1) defines what kinds of exceptions to report,
*>          using INFO and possibly also INFO_ARRAY for more details.
*>          FLAG_REPORT(1) 
*>           <= -1 turns off all error checking, so INFO=0 is 
*>                 returned.
*>            =  0 does standard argument checking:
*>                 INFO = 0  means successful exit
*>                 INFO = -i means the i-th (non-floating point) 
*>                           argument had an illegal value 
*>                           (first error found is reported)
*>                 INFO = i means U(i,i) = 0. The factorization
*>                          has been completed, but the factor U is 
*>                          exactly singular, and division by zero 
*>                          will occur if it is used to solve a system 
*>                          of equations.
*>                 Using INFO to report the above errors has priority 
*>                 over reporting any of the errors described below. 
*>                 More generally, an error that would be found with a 
*>                 lower value of FLAG_REPORT(1) has priority to 
*>                 report using INFO than an error that would only
*>                 be found with a higher value of FLAG_REPORT(1).  
*>            =  1 also checks for Infs and NaNs in inputs and  
*>                 outputs, if INFO not already nonzero:
*>                 INFO = -3 means A contained an Inf or NaN on 
*>                           input; execution continues.
*>                 INFO = N+1 means that A contained an Inf or NaN on 
*>                        output but not input.
*>                 A will also be checked on input and output if 
*>                 FLAG_REPORT(2) = 1, 2 or 3 and reported in 
*>                 INFO_ARRAY as described below.
*>            >= 2 also checks for Infs and NaNs as inputs or outputs 
*>                 in all internal LAPACK calls in the call chain, if 
*>                 INFO is not already nonzero, and 
*>                 FLAG_REPORT(2) = 1, 2 or 3. In this case: 
*>                 INFO = N+2 means that either the first call
*>                        to SGETRF2_EC had an Inf or NaN as an 
*>                        input or output as above, or a subroutine
*>                        in its call tree did.
*>                 INFO = N+3 means that either the second call
*>                        to SGETRF2_EC had an Inf or NaN as an 
*>                        input or output as above, or a subroutine
*>                        in its call tree did.
*>                 Each input, output and internal LAPACK call will
*>                 also be checked if FLAG_REPORT(2) = 1, 2 or 3 and 
*>                 reported in INFO_ARRAY as described below.
*>
*>          FLAG_REPORT(2) defines how to report the exceptions 
*>          requested by FLAG_REPORT(1).
*>            If FLAG_REPORT(1) <= -1, FLAG_REPORT(2) is ignored and 
*>            INFO=0 is returned. 
*>            Otherwise, FLAG_REPORT(2)
*>             <=  0 only returns the value of INFO described above.
*>              =  1 also returns INFO_ARRAY, as described below.
*>              =  2 means that SGETRF_EC will also call 
*>                   REPORT_EXCEPTIONS to report INFO_ARRAY, if INFO 
*>                   is nonzero.
*>              =  3 means that all calls in the call tree will also
*>                   call REPORT_EXCEPTIONS, if the value of INFO they
*>                   return is nonzero.
*>             >=  4 means that SGETRF_EC will call 
*>                   GET_FLAGS_TO_REPORT(CONTEXT,FLAG_REPORT) 
*>                   to get values of FLAG_REPORT to use, overriding 
*>                   input values. The user needs to have called 
*>                   SET_FLAGS_TO_REPORT(CONTEXT,FLAG_REPORT) 
*>                   before calling SGETRF_EC in order to set 
*>                   FLAG_REPORT, otherwise the default is 
*>                   FLAG_REPORT = [0, 0]. The input array FLAG_REPORT 
*>                   will not be overwritten.
*> \endverbatim
*>
*> \param[out] INFO_ARRAY
*> \verbatim
*>          INFO_ARRAY is INTEGER array, dimension( 9 )
*>          If FLAG_REPORT(1) <= -1 or FLAG_REPORT(2) <= 0, 
*>          INFO_ARRAY is not accessed. Otherwise:
*>          INFO_ARRAY(1)
*>              = value of INFO from standard argument checking 
*>                (as defined by FLAG_REPORT(1) = 0)
*>          INFO_ARRAY(2)
*>              = value of FLAG_REPORT(1) used to determine the rest 
*>                of INFO_ARRAY
*>          INFO_ARRAY(3)
*>              = value of FLAG_REPORT(2) used to determine the rest 
*>                of INFO_ARRAY
*>          INFO_SGETRF(4)
*>              = value of INFO as specified by FLAG_REPORT(1) above
*>          INFO_ARRAY(5)
*>              = 1 = number of input/output arguments reported on
*>          INFO_ARRAY(6)
*>              = 2 = number of internal LAPACK calls reported on
*>          INFO_ARRAY(7) reports exceptions in A, as specified by 
*>            FLAG_REPORT
*>              = -1 if not checked (default)
*>              =  0 if checked and contains no Infs or NaNs
*>              =  1 if checked and contains an Inf or NaN on input
*>                   but not output
*>              =  2 if checked and contains an Inf or NaN on output 
*>                   but not input
*>              =  3 if checked and contains an Inf or NaN on input 
*>                   and output
*>            If INFO_ARRAY(7) = 0 or 1 on input, then A will not
*>            be checked again. Input values < -1 or > 1 will be 
*>            treated the same as -1, i.e. not checked.
*>          INFO_ARRAY(8) reports exceptions in the first
*>            call to SGETRF2_EC, as specified by FLAG_REPORT
*>              = -1 if not checked (default)
*>              =  0 if checked and no Infs or NaNs reported
*>              =  1 if checked and no input or output contains an Inf 
*>                   or NaN, but some LAPACK call deeper in the call 
*>                   chain signaled an Inf or NaN
*>              =  2 if checked and an input argument contains an Inf 
*>                   or NaN, but not an output
*>              =  3 if checked and an output argument contains an Inf 
*>                   or NaN, but not an input
*>              =  4 if checked and an argument contains an Inf or NaN 
*>                   on input and output
*>          INFO_ARRAY(9) reports exceptions in the second
*>            call to SGETRF2_EC, analogously to INFO_ARRAY(8),
*>            reporting the maximum over all calls. 
*> \endverbatim
*>
*> \param[in] CONTEXT
*> \verbatim
*>           CONTEXT is POINTER to an "opaque object"
*> \endverbatim
*>
*
*  Authors:
*  ========
*
*> \author Univ. of Tennessee
*> \author Univ. of California Berkeley
*> \author Univ. of Colorado Denver
*> \author NAG Ltd.
*
*> \ingroup realGEcomputational
*
*  =====================================================================
      SUBROUTINE SGETRF_EC( M, N, A, LDA, IPIV,
     $                      INFO, FLAG_REPORT, INFO_ARRAY, CONTEXT )
*
*  -- LAPACK computational routine --
*  -- LAPACK is a software package provided by Univ. of Tennessee,    --
*  -- Univ. of California Berkeley, Univ. of Colorado Denver and NAG Ltd..--
*
*     .. Scalar Arguments ..
      INTEGER            INFO, LDA, M, N
*     ..
*     .. Array Arguments ..
      INTEGER            IPIV( * )
      REAL               A( LDA, * )
      INTEGER            FLAG_REPORT( 2 ), INFO_ARRAY( * )
*     ..
*     .. Pointer Arguments
      POINTER            CONTEXT
*
*  =====================================================================
*
*     .. Parameters ..
      REAL               ONE
      PARAMETER          ( ONE = 1.0E+0 )
      CHARACTER, DIMENSION(6), PARAMETER :: 
     $   ROUTINE_NAME = (/ 'S','G','E','T','R','F' /)
*     ..
*     .. Local Scalars ..
      LOGICAL            CALL_REPORT_EXCEPTIONS
      INTEGER            I, IINFO, J, JB, NB
      INTEGER            WHAT, HOW
*     ..
*     .. Local Arrays ..
      INTEGER            FLAG_REPORT_INTERNAL(2), FLAG_REPORT_CALL(2)
      INTEGER            INFO_SGETRF2_TMP1(9), INFO_SGETRF2_TMP2(9)
      INTEGER            INFO_INTERNAL(2)
*     ..
*     .. External Subroutines ..
      EXTERNAL           SGEMM, SGETRF2_EC, SLASWP, STRSM, XERBLA
      EXTERNAL           CHECKINIT1, CHECKINIT2
      EXTERNAL           SGECHECKARG, CHECKCALL
      EXTERNAL           UPDATE_INFO
*     ..
*     .. External Functions ..
      INTEGER            ILAENV
      EXTERNAL           ILAENV
*     ..
*     .. Intrinsic Functions ..
      INTRINSIC          MAX, MIN
*     ..
*     .. Executable Statements ..
*
*     Test the input parameters.
*
      INFO = 0
*     Initialize error checking flags
      CALL CHECKINIT1(FLAG_REPORT, FLAG_REPORT_INTERNAL, 
     $                FLAG_REPORT_CALL, CALL_REPORT_EXCEPTIONS, 
     $                CONTEXT ) 
      WHAT = FLAG_REPORT_INTERNAL( 1 )
      HOW = FLAG_REPORT_INTERNAL( 2 )      
      IF (WHAT .EQ. -1 ) GOTO 100
*
*     Check for standard input errors
      IF( M.LT.0 ) THEN
         INFO = -1
      ELSE IF( N.LT.0 ) THEN
         INFO = -2
      ELSE IF( LDA.LT.MAX( 1, M ) ) THEN
         INFO = -4
      END IF
*
*     Initialize error flags
      CALL CHECKINIT2( FLAG_REPORT_INTERNAL, INFO, INFO_INTERNAL, 
     $                 INFO_ARRAY, 1, 2)  
*
      IF( INFO.NE.0 ) THEN
         IF (CALL_REPORT_EXCEPTIONS) 
     $     CALL REPORT_EXCEPTIONS(CONTEXT,6,ROUTINE_NAME,INFO_ARRAY)
         CALL XERBLA( 'SGETRF', -INFO )
         RETURN
      END IF
*
*     Quick return if possible
*
      IF( M.EQ.0 .OR. N.EQ.0 )
     $   RETURN
*
*     Check for exceptional inputs in A
      CALL SGECHECKARG( FLAG_REPORT_INTERNAL, M, N, A, LDA, 
     $                  INFO, INFO_INTERNAL, INFO_ARRAY, 3, 2, 0, 7 )
100   CONTINUE
*
*     Quick return if possible
*
      IF( M.EQ.0 .OR. N.EQ.0 )
     $   RETURN
*
*     Determine the block size for this environment.
*
      NB = ILAENV( 1, 'SGETRF', ' ', M, N, -1, -1 )
      IF( NB.LE.1 .OR. NB.GE.MIN( M, N ) ) THEN
*
*        Use unblocked code.
*
*        Indicate if input already checked for Infs and NaNs
         IF (WHAT .GE. 1 .AND. HOW .GE. 1) THEN
            INFO_SGETRF2_TMP1(7) = -1
            IF (INFO_ARRAY(7) .NE. 0)
     $          INFO_SGETRF2_TMP1(7) = INFO_ARRAY(7)
         ENDIF
*
         CALL SGETRF2_EC( M, N, A, LDA, IPIV, 
     $             INFO, FLAG_REPORT_CALL, INFO_SGETRF2_TMP1, CONTEXT)
*   
*        Check inputs, outputs and internal calls of
*        first call to SGETRF2_EC
         CALL CHECKCALL( FLAG_REPORT_INTERNAL, INFO_INTERNAL,
     $                   INFO_SGETRF2_TMP1, INFO_ARRAY, N+2, 8 )
      ELSE
*
*        Use blocked code.
*
         DO 20 J = 1, MIN( M, N ), NB
            JB = MIN( MIN( M, N )-J+1, NB )
*
*           Factor diagonal and subdiagonal blocks and test for exact
*           singularity.
*
*           Indicate that input not already checked
            IF (WHAT .GE. 1 .AND. HOW .GE. 1) 
     $         INFO_SGETRF2_TMP2(7) = -1
*
            CALL SGETRF2_EC( M-J+1, JB, A(J,J), LDA, IPIV(J), 
     $            IINFO, FLAG_REPORT_CALL, INFO_SGETRF2_TMP2, CONTEXT)
*   
*           Check inputs, outputs and internal calls of
*           second call to SGETRF2_EC
            CALL CHECKCALL( FLAG_REPORT_INTERNAL, INFO_INTERNAL,
     $                      INFO_SGETRF2_TMP2, INFO_ARRAY, N+3, 9 )
*
*           Adjust INFO and the pivot indices.
*
            IF( INFO.EQ.0 .AND. IINFO.GT.0 )
     $         INFO = IINFO + J - 1
            DO 10 I = J, MIN( M, J+JB-1 )
               IPIV( I ) = J - 1 + IPIV( I )
   10       CONTINUE
*
*           Apply interchanges to columns 1:J-1.
*
            CALL SLASWP( J-1, A, LDA, J, J+JB-1, IPIV, 1 )
*
            IF( J+JB.LE.N ) THEN
*
*              Apply interchanges to columns J+JB:N.
*
               CALL SLASWP( N-J-JB+1, A( 1, J+JB ), LDA, J, J+JB-1,
     $                      IPIV, 1 )
*
*              Compute block row of U.
*
               CALL STRSM( 'Left', 'Lower', 'No transpose', 'Unit',JB,
     $                     N-J-JB+1, ONE, A( J, J ), LDA, A(J, J+JB),
     $                     LDA )
               IF( J+JB.LE.M ) THEN
*
*                 Update trailing submatrix.
*
                  CALL SGEMM( 'No transpose','No transpose', M-J-JB+1,
     $                        N-J-JB+1, JB, -ONE, A( J+JB, J ), LDA,
     $                        A( J, J+JB ), LDA, ONE, A( J+JB, J+JB ),
     $                        LDA )
               END IF
            END IF
   20    CONTINUE
      END IF
*
*     Check for errors before returning
*
      IF (WHAT.EQ.-1) RETURN
*     Check for exceptional outputs in A
      CALL SGECHECKARG(FLAG_REPORT_INTERNAL, M, N, A, LDA, 
     $                 INFO, INFO_INTERNAL, INFO_ARRAY, 3, 3, N+1, 7)
*
*     Update INFO and INFO_ARRAY
      CALL UPDATE_INFO( INFO, INFO_ARRAY, INFO_INTERNAL )
      IF (CALL_REPORT_EXCEPTIONS .AND. INFO .NE. 0) 
     $   CALL REPORT_EXCEPTIONS(CONTEXT, 6, ROUTINE_NAME, INFO_ARRAY)
      RETURN
*
*     End of SGETRF_EC
*
      END
\end{verbatim}

\newpage
\begin{verbatim}
*> \brief \b SGETRF2_EC
*
*  =========== DOCUMENTATION ===========
*
* Online html documentation available at
*            http://www.netlib.org/lapack/explore-html/
*
*  Definition:
*  ===========
*
*       RECURSIVE SUBROUTINE SGETRF2_EC( M, N, A, LDA, IPIV, 
*      $                     INFO, FLAG_REPORT, INFO_ARRAY, CONTEXT ) 
*
*       .. Scalar Arguments ..
*       INTEGER            INFO, LDA, M, N
*       ..
*       .. Array Arguments ..
*       INTEGER            IPIV( * )
*       INTEGER            FLAG_REPORT( 2 ), INFO_ARRAY( 10 )
*       REAL               A( LDA, * )
*       ..
*       .. Pointer Arguments ..
*       POINTER            CONTEXT ... advice requested
*
*
*> \par Purpose:
*  =============
*>
*> \verbatim
*>
*> SGETRF2_EC computes an LU factorization of a general M-by-N matrix
*> A using partial pivoting with row interchanges.
*>
*> The factorization has the form
*>    A = P * L * U
*> where P is a permutation matrix, L is lower triangular with unit
*> diagonal elements (lower trapezoidal if m > n), and U is upper
*> triangular (upper trapezoidal if m < n).
*>
*> This is the recursive version of the algorithm. It divides
*> the matrix into four submatrices:
*>
*>        [  A11 | A12  ]  where A11 is n1 by n1 and A22 is n2 by n2
*>    A = [ -----|----- ]  with n1 = min(m,n)/2
*>        [  A21 | A22  ]       n2 = n-n1
*>
*>                                       [ A11 ]
*> The subroutine calls itself to factor [ --- ],
*>                                       [ A12 ]
*>                 [ A12 ]
*> do the swaps on [ --- ], solve A12, update A22,
*>                 [ A22 ]
*>
*> then calls itself to factor A22 and do the swaps on A21.
*>
*> SGETRF2_EC also provides new exception handling and 
*> reporting capabilities.
*>
*> \endverbatim
*
*  Arguments:
*  ==========
*
*> \param[in] M
*> \verbatim
*>          M is INTEGER
*>          The number of rows of the matrix A.  M >= 0.
*> \endverbatim
*>
*> \param[in] N
*> \verbatim
*>          N is INTEGER
*>          The number of columns of the matrix A.  N >= 0.
*> \endverbatim
*>
*> \param[in,out] A
*> \verbatim
*>          A is REAL array, dimension (LDA,N)
*>          On entry, the M-by-N matrix to be factored.
*>          On exit, the factors L and U from the factorization
*>          A = P*L*U; the unit diagonal elements of L are not stored.
*> \endverbatim
*>
*> \param[in] LDA
*> \verbatim
*>          LDA is INTEGER
*>          The leading dimension of the array A.  LDA >= max(1,M).
*> \endverbatim
*>
*> \param[out] IPIV
*> \verbatim
*>          IPIV is INTEGER array, dimension (min(M,N))
*>          The pivot indices; for 1 <= i <= min(M,N), row i of the
*>          matrix was interchanged with row IPIV(i).
*> \endverbatim
*>
*> \param[out] INFO
*> \verbatim
*>          INFO is INTEGER
*>          INFO is defined below depending on FLAG_REPORT
*> \endverbatim
*>
*> \param[in] FLAG_REPORT
*> \verbatim
*>          FLAG_REPORT is INTEGER array, dimension(2)
*>          FLAG_REPORT(1) defines what kinds of exceptions to report,
*>          report, using INFO and possibly also INFO_ARRAY for more 
*>          details. FLAG_REPORT(1)
*>           <= -1 turns off all error checking, so INFO=0 is 
*>                 returned.
*>            =  0 does standard argument checking:
*>                 INFO = 0  means successful exit
*>                 INFO = -i means the i-th (non-floating point) 
*>                           argument had an illegal value 
*>                           (first error found is reported)
*>                 INFO = i means U(i,i) = 0. The factorization
*>                          has been completed, but the factor U is 
*>                          exactly singular, and division by zero 
*>                          will occur if it is used to solve a 
*>                          system of equations.
*>                 Using INFO to report the above errors has priority 
*>                 over reporting any of the errors described below. 
*>                 More generally, an error that would be found with
*>                 a lower value of FLAG_REPORT(1) has priority to 
*>                 report using INFO than an error that would only
*>                 be found with a higher value of FLAG_REPORT(1).  
*>            =  1 also checks for Infs and NaNs in inputs and  
*>                 outputs, if INFO not already nonzero:
*>                 INFO = -3 means A contained an Inf or NaN on 
*>                           input; execution continues.
*>                 INFO = N+1 means that A contained an Inf or NaN on 
*>                        output but not input.
*>                 A will also be checked on input and output if 
*>                 FLAG_REPORT(2) = 1, 2 or 3 and reported in 
*>                 INFO_ARRAY as described below.
*>            >= 2 also checks for Infs and NaNs as inputs or outputs 
*>                 in all internal LAPACK calls in the call chain, if 
*>                 INFO is not already nonzero, and 
*>                 FLAG_REPORT(2) = 1, 2 or 3. In this case: 
*>                 INFO = N+2 means that either the first recursive
*>                        call to SGETRF2_EC had an Inf or NaN as an 
*>                        input or output as above, or a subroutine
*>                        in its call tree did.
*>                 INFO = N+3 means that either the second recursive
*>                        call to SGETRF2_EC had an Inf or NaN as an 
*>                        input or output as above, or a subroutine
*>                        in its call tree did.
*>                 Each input, output and internal LAPACK call will
*>                 also be checked if FLAG_REPORT(2) = 1, 2 or 3 and 
*>                 reported in INFO_ARRAY as described below.
*>
*>          FLAG_REPORT(2) defines how to report the exceptions 
*>          requested by FLAG_REPORT(1).
*>            If FLAG_REPORT(1) <= -1, FLAG_REPORT(2) is ignored and 
*>            INFO=0 is returned. 
*>            Otherwise, FLAG_REPORT(2)
*>             <=  0 only returns the value of INFO described above.
*>              =  1 also returns INFO_ARRAY, as described below.
*>              =  2 means that SGETRF2_EC will also call 
*>                   REPORT_EXCEPTIONS to report INFO_ARRAY, if INFO 
*>                   is nonzero.
*>              =  3 means that all calls in the call tree will also
*>                   call REPORT_EXCEPTIONS, if the value of INFO
*>                   they return is nonzero.
*>             >=  4 means that SGETRF2_EC will call 
*>                   GET_FLAGS_TO_REPORT(CONTEXT,FLAG_REPORT) 
*>                   to get values of FLAG_REPORT to use, overriding 
*>                   input values. The user needs to have called 
*>                   SET_FLAGS_TO_REPORT(CONTEXT,FLAG_REPORT) 
*>                   before calling SGETRF2_EC in order to set 
*>                   FLAG_REPORT, otherwise the default is 
*>                   FLAG_REPORT = [0, 0]. The input array 
*>                   FLAG_REPORT will not be overwritten.
*> \endverbatim
*>
*> \param[out] INFO_ARRAY
*> \verbatim
*>          INFO_ARRAY is INTEGER array, dimension( 9 )
*>          If FLAG_REPORT(1) <= -1 or FLAG_REPORT(2) <= 0, 
*>          INFO_ARRAY is not accessed. Otherwise:
*>          INFO_ARRAY(1)
*>              = value of INFO from standard argument checking 
*>                (as defined by FLAG_REPORT(1) = 0)
*>          INFO_ARRAY(2)
*>              = value of FLAG_REPORT(1) used to determine the rest 
*>                of INFO_ARRAY
*>          INFO_ARRAY(3)
*>              = value of FLAG_REPORT(2) used to determine the rest 
*>                of INFO_ARRAY
*>          INFO_ARRAY(4)
*>              = value of INFO as specified by FLAG_REPORT(1) above
*>          INFO_ARRAY(5)
*>              = 1 = number of input/output arguments reported on
*>          INFO_ARRAY(6)
*>              = 2 = number of internal LAPACK calls reported on
*>          INFO_ARRAY(7) reports exceptions in A, as specified by 
*>            FLAG_REPORT
*>              = -1 if not checked (default)
*>              =  0 if checked and contains no Infs or NaNs
*>              =  1 if checked and contains an Inf or NaN on input
*>                   but not output
*>              =  2 if checked and contains an Inf or NaN on output 
*>                   but not input
*>              =  3 if checked and contains an Inf or NaN on input 
*>                   and output
*>            If INFO_ARRAY(7) = 0 or 1 on input, then A will not
*>            be checked again. Input values < -1 or > 1 will be 
*>            treated the same as -1, i.e. not checked.
*>          INFO_ARRAY(8) reports exceptions in the first recursive
*>            call to SGETRF2_EC, as specified by FLAG_REPORT
*>              = -1 if not checked (default)
*>              =  0 if checked and no Infs or NaNs reported
*>              =  1 if checked and no input or output contains an
*>                   Inf or NaN, but some LAPACK call deeper in the 
*>                   call chain signaled an Inf or NaN
*>              =  2 if checked and an input argument contains an
*>                   Inf or NaN, but not an output
*>              =  3 if checked and an output argument contains an 
*>                   Inf or NaN, but not an input
*>              =  4 if checked and an argument contains an 
*>                   Inf or NaN on input and output
*>          INFO_ARRAY(9) reports exceptions in the second recursive
*>            call to SGETRF2_EC, analogously to INFO_ARRAY(8) 
*> \endverbatim
*>
*> \param[in] CONTEXT
*> \verbatim
*>           CONTEXT is POINTER to an "opaque object"
*> \endverbatim
*>
*
*  Authors:
*  ========
*
*> \author Univ. of Tennessee
*> \author Univ. of California Berkeley
*> \author Univ. of Colorado Denver
*> \author NAG Ltd.
*
*> \ingroup realGEcomputational
*
*  =====================================================================
      RECURSIVE SUBROUTINE SGETRF2_EC( M, N, A, LDA, IPIV, 
     $                     INFO, FLAG_REPORT, INFO_ARRAY, CONTEXT )
*
*  -- LAPACK computational routine --
*  -- LAPACK is a software package provided by Univ. of Tennessee,    --
*  -- Univ. of California Berkeley, Univ. of Colorado Denver and NAG Ltd..--
*
*     .. Scalar Arguments ..
      INTEGER            INFO, LDA, M, N
*     ..
*     .. Array Arguments ..
      INTEGER            IPIV( * )
      REAL               A( LDA, * )
      INTEGER            FLAG_REPORT( 2 ), INFO_ARRAY( * )
*     ..
*     .. Pointer Arguments
      POINTER            CONTEXT
*
*  =====================================================================
*
*     .. Parameters ..
      REAL               ONE, ZERO
      PARAMETER          ( ONE = 1.0E+0, ZERO = 0.0E+0 )
      CHARACTER, DIMENSION(7), PARAMETER :: 
     $   ROUTINE_NAME = (/ 'S','G','E','T','R','F','2' /)
*     ..
*     .. Local Scalars ..
      LOGICAL            CALL_REPORT_EXCEPTIONS
      REAL               SFMIN, TEMP
      INTEGER            I, IINFO, N1, N2
      INTEGER            WHAT, HOW
*     ..
*     .. Local Arrays ..
      INTEGER            FLAG_REPORT_INTERNAL(2), FLAG_REPORT_CALL(2) 
      INTEGER            INFO_SGETRF2_TMP1(9), INFO_SGETRF2_TMP2(9)
      INTEGER            INFO_INTERNAL(2) 
*     ..
*     .. External Functions ..
      REAL               SLAMCH
      INTEGER            ISAMAX 
      EXTERNAL           SLAMCH, ISAMAX
*     ..
*     .. External Subroutines ..
      EXTERNAL           SGEMM, SSCAL, SLASWP, STRSM, XERBLA
      EXTERNAL           CHECKINIT1, CHECKINIT2
      EXTERNAL           SGECHECKARG, CHECKCALL
      EXTERNAL           UPDATE_INFO
*     ..
*     .. Intrinsic Functions ..
      INTRINSIC          MAX, MIN
*     ..
*     .. Executable Statements ..
*
*     Test the input parameters
*
      INFO = 0
*     Initialize error checking flags
      CALL CHECKINIT1(FLAG_REPORT, FLAG_REPORT_INTERNAL, 
     $                FLAG_REPORT_CALL, CALL_REPORT_EXCEPTIONS, 
     $                CONTEXT ) 
      WHAT = FLAG_REPORT_INTERNAL( 1 )
      HOW = FLAG_REPORT_INTERNAL( 2 )      
      IF (WHAT .EQ. -1 ) GOTO 100
*
*     Check for standard input errors
      IF( M.LT.0 ) THEN
         INFO = -1
      ELSE IF( N.LT.0 ) THEN
         INFO = -2
      ELSE IF( LDA.LT.MAX( 1, M ) ) THEN
         INFO = -4
      END IF
*
*     Initialize error flags
      CALL CHECKINIT2( FLAG_REPORT_INTERNAL, INFO, INFO_INTERNAL, 
     $                 INFO_ARRAY, 1, 2)  
*
      IF( INFO.NE.0 ) THEN
         IF (CALL_REPORT_EXCEPTIONS) 
     $     CALL REPORT_EXCEPTIONS(CONTEXT,7,ROUTINE_NAME,INFO_ARRAY)
         CALL XERBLA( 'SGETRF2', -INFO )
         RETURN
      END IF
*
*     Quick return if possible
*
      IF( M.EQ.0 .OR. N.EQ.0 )
     $   RETURN
*
*     Check for exceptional inputs in A
      CALL SGECHECKARG( FLAG_REPORT_INTERNAL, M, N, A, LDA, 
     $                  INFO, INFO_INTERNAL, INFO_ARRAY, 3, 2, 0, 7 )
*
100   CONTINUE
*
*     Quick return if possible
*
      IF( M.EQ.0 .OR. N.EQ.0 )
     $   RETURN
*
      IF ( M.EQ.1 ) THEN
*
*        Use unblocked code for one row case
*        Just need to handle IPIV and INFO
*
         IPIV( 1 ) = 1
         IF ( A(1,1).EQ.ZERO )
     $      INFO = 1
*
      ELSE IF( N.EQ.1 ) THEN
*
*        Use unblocked code for one column case
*
*
*        Compute machine safe minimum
*
         SFMIN = SLAMCH('S')
*
*        Find pivot and test for singularity
*
         I = ISAMAX( M, A( 1, 1 ), 1 )
         IPIV( 1 ) = I
         IF( A( I, 1 ).NE.ZERO ) THEN
*
*           Apply the interchange
*
            IF( I.NE.1 ) THEN
               TEMP = A( 1, 1 )
               A( 1, 1 ) = A( I, 1 )
               A( I, 1 ) = TEMP
            END IF
*
*           Compute elements 2:M of the column
*
            IF( ABS(A( 1, 1 )) .GE. SFMIN ) THEN
               CALL SSCAL( M-1, ONE / A( 1, 1 ), A( 2, 1 ), 1 )
            ELSE
               DO 10 I = 1, M-1
                  A( 1+I, 1 ) = A( 1+I, 1 ) / A( 1, 1 )
   10          CONTINUE
            END IF
*
         ELSE
            INFO = 1
         END IF
*
      ELSE
*
*        Use recursive code
*
         N1 = MIN( M, N ) / 2
         N2 = N-N1
*
*               [ A11 ]
*        Factor [ --- ]
*               [ A21 ]
*
*        Indicate if input already checked, with no Infs and NaNs
         IF (WHAT .GE. 1 .AND. HOW .GE. 1) THEN
            INFO_SGETRF2_TMP1(7) = -1
            IF (INFO_ARRAY(7) .NE. -1) 
     $          INFO_SGETRF2_TMP1(7) = INFO_ARRAY(7)
         ENDIF
*
         CALL SGETRF2_EC( M, N1, A, LDA, IPIV, 
     $        IINFO, FLAG_REPORT_CALL, INFO_SGETRF2_TMP1, CONTEXT )
*   
         IF ( INFO.EQ.0 .AND. IINFO.GT.0 ) THEN
            INFO = IINFO
         ENDIF
*   
*        Check inputs, outputs and internal calls of
*        first call to SGETRF2_EC
         CALL CHECKCALL( FLAG_REPORT_INTERNAL, INFO_INTERNAL,
     $                   INFO_SGETRF2_TMP1, INFO_ARRAY, N+2, 8 )
*
*                              [ A12 ]
*        Apply interchanges to [ --- ]
*                              [ A22 ]
*
         CALL SLASWP( N2, A( 1, N1+1 ), LDA, 1, N1, IPIV, 1 )
*
*        Solve A12
*
         CALL STRSM( 'L', 'L', 'N', 'U', N1, N2, ONE, A, LDA,
     $               A( 1, N1+1 ), LDA )
*
*        Update A22
*
         CALL SGEMM( 'N', 'N', M-N1, N2, N1, -ONE, A( N1+1, 1 ), LDA,
     $               A( 1, N1+1 ), LDA, ONE, A( N1+1, N1+1 ), LDA )
*
*        Factor A22
*
*        Indicate that input not already checked
         IF (WHAT .GE. 1 .AND. HOW .GE. 1) INFO_SGETRF2_TMP2(7) = -1
*
         CALL SGETRF2_EC( M-N1, N2, A( N1+1, N1+1 ), LDA, IPIV(N1+1),
     $         IINFO, FLAG_REPORT_CALL, INFO_SGETRF2_TMP2, CONTEXT )
*   
*        Check inputs, outputs and internal calls of
*        second call to SGETRF2_EC
         CALL CHECKCALL( FLAG_REPORT_INTERNAL, INFO_INTERNAL,
     $                   INFO_SGETRF2_TMP2, INFO_ARRAY, N+3, 9 )
*
*        Adjust INFO and the pivot indices
*
         IF ( INFO.EQ.0 .AND. IINFO.GT.0 )
     $      INFO = IINFO + N1
         DO 20 I = N1+1, MIN( M, N )
            IPIV( I ) = IPIV( I ) + N1
   20    CONTINUE
*
*        Apply interchanges to A21
*
         CALL SLASWP( N1, A( 1, 1 ), LDA, N1+1, MIN( M, N), IPIV, 1 )
*
      END IF
*
*     Check for errors before returning
*
      IF (WHAT.EQ.-1) RETURN
*     Check for exceptional outputs in A
      CALL SGECHECKARG(FLAG_REPORT_INTERNAL, M, N, A, LDA, 
     $                 INFO, INFO_INTERNAL, INFO_ARRAY, 3, 3, N+1, 7)
*
*     Update INFO and INFO_ARRAY
      CALL UPDATE_INFO( INFO, INFO_ARRAY, INFO_INTERNAL )
      IF (CALL_REPORT_EXCEPTIONS .AND. INFO .NE. 0) 
     $   CALL REPORT_EXCEPTIONS(CONTEXT, 7, ROUTINE_NAME, INFO_ARRAY)
      RETURN
*
*     End of SGETRF2_EC
*
      END
\end{verbatim}

\newpage
\begin{verbatim}
*> \brief \b SGETRS_EC
*
*  =========== DOCUMENTATION ===========
*
* Online html documentation available at
*            http://www.netlib.org/lapack/explore-html/
*
*  Definition:
*  ===========
*
*       SUBROUTINE SGETRS_EC( TRANS, N, NRHS, A, LDA, IPIV, B, LDB, 
*      $                     INFO, FLAG_REPORT, INFO_ARRAY, CONTEXT )
*
*       .. Scalar Arguments ..
*       CHARACTER          TRANS
*       INTEGER            INFO, LDA, LDB, N, NRHS
*       ..
*       .. Array Arguments ..
*       INTEGER            IPIV( * )
*       INTEGER            FLAG_REPORT( 2 ), INFO_ARRAY( 7 )
*       REAL               A( LDA, * ), B( LDB, * )
*       ..
*       .. Pointer Arguments ..
*       POINTER            CONTEXT ... advice requested
*
*
*> \par Purpose:
*  =============
*>
*> \verbatim
*>
*> SGETRS_EC solves a system of linear equations
*>    A * X = B  or  A**T * X = B
*> with a general N-by-N matrix A using the LU factorization computed
*> by SGETRF_EC.
*>
*> SGETRS_EC also provides new exception handling and 
*> reporting capabilities.
*>
*> \endverbatim
*
*  Arguments:
*  ==========
*
*> \param[in] TRANS
*> \verbatim
*>          TRANS is CHARACTER*1
*>          Specifies the form of the system of equations:
*>          = 'N':  A * X = B  (No transpose)
*>          = 'T':  A**T* X = B  (Transpose)
*>          = 'C':  A**T* X = B  (Conjugate transpose = Transpose)
*> \endverbatim
*>
*> \param[in] N
*> \verbatim
*>          N is INTEGER
*>          The order of the matrix A.  N >= 0.
*> \endverbatim
*>
*> \param[in] NRHS
*> \verbatim
*>          NRHS is INTEGER
*>          The number of right hand sides, i.e., the number of columns
*>          of the matrix B.  NRHS >= 0.
*> \endverbatim
*>
*> \param[in] A
*> \verbatim
*>          A is REAL array, dimension (LDA,N)
*>          The factors L and U from the factorization A = P*L*U
*>          as computed by SGETRF.
*> \endverbatim
*>
*> \param[in] LDA
*> \verbatim
*>          LDA is INTEGER
*>          The leading dimension of the array A.  LDA >= max(1,N).
*> \endverbatim
*>
*> \param[in] IPIV
*> \verbatim
*>          IPIV is INTEGER array, dimension (N)
*>          The pivot indices from SGETRF; for 1<=i<=N, row i of the
*>          matrix was interchanged with row IPIV(i).
*> \endverbatim
*>
*> \param[in,out] B
*> \verbatim
*>          B is REAL array, dimension (LDB,NRHS)
*>          On entry, the right hand side matrix B.
*>          On exit, the solution matrix X.
*> \endverbatim
*>
*> \param[in] LDB
*> \verbatim
*>          LDB is INTEGER
*>          The leading dimension of the array B.  LDB >= max(1,N).
*> \endverbatim
*>
*> \param[out] INFO
*> \verbatim
*>          INFO is INTEGER
*>          INFO is defined below depending on FLAG_REPORT
*> \endverbatim
*>
*> \param[in] FLAG_REPORT
*> \verbatim
*>          FLAG_REPORT is INTEGER array, dimension(2)
*>          FLAG_REPORT(1) defines what kinds of exceptions to report 
*>          using INFO and possibly also INFO_ARRAY for more details.
*>          FLAG_REPORT(1)
*>           <= -1 turns off all error checking, so INFO=0 is 
*>                 returned.
*>            =  0 does standard argument checking:
*>                 INFO = 0  means successful exit
*>                 INFO = -i means the i-th (non-floating point) 
*>                           argument had an illegal value 
*>                           (first error found is reported)
*>                 Using INFO to report the above errors has priority 
*>                 over reporting any of the errors described below. 
*>                 More generally, an error that would be found with 
*>                 a lower value of FLAG_REPORT(1) has priority to 
*>                 report using INFO than an error that would only
*>                 be found with a higher value of FLAG_REPORT(1).  
*>            =  1 also checks for Infs and NaNs in inputs and 
*>                 outputs, if INFO is not already nonzero:
*>                 INFO = -4 means A contained an Inf or NaN on 
*>                           input; execution continues.
*>                 INFO = -7 means B contained an Inf or NaN on 
*>                           input but A did not; 
*>                           execution continues.
*>                 INFO = 1  means B contained an Inf or NaN on 
*>                           output but neither A nor B did on input. 
*>                 Since A is an input variable, it is not checked
*>                 on output.
*>                 Each input and output will also be checked if 
*>                 FLAG_REPORT(2) = 1, 2, or 3 and reported in 
*>                 INFO_ARRAY as described below.
*>           >=  2 has the same behavior as 1, since there are no 
*>                 internal calls to LAPACK routines with INFO 
*>                 parameters to be checked.
*>
*>          FLAG_REPORT(2) defines how to report the exceptions 
*>          requested by FLAG_REPORT(1).
*>            If FLAG_REPORT(1) <= -1, FLAG_REPORT(2) is ignored and 
*>            INFO=0 is returned. 
*>            Otherwise, FLAG_REPORT(2)
*>             <=  0 only returns the value of INFO described above.
*>              =  1 also returns INFO_ARRAY, as described below.
*>              =  2 means that SGETRS_EC will also call 
*>                   REPORT_EXCEPTIONS to report INFO_ARRAY, if INFO 
*>                   is nonzero.
*>              =  3 has the same behavior as 2. (If SGETRS called
*>                   any LAPACK routines with INFO parameters 
*>                   internally then they would call
*>                   REPORT_EXCEPTIONS too, but there are none.)
*>             >=  4 means that SGETRS_EC will call 
*>                   GET_FLAGS_TO_REPORT(CONTEXT,FLAG_REPORT) 
*>                   to get values of FLAG_REPORT to use, overriding 
*>                   input values. The user needs to have called 
*>                   SET_FLAGS_TO_REPORT(CONTEXT,FLAG_REPORT) 
*>                   before calling SGETRS_EC in order to set 
*>                   FLAG_REPORT, otherwise the default is 
*>                   FLAG_REPORT = [0, 0]. The input array
*>                   FLAG_REPORT will not be overwritten.
*> \endverbatim
*>
*> \param[in,out] INFO_ARRAY
*> \verbatim
*>          INFO_ARRAY is INTEGER array, dimension( 8 )
*>          If FLAG_REPORT(1) <= -1 or FLAG_REPORT(2) <= 0, 
*>          INFO_ARRAY is not accessed. Otherwise:
*>          INFO_ARRAY(1)
*>              = value of INFO from standard argument checking 
*>                (as defined by FLAG_REPORT(1) = 0)
*>          INFO_ARRAY(2)
*>              = value of FLAG_REPORT(1) used to determine the rest 
*>                of INFO_ARRAY
*>          INFO_ARRAY(3)
*>              = value of FLAG_REPORT(2) used to determine the rest 
*>                of INFO_ARRAY
*>          INFO_ARRAY(4)
*>              = value of INFO as specified by FLAG_REPORT(1) above
*>          INFO_ARRAY(5)
*>              = 2 = number of input/output arguments reported on
*>          INFO_ARRAY(6)
*>              = 0 = number of internal LAPACK calls reported on
*>          INFO_ARRAY(7) reports exceptions in A, as specified by 
*>            FLAG_REPORT
*>              = -1 if not checked (default)
*>              =  0 if checked and contains no Infs or NaNs
*>              =  1 if checked and contains an Inf or NaN on input
*>            If INFO_ARRAY(7) = 0 or 1 on input, then A will not
*>            be checked again on input. Input values < -1 or > 1 
*>            will be treated the same as -1, i.e. not checked.
*>          INFO_ARRAY(8) reports exceptions in B, as specified by
*>            FLAG_REPORT
*>              = -1 if not checked (default)
*>              =  0 if checked and contains no Infs or NaNs
*>              =  1 if checked and contains an Inf or NaN on input
*>                   but not output
*>              =  2 if checked and contains an Inf or NaN on output 
*>                   but not input
*>              =  3 if checked and contains an Inf or NaN on input 
*>                   and output
*>            As above, if INFO_ARRAY(8) = 0 or 1 on input, then B
*>            will not be checked again on input. Input values < -1
*>            or > 1 will be treated the same as -1, 
*>            i.e. not checked.
*> \endverbatim
*>
*> \param[in] CONTEXT
*> \verbatim
*>           CONTEXT is POINTER to an "opaque object"
*> \endverbatim
*
*  Authors:
*  ========
*
*> \author Univ. of Tennessee
*> \author Univ. of California Berkeley
*> \author Univ. of Colorado Denver
*> \author NAG Ltd.
*
*> \ingroup realGEcomputational
*
*  =====================================================================
      SUBROUTINE SGETRS_EC( TRANS, N, NRHS, A, LDA, IPIV, B, LDB, 
     $                      INFO, FLAG_REPORT, INFO_ARRAY, CONTEXT )
*
*  -- LAPACK computational routine --
*  -- LAPACK is a software package provided by Univ. of Tennessee,    --
*  -- Univ. of California Berkeley, Univ. of Colorado Denver and NAG Ltd..--
*
*     .. Scalar Arguments ..
      CHARACTER          TRANS
      INTEGER            INFO, LDA, LDB, N, NRHS
*     ..
*     .. Array Arguments ..
      INTEGER            IPIV( * )
      REAL               A( LDA, * ), B( LDB, * )
      INTEGER            FLAG_REPORT( 2 ), INFO_ARRAY( * )
*     ..
*     .. Pointer Arguments
      POINTER            CONTEXT
*  =====================================================================
*
*     .. Parameters ..
      REAL               ONE
      PARAMETER          ( ONE = 1.0E+0 )
      CHARACTER, DIMENSION(6), PARAMETER :: 
     $   ROUTINE_NAME = (/ 'S','G','E','T','R','S' /)
*     ..
*     .. Local Scalars ..
      LOGICAL            NOTRAN, CALL_REPORT_EXCEPTIONS
      INTEGER            WHAT, HOW
*     ..
*     .. Local Arrays ..
      INTEGER            FLAG_REPORT_INTERNAL(2), FLAG_REPORT_CALL(2)    
      INTEGER            INFO_INTERNAL( 2 )
*     ..
*     .. External Functions ..
      LOGICAL            LSAME
      EXTERNAL           LSAME
*     ..
*     .. External Subroutines ..
      EXTERNAL           SLASWP, STRSM, XERBLA
      EXTERNAL           CHECKINIT1, CHECKINIT2
      EXTERNAL           SGECHECKARG
      EXTERNAL           UPDATE_INFO
*     ..
*     .. Intrinsic Functions ..
      INTRINSIC          MAX, MIN
*     ..
*     .. Executable Statements ..
*
*     Test the input parameters.
*
      INFO = 0
*     Initialize error checking flags
      CALL CHECKINIT1(FLAG_REPORT, FLAG_REPORT_INTERNAL,  
     $                FLAG_REPORT_CALL, CALL_REPORT_EXCEPTIONS, 
     $                CONTEXT ) 
      WHAT = FLAG_REPORT_INTERNAL( 1 )
      HOW = FLAG_REPORT_INTERNAL( 2 )     
      IF (WHAT .EQ. -1 ) GOTO 100
*
*     Check for standard input errors
      NOTRAN = LSAME( TRANS, 'N' )
      IF( .NOT.NOTRAN .AND. .NOT.LSAME( TRANS, 'T' ) .AND. .NOT.
     $    LSAME( TRANS, 'C' ) ) THEN
         INFO = -1
      ELSE IF( N.LT.0 ) THEN
         INFO = -2
      ELSE IF( NRHS.LT.0 ) THEN
         INFO = -3
      ELSE IF( LDA.LT.MAX( 1, N ) ) THEN
         INFO = -5
      ELSE IF( LDB.LT.MAX( 1, N ) ) THEN
         INFO = -8
      END IF
*
*     Initialize error flags
      CALL CHECKINIT2( FLAG_REPORT_INTERNAL, INFO, INFO_INTERNAL, 
     $                 INFO_ARRAY, 2, 0)  
*
      IF( INFO.NE.0 ) THEN
         IF (CALL_REPORT_EXCEPTIONS) 
     $      CALL REPORT_EXCEPTIONS(CONTEXT,6,ROUTINE_NAME,INFO_ARRAY)
         CALL XERBLA( 'SGETRS', -INFO )
         RETURN
      END IF
*
*     Quick return if possible
*
      IF( N.EQ.0 .OR. NRHS.EQ.0 )
     $   RETURN
*
*     Check for exceptional inputs in A 
      CALL SGECHECKARG(FLAG_REPORT_INTERNAL, N, N, A, LDA, 
     $                 INFO, INFO_INTERNAL, INFO_ARRAY, 4, 0, 0, 7)
*     Check for exceptional inputs in B 
      CALL SGECHECKARG(FLAG_REPORT_INTERNAL, N, NRHS, B, LDB, 
     $                 INFO, INFO_INTERNAL, INFO_ARRAY, 7, 2, 0, 8)
*
100   CONTINUE
*
*     Quick return if possible
*
      IF( N.EQ.0 .OR. NRHS.EQ.0 )
     $   RETURN
*
      IF( NOTRAN ) THEN
*
*        Solve A * X = B.
*
*        Apply row interchanges to the right hand sides.
*
         CALL SLASWP( NRHS, B, LDB, 1, N, IPIV, 1 )
*
*        Solve L*X = B, overwriting B with X.
*
         CALL STRSM('Left', 'Lower', 'No transpose', 'Unit', N, NRHS,
     $               ONE, A, LDA, B, LDB )
*
*        Solve U*X = B, overwriting B with X.
*
         CALL STRSM( 'Left', 'Upper', 'No transpose', 'Non-unit', N,
     $               NRHS, ONE, A, LDA, B, LDB )
      ELSE
*
*        Solve A**T * X = B.
*
*        Solve U**T *X = B, overwriting B with X.
*
         CALL STRSM( 'Left', 'Upper', 'Transpose', 'Non-unit', N,
     $               NRHS, ONE, A, LDA, B, LDB )
*
*        Solve L**T *X = B, overwriting B with X.
*
         CALL STRSM( 'Left', 'Lower', 'Transpose', 'Unit', N, NRHS,
     $               ONE, A, LDA, B, LDB )
*
*        Apply row interchanges to the solution vectors.
*
         CALL SLASWP( NRHS, B, LDB, 1, N, IPIV, -1 )
      END IF
*
*     Check for errors before returning
*
      IF (WHAT .EQ. -1) RETURN
*     Check for exceptional outputs in B 
      CALL SGECHECKARG(FLAG_REPORT_INTERNAL, N, NRHS, B, LDB, 
     $                 INFO, INFO_INTERNAL, INFO_ARRAY, 7, 3, 1, 8)  
*
*     Update INFO and INFO_ARRAY
      CALL UPDATE_INFO( INFO, INFO_ARRAY, INFO_INTERNAL )
      IF (CALL_REPORT_EXCEPTIONS .AND. INFO .NE. 0) 
     $   CALL REPORT_EXCEPTIONS(CONTEXT, 6, ROUTINE_NAME, INFO_ARRAY)
      RETURN
*
*     End of SGETRS_EC
*
      END
\end{verbatim}

\newpage
\bibliographystyle{unsrt}
\bibliography{twrdcnstexcphndlnmrl}

\end{document}